\documentclass{elsarticle}

\let\today\relax
\makeatletter
\def\ps@pprintTitle{%
    \let\@oddhead\@empty
    \let\@evenhead\@empty
    \def\@oddfoot{\footnotesize\itshape
         {Submitted preprint} \hfill\today}%
    \let\@evenfoot\@oddfoot
    }
\makeatother

\usepackage{graphicx} 
\usepackage{array}
\usepackage{amsmath}
\usepackage{amssymb}
\usepackage{amsfonts}
\usepackage{dsfont}
\usepackage{dcolumn}
\newcolumntype{d}[1]{D{.}{.}{#1}}
\usepackage{bbm}
\usepackage[dvipsnames]{xcolor}
\usepackage{url}
\usepackage{booktabs}
\usepackage{hyperref}
\usepackage{siunitx}
\usepackage{float}
\DeclareSIUnit\MWh{MWh}
\DeclareSIUnit\GWh{GWh}

\hypersetup{breaklinks,hidelinks}

\usepackage[belowskip = -10pt, labelfont=bf]{caption}

\begin{document}

\title{Evaluating Advanced Nuclear Fission Technologies for Future Decarbonized Power Grids}

\author{Emilio Cano Renteria}
\ead{renteria@alumni.princeton.edu}
\author{Jacob A. Schwartz\corref{cor1}\fnref{aff2}}
\ead{jschwart@pppl.gov}
\author{Jesse Jenkins\fnref{aff1}}
\ead{jdj2@princeton.edu}

\affiliation[aff2]{ 
    organization = {Princeton Plasma Physics Laboratory},
    postcode={08540},
    city={Princeton, NJ},
    country={USA.}
}
\affiliation[aff1]{
    organization={Andlinger Center for Energy and the Environment, Princeton University},
    postcode={08540},
    city={Princeton, NJ},
    country={USA.}
}
\cortext[cor1]{Corresponding author}

\begin{abstract}
Advanced nuclear fission, which encompasses various innovative nuclear reactor designs, could contribute to the decarbonization of the United States electricity sector.
However, little is known about how cost-competitive these reactors would be compared to other technologies, or about which aspects of their designs offer the most value to a decarbonized power grid.
We employ an electricity system optimization model and a case study of a decarbonized U.S.\ Eastern Interconnection circa 2050 to generate initial indicators of future economic value for advanced reactors and the sensitivity of future value to various design parameters, the availability of competing technologies, and the underlying policy environment.
These results can inform long-term cost targets and guide near-term innovation priorities, investments, and reactor design decisions.
We find that advanced reactors should cost \$\qty{5.7}{\per\watt}--\$\qty{7.3}{\per\watt}
to gain an initial market share (assuming 30 year asset life and \qtyrange{3.5}{6.5}{\percent} real weighted average cost of capital), while those that include thermal storage in their designs can cost up to \$\qty{6.0}{\per\watt}--\$\qty{7.7}{\per\watt} (not including cost of storage).
Since the marginal value of advanced fission reactors declines as market penetration increases, break-even costs fall $\sim$32\% at 100 GW of cumulative capacity and $\sim$51\% at 300 GW.
Additionally, policies that provide investment tax credits for nuclear energy create the most favorable environment for advanced nuclear fission.
These findings can inform near-term resource allocation decisions by stakeholders, innovators and investors working in the energy technology sector.
\end{abstract} 
\maketitle

\section{Introduction}
There is an imperative to reduce emissions from the energy sector to mitigate future climate change~\cite{RN8287, ipcc2023}.
In the United States, fossil fuel sources currently supply more than half of the country's electrical power.
There exist several ``pathways'' toward decarbonization that the U.S.\ could pursue, each of which includes different mixes of clean technologies.
There is general consensus in the literature that these pathways will all likely include large shares of solar and wind energy complemented by short-duration battery storage (e.g. lithium-ion batteries) and some level of demand-side flexibility, but there is no consensus about what other resource can be used to balance out the limitations of weather-dependent, variable renewable energy sources and energy-constrained storage \cite{sepulveda_role_2018, larson_net-zero_2021, motalebi_role_2022, sepulveda_design_2021, ricks_value_2022}.
This other resource must be firm (that is, able to generate power whenever needed, for as long as needed) and must have net-zero emissions to allow the U.S.\ to reach full decarbonization.
Firm, zero-carbon resources include nuclear, geothermal, natural gas with full carbon capture and storage, and combustion or oxidation of hydrogen or other zero-carbon or carbon-neutral fuels \cite{sepulveda_role_2018}, and they may be complemented by (and partially substituted with) certain energy storage technologies with very low-cost of storage capacity that are techno-economically suitable for multi-day discharge periods \cite{sepulveda_design_2021}. 

Of these resources, nuclear energy has historically been the most widespread in the U.S., providing about 20\% of the country's total generation \cite{energy_information_administration_nuclear_2022}.
However, investment into new nuclear projects has been stagnant, and the Energy Information Administration (EIA) expects this to remain the case in the future because new nuclear power plants are not currently cost-competitive with other zero-carbon sources of energy \cite{energy_information_administration_annual_2023}.
Nuclear innovation companies are seeking to develop and commercialize newer, so-called “advanced” reactor designs that are potentially less costly and/or offer greater value by bringing improvements in reactor safety, efficiency, and flexibility over traditional designs~ \cite{nuclear_innovation_alliance_nia_advanced_2023}.
In this paper, we perform an economic analysis of these advanced nuclear fission technologies to evaluate whether they can serve as cost-competitive alternatives to other firm, zero-carbon resources in a decarbonizing electricity system.
Specifically, we employ an electricity system optimization model and a case study of a decarbonized U.S.\ Eastern Interconnection circa 2050 to generate initial indicators of future economic value for advanced reactors and the sensitivity of future value to various design parameters, the availability of competing technologies, and the underlying policy environment.
These results can inform long-term cost targets and guide near-term innovation priorities, investments, and reactor design decisions.

\subsection{Advanced nuclear fission}
We focus on advanced nuclear fission reactors which bring potential improvements over traditional designs in four main categories: cost, size, flexibility, and efficiency. Table \ref{tab:adv} summarizes the different types of advanced nuclear reactor designs currently under development. 

\subsubsection{Cost}
For the reasons outlined previously, any advanced reactor that wishes to be competitive must prioritize lowering its costs in order to incentivize investment. This cost reduction can be achieved primarily through innovations in design (discussed in this section) or reductions in reactor size (discussed in the following section). 

The BWRX-300, an advanced reactor by GE Hitachi, exemplifies a simplified nuclear reactor design focused centrally on reducing costs. The BWRX-300 is a simplified version of the tried-and-tested boiling water reactor (BWR), thereby bypassing many of the regulatory obstacles of approving a nuclear reactor \cite{office_of_nuclear_energy_first_2020}.
Compared to traditional BWRs, the simplified design of the BWRX-300 aims to reduce material and land area costs \cite{office_of_nuclear_energy_first_2020}. 

Improvements in the safety design of advanced reactors have the potential to reduce costs as well.
Passive safety mechanisms make it physically impossible for reactors to reach the temperatures necessary for a nuclear meltdown by implementing cooling through natural processes such as convection \cite{glaser_small_2015} and gravity \cite{nuclear_innovation_alliance_nia_advanced_2023}.
Additionally, some reactors, like the NuScale VOYGR small modular pressurized-water reactor, are designed to be contained underground to reduce exposure risk.
These passive safety mechanisms can reduce costs because they may require less redundancy in design, less frequent and intense monitoring by safety staff, and easier licensing and approval \cite{world_nuclear_association_small_2023, chris_charles_nrc_2018}.

Finally, innovations in the design of the complete nuclear plant (reactor and turbine) are being pursued to reduce costs as well. As coal plants shut down across the United States due to poor economics, their steam turbine infrastructure can be re-purposed to pair with advanced reactors, eliminating the need to finance a new turbine alongside a new reactor. The Department of Energy estimates that over 80\% of coal power plants in the U.S.\ could be re-used for advanced reactors, totaling \qty{263}{GWe} of capacity --- more than double the nuclear capacity that exists today \cite{nuclear_innovation_alliance_resources_2023}. In addition, reactors built at retired coal plants would qualify as sited in ``energy communities'' which have traditionally relied on fossil fuels and where additional federal subsidies can be obtained for new clean energy projects \cite{nuclear_innovation_alliance_resources_2023}. TerraPower, developer of the Natrium sodium-cooled fast reactor, is currently planning to construct its first plant at the site of a coal power plant in Wyoming \cite{terrapower_terrapower_2023}.

\subsubsection{Size} 
In addition to design improvements, reductions in size can also help lower the cost of new reactors. The typical nuclear reactor in the U.S.\ has about \qty{1000}{MW} of capacity \cite{energy_information_administration_nuclear_2022}. The costs and timelines for constructing a power plant of this size are much larger than for smaller generators and are often subject to great uncertainty when the project begins. For example, the recent construction of two AP1000 reactors at Plant Vogtle in Georgia exceeded the predicted timeline by seven years, and construction costs more than doubled initial estimates \cite{amy_georgia_2023}. Similar excessive delays and cost overruns have also been experienced at recent large-scale reactor projects in Europe. These uncertainties around construction cost and timelines raise the cost of capital for nuclear projects and have contributed to the reduced investment into new gigawatt-scale nuclear reactors in the United States \cite{department_of_energy_quadrennial_2015}. Small modular reactors (SMRs), the name given to advanced designs with capacities of approximately \qty{300}{MW} or smaller, offer compact and standardized designs that may allow for factory construction of many of their important components, leaving only the task of assembly to be performed on site \cite{nuclear_innovation_alliance_nia_advanced_2023}. This ultimately allows for more streamlined and simplified construction, potentially reducing risk and increasing predictability and thus lowering capital costs and their associated financing. Furthermore, their standardized and modular designs allow for serial production and deployment of multiple individual small units at single sites to form larger capacity generators (e.g. at many 100s or 1000s of MW scale) while facilitating local learning-by-doing (e.g. cumulative experience of local engineering and construction labor and supply chains). Design standardization and repeated construction at multi-unit plants appear to have been critical factors behind lower historical nuclear construction costs experienced in certain countries and time periods \cite{lovering_historical_2016, mit_energy_initiative_future_2018}.
\begin{table}[t]
    \centering
    \begin{small}
    \begin{tabular}[t]{>{\raggedright}p{0.20\linewidth}>{\raggedright}p{0.30\linewidth}>{\raggedright\arraybackslash}p{0.38\linewidth}}
         \toprule
         \textbf{Reactor Type} & \textbf{Description}  & \textbf{Value} \\
         \midrule
        Microreactors & Reactors with capacities smaller than \qty{10}{MWe}. & Decarbonization of remote communities and natural disaster response. \\
        Small modular reactors & Reactors with capacities between \qty{10}{MWe} and \qty{300}{MWe}. & Streamlined reactor construction pipeline with potentially lower costs, faster timelines, and less construction finance risk. \\
        Fast reactors & Reactors with no moderator that have faster moving neutrons. & Higher fuel efficiency and longer refueling cycles leading to lower operational costs. \\
        Non-water-cooled reactors & Reactors with metal or gas coolants instead of water. & Higher operating temperatures leading to greater turbine efficiency and co-generation opportunities.\\
        \bottomrule
    \end{tabular}
    \end{small}
    \caption{Advanced nuclear reactor designs and the value they would bring to a decarbonized power grid \cite{nuclear_innovation_alliance_nia_advanced_2023, nichol_cost_2019, smr_start_economics_2021, glaser_small_2015}.}
    \label{tab:adv}
\end{table}
\subsubsection{Flexibility}
To better integrate into power grids with high variable renewable energy (VRE) penetration, many advanced fission reactors are being developed with increased flexibility as a primary goal. Some designs incorporate this flexibility by allowing for faster ramp rates: the speeds at which power output can be increased or decreased in the core of the nuclear reactor \cite{nuclear_innovation_alliance_nia_advanced_2023}. Faster ramping would allow advanced reactors to better respond to quick and unpredictable fluctuations in VRE generation and potentially contribute to important ancillary services like frequency regulation and operating reserves \cite{jenkins_benefits_2018}.
\par Other advanced reactor designs are instead pursuing flexibility by coupling their reactors with thermal storage, allowing heat from the nuclear core to be diverted to storage as necessary. This allows the nuclear core to operate at a steady state, while providing flexibility in electrical production by controlling how much heat goes to the storage and the turbine. For example, at times when VRE generation is high, more heat can flow to the thermal storage medium and less to the turbine, reducing total electrical output, while the opposite can be done to increase power generation. Molten salts have already been used to store thermal energy and increase flexibility for other resources such as concentrated solar power \cite{gonzalez-roubaud_review_2017}, and they have also been proposed for some reactor designs, such as the TerraPower Natrium reactor. Ceramic `firebrick' \cite{stack_performance_2019}, crushed rock \cite{forsberg_low-cost_2023}, and other thermally conductive materials can also be used to store heat from nuclear power plants.  

\subsubsection{Efficiency}
The overall efficiency of a nuclear power plant is defined as the amount of electricity produced for every unit of fuel consumed.
For nuclear energy, this overall efficiency depends on the efficiencies of both the core and the turbine. 

The efficiency of the core is defined by the amount of heat that can be extracted per unit of fuel.
Some advanced reactors run on High-Assay Low-Enriched Uranium (HALEU) fuel, which allows them to reach higher core efficiencies than traditional reactors running on Low-Enriched Uranium (LEU) fuel \cite{glaser_small_2015}.

On the other hand, the efficiency of the turbine is defined by the amount of electricity generated for every unit of heat consumed. For thermodynamic reasons, this value is largely dependent on the temperature of the fluid used to power the turbine.
Traditional reactors use water as a coolant and produce power using super-heated steam at temperatures of \qtyrange{300}{400}{\degreeCelsius}, corresponding to an efficiency of about 30\% \cite{department_of_energy_quadrennial_2015}.
However, advanced reactors are being designed with different coolants and working fluids that can reach higher temperatures than water/steam.
For example, molten salt or metal cooled reactors can achieve operating temperatures of \qtyrange{600}{700}{\degreeCelsius}, corresponding to around 40\% efficiency.
Even higher temperatures can be achieved using coolants such as helium gas, with so-called high-temperature gas-cooled reactors capable of reaching \qty{850}{\degreeCelsius}.
At these temperatures, conversion efficiency reaches \qtyrange{45}{48}{\percent} \cite{glaser_small_2015, guidez_fast_2023}.
These high efficiencies are beneficial because they reduce the amount of fuel necessary to produce electricity, lowering the variable costs of the reactor.
They can also lower the capital costs of a nuclear power plant by reducing the reactor capacity necessary to generate the same amount of electric power.

\subsection{Contribution of this work}
Although the potential benefits of advanced nuclear reactors are well-understood, few studies have assessed whether this added value improves their cost competitiveness against other firm, zero-carbon resources.
Previous studies that have considered the economics of advanced reactors have focused exclusively on SMRs, and they have simply estimated the isolated costs of these reactors, without considering how their value changes as they interact with other resources at a systems level \cite{glaser_small_2015, nichol_cost_2019, smr_start_economics_2021}. 

In this paper, we address this gap by using a detailed electricity system planning model and a case study representing the U.S.\ Eastern Interconnection circa 2050 to determine the break-even cost of advanced reactors in future decarbonized power grids, while accounting for variations in plant design, competing technologies, and policy environment.
We employ GenX, an open-source electricity system optimization model, to plan the least-cost capacity mix and capture the hourly interactions between different resources \cite{jenkins_enhanced_2017}.\footnote{See https://github.com/GenXProject/GenX/.} By minimizing the overall system cost and allowing the model to pick which resources to dispatch at hourly resolution, we can capture the break-even cost at which advanced reactors become economically competitive with other resources at different nuclear penetration levels (e.g. increasing capacity deployment).
Finally, we analyze how this price changes under different policies (e.g. stringency of CO$_2$ emissions limits or presence of tax incentives), parametric uncertainties (e.g. competing technology, fuel, and financing costs) and plant design decisions (e.g. operating flexibility, efficiency, and integration of thermal storage).
Estimated break-even costs can be used to establish cost targets that nuclear developers must meet if they wish their reactor designs to be competitive at scale, while understanding the sensitivity of break-even costs to reactor design parameters can help inform near-term innovation priorities and design decisions.
These results can also provide reference points for investors and government agencies considering investing in innovative nuclear fission concepts, projects, or companies.

\section{Methods}
The GenX electricity system model optimizes investment and dispatch decisions for generators, energy storage, demand-side flexibility, and inter-regional transmission in order to produce the lowest-cost system that can meet a given temporal and geographical demand profile \cite{jenkins_enhanced_2017}. It is highly configurable to different policy scenarios and technology mixes, allowing for thorough economic analyses of emerging clean technologies \cite{sepulveda_design_2021, ricks_value_2022, ricks_minimizing_2023, schwartz_value_2023}.

\subsection{Advanced fission plant representation}
\begin{figure}
\begin{center}
\includegraphics[width=\textwidth]{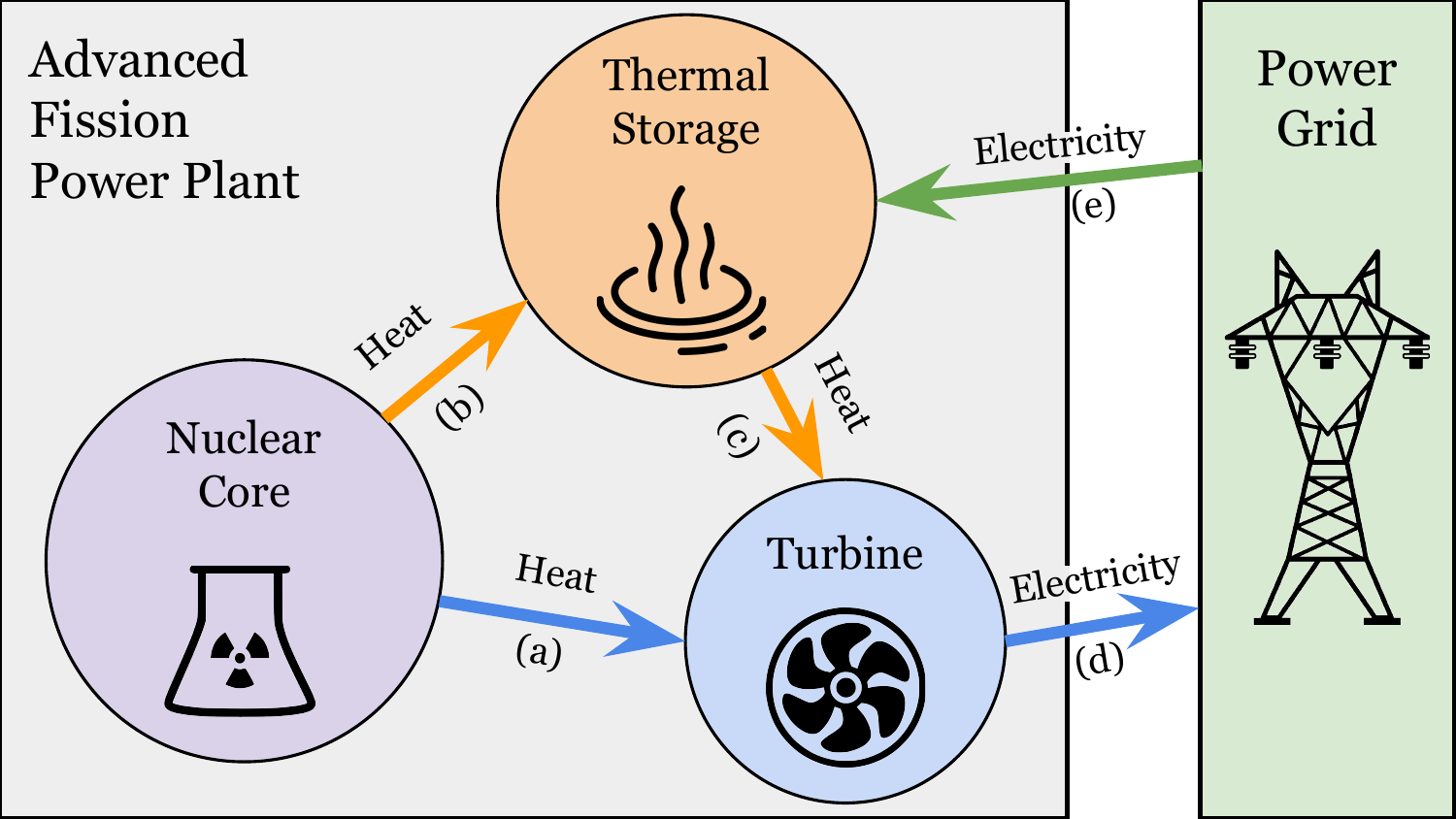}
\caption[Advanced Fission Plant GenX Implementation]{Schematic representation of energy flow in our advanced fission plant implementation in GenX. Unlike standard nuclear plants in GenX, this implementation separates the nuclear core from the power conversion system (glossed as ``turbine''), allowing for distinct operational parameters and coupled thermal storage. Arrows labeled (a, d) represent the flow in a reactor with no thermal storage. Arrows labeled (a, b, c, d) represent the flow in a reactor coupled with thermal storage. Arrows labeled (a, b, c, d, e) represent the flow in a reactor coupled with thermal storage and resistive heating.}\label{fig:schematic}
\end{center}
\end{figure}

To represent advanced fission reactors in GenX, we build upon an existing implementation for next-generation fusion power plants by Schwartz et al.\ \cite{schwartz_value_2023}. Figure~\ref{fig:schematic} provides a visual representation of how we treat energy flow in this implementation.
Instead of considering nuclear plants as only generators of electricity, as GenX does for other resources, we instead model the intermediate heat stage associated with nuclear electricity production; in our implementation, all fuel is first converted to heat, and then heat is converted to electricity.
This allows us to model different sizes of nuclear core and power conversion system (or ``turbine''), core ramping rates, varying conversion efficiencies, and the integration of thermal storage and resistive heating\footnote{ Resistive heating provides the ability to consume grid electricity when power prices are low and convert it into heat for storage in thermal storage media and then dispatch heat to the plant's generator to produce electricity when prices are higher.}.
In essence, we can independently model and co-optimize the capacity sizing and operations of each sub-component of the integrated plant.
In doing so, we can better capture the impacts of operating flexibility and the efficiency improvements associated with advanced reactors.
\subsubsection{Implementation in GenX}
The mathematical formulation of this advanced fission plant implementation consists of added decision variables and constraints to the GenX model.
The description in this section assumes a working familiarity with both constrained optimization electricity systems models as well as with the specific design of GenX. For a description of these concepts, refer to the GenX model definition \cite{jenkins_enhanced_2017}.

First, we defined non-negative decision variables for the amount of nuclear core, thermal storage, and resistive heating capacity to be built:
\begin{equation*}\label{eq:capvars}
    vCoreCap_g,\  vStorCap_s,\  vRHCap_h\ \ge 0 \quad \forall g \in G, s \in TS, h \in RH.
\end{equation*}
Here, G defines the set of all fission plants, TS defines the subset of those with coupled thermal storage, and RH defines the subset of those with resistive heating. Next, we added the costs associated with building this capacity:
\begin{equation*}\label{eq:capcosts}
\begin{split}
            eSystemCost \mathrel{+}=
    \sum_{g\in G} vCoreCap_g * cFixCoreCap_g + \\ \sum_{s\in TS} vStorCap_s * cFixStorCap_s + \sum_{h\in RH} vRHCap_h * cFixRHCap_h .
\end{split}
\end{equation*}
Decision variables for the turbine (including the generator block) already exist in GenX and need not be redefined. Next, we added hourly operational decision variables for the amount of heat produced in the core, the state of charge of the thermal tanks, and the amount of electricity used for resistive heating:
\begin{equation*}
    vCoreP_{g, t},\ vSOC_{s, t},\ vRH_{h, t}\ \ge 0 \quad \forall\ t \in T, g \in G, s \in TS, h \in RH.
\end{equation*}
These operational parameters were upper bounded by the amount of capacity built for each. Upper bound constraints for the nuclear core are defined by unit commitment constraints.
\begin{equation*}\label{eq:upper}
\begin{split}
        \forall t \in T: \quad vSOC_{s, t} \leq vStorCap_s, s\in TS;& \ 
        vRH_{h, t} \leq vRHCap_h,\ h\in RH.
\end{split}
\end{equation*}

We linked these thermal variables to the turbine already implemented in GenX and defined how energy can flow between the separate components.
For plants with no storage, energy can only flow from the core to the turbine, where it is converted to electricity:
\begin{equation*}
    vCoreP_{g, t} * pEff_g = vTurbP_{g, t} \quad \forall t \in T, g \in G, g \notin TS.
\end{equation*}
For those with storage, we define the flow of energy by the change in state of charge between two consecutive time periods. Heat is gained via heat from the core and resistive heating and lost via thermal heat decay and generation at the turbine:
\begin{equation*}
\begin{split}
        vSOC_{s, t} &= vSOC_{s, t-1}\ + \ vCoreP_{s, t}\ + \ vRH_{s, t} * pEffRH_s * \mathbbm{1}_{s \in RH} \\ &- \  vSOC_{s, t-1} * pDecay_s \;\; - \;\; vTurbP_{s, t} \;/\; pEff_s \quad \forall s \in TS.
\end{split}
\end{equation*}
In cases where resistive heating is used, we added this electrical load to the supply and demand balance at each location in the system:
\begin{equation*}
    ePowerBalance_{z, t}  \mathrel{-}=   \sum_{h \in RH} vRH_{h, t} * \mathds{1}_{h.zone \in z} \;\;\;\; \forall z \in Z, t \in T.
\end{equation*}
Finally, we added unit commitment and ramping constraints to the nuclear core in the same manner as these are applied to other generators in GenX. The mathematical definition for these constraints is available in the GenX documentation \cite{jenkins_enhanced_2017}.

\subsection{Modeling system setup}
This study models the Eastern Interconnection of the United States using 20 zones (Figure \ref{fig:map}). These zones are either identical to, or conglomerates of, those used in the Environmental Protection Agency (EPA) Integrated Planning Model (IPM) for the U.S.\ power system  \cite{environmental_protection_agency_documentation_2021}.
We focus on the Eastern Interconnection because it currently includes over 90\% of nuclear power capacity in the U.S., so it has historically been a more favorable environment for nuclear energy deployment, and it has larger potential for replacement (or re-powering) of retired reactors in the future \cite{energy_information_administration_nuclear_2022}. 

Each of the 20 zones is modeled as a ``copper plate.''
That is, perfect transmission is permitted within the zones, allowing for aggregation of generation and demand across the zone, although the costs of connecting different candidate renewable energy development sites to demand centers within each zone are estimated and included in the capital costs for renewable energy generators.
Transmission capacity sizing and power flows between different zones are explicitly modeled with existing inter-regional transfer capacity based on EPA's IPM model (2021 edition) \cite{environmental_protection_agency_documentation_2021}.
Where it deems economical, the model can expand transmission capacity, incurring associated transmission expansion costs. See \ref{sec:transmission} for transmission line data. 

We model the Eastern Interconnection system for the target year of 2050; this year is chosen to allow enough time for advanced reactors to reach commercial maturity.
Many of the advanced reactor designs currently under development will not be prototyped until the late 2020s and early 2030s \cite{nuclear_innovation_alliance_nia_advanced_2023}.
Thus, it is unreasonable to expect significant advanced reactor deployment by 2035, while 2050 permits sufficient time for considerable scale-up.

We model the entire 2050 year using hourly data (8760 total hours) for electricity demand in each zone, as well as capacity factor profiles for a total of 304 wind and 216 solar resource clusters spread out over the 20 zones.
We compile input data using PowerGenome, an open source tool that generates GenX-specific input data for a desired model configuration \cite{schivley_powergenomepowergenome_2022}.
Estimated demand growth data are from Princeton University's REPEAT (2024) project's mid-range scenario \cite{jenkins_repeat_2024}.
PowerGenome estimates electricity demand in the following manner: 
present-day demand, EV charging, electric space and water heating profiles are taken from the National Renewable Energy Laboratory (NREL) \textit{Electrification Futures Study} (EFS) \cite{mai_electrification_2020};
 the demand other than for EVs and heating equipment is scaled based on 2050 projections from the U.S. Energy Information Administration (EIA) \textit{Annual Energy Outlook 2022} Reference Case \cite{energy_information_administration_annual_2022};
then the EV~charging, space and water heating loads are re-added based on equipment/vehicle stock projections from the REPEAT (2024) study \cite{jenkins_repeat_2024} and the per unit profiles from the NREL~EFS.
Our final demand profile has an average hourly load of \qty{540}{GW} and a peak yearly load of \qty{954}{GW}. Full demand profiles are presented in \ref{sec:demand}.
PowerGenome determines the costs and maximum capacities of renewable clusters in each zone based on a methodology described in the Supplemental Information of Patankar et al \cite{patankar_landuse_2023}.
It further provides hourly VRE capacity factors for each wind and solar cluster based on the 2012 sample weather year (consistent with the meteorological year used in the NREL EFS for electric heating and EV loads).
The annual averages for each zone are plotted in \ref{sec:vre}.

\begin{figure}
\begin{center}
\includegraphics[width=\textwidth]{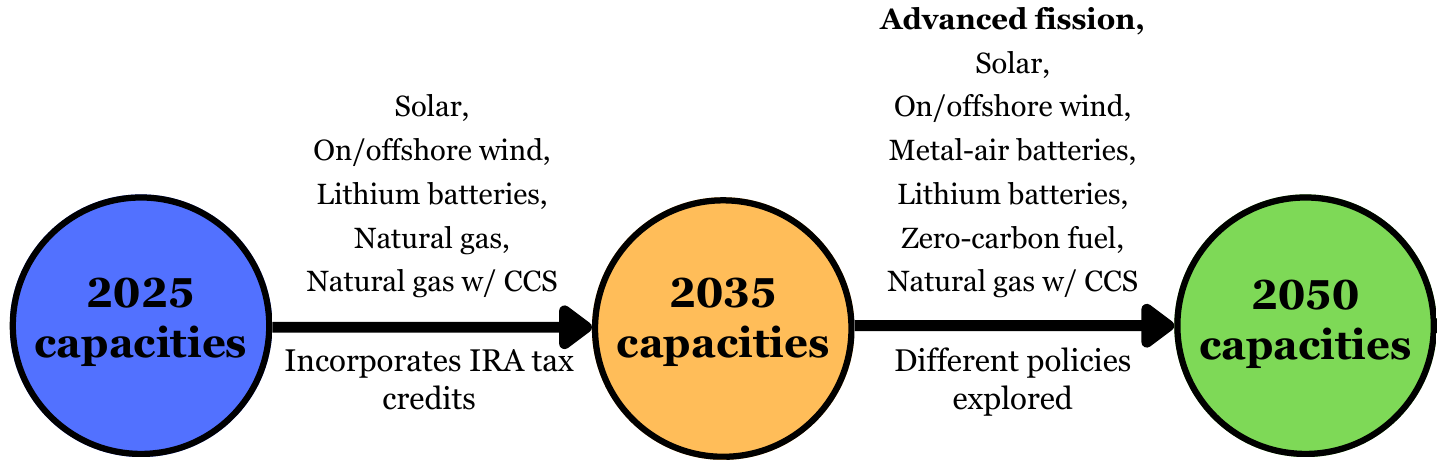}
\caption[Two Stage Capacity Expansion]{Two stage capacity expansion from 2025 to 2035 and from 2035 to 2050. Technologies included for each expansion are pictured, with the notable addition of advanced fission for the 2035 to 2050 expansion. From 2025 to 2035, IRA tax credits are included, and from 2035 to 2050 different policies are explored.}\label{fig:time}
\end{center}
\end{figure}

Furthermore, we model the build-out of capacity to meet 2050 demand in two stages (Figure \ref{fig:time}).
We first perform a capacity expansion from 2025 capacities to 2035, and we then use those 2035 capacities as starting points for the 2050 expansion.
The optimizations of each stage are independent of each other, so the first expansion only optimizes for 2035 costs, and the second only for 2050 costs.
For the existing resources in 2025, we include natural gas plants, coal plants, solar, wind, traditional nuclear, biomass generators, and both hydroelectric power and storage. Values for these capacities are sourced from the EIA Form 860 \cite{energy_information_administration_form_2023} by PowerGenome.
Out of this starting set of technologies, only some of them are eligible for new construction based on expected power system trends. These include solar PV, onshore wind, and natural gas. Notably, we do not allow new nuclear to be constructed between 2025 to 2035 under the assumption that all new nuclear construction will consist of advanced reactors in the 2035 to 2050 stage. While some first-of-a-kind reactors may be constructed prior to 2035, we omit this capacity for simplicity.

In addition to this existing technology mix, we consider new technologies for construction that do not currently have significant capacity penetration. These include lithium batteries, offshore wind, and natural gas combined cycle power plants with carbon capture and storage (NG-CCS).
Costs and operational parameters for these technologies come from the 2023 NREL Annual Technology Baseline (ATB) data set \cite{national_renewable_energy_laboratory_annual_2023}, with the NG-CCS data adjusted to 100\% carbon capture as in Schwartz et al.\ \cite{schwartz_value_2023}.
Technology-related cost parameters are listed in \ref{sec:techcosts}. 

To model current policy incentives for decarbonized technologies, we include Production Tax Credits (PTCs) and Investment Tax Credits (ITCs) that were signed into law with the 2022 Inflation Reduction Act (IRA). All new sources of carbon-free electricity generation can elect either a PTC awarded on a per-MWh-produced basis or an ITC based on capital investment costs; both of these policies will remain active under current law for projects that commence construction before the end of 2033 or the year after U.S. electricity sector CO$_2$ emissions fall to 25\% of 2022 levels, whichever comes later \cite{environmental_protection_agency_inflation_2023}. As such, we assume these credits are available for all carbon-free generators coming online before 2035.
Energy storage devices are also eligible for the ITC during this period. These tax incentives are modeled by lowering technology production and investment costs respectively while adjusting for their net-present value and a transferability cost of 7.5\% of the tax credit value (see \ref{sec:ira}). Additionally, the 2035 expansion stage includes current state renewable portfolio standards as well as the offshore wind mandates enacted in several Atlantic states \cite{national_conference_of_state_legislatures_state_2021}, but these state policies are omitted in the 2050 expansion, as full decarbonization is enforced instead. Finally, capacity reserve margins, which require available generation and storage capacity to exceed peak demand by a set planning reserve margin, are implemented to ensure system reliability (\ref{sec:reserves}).

Once the model determines the optimal generator and transmission capacity amounts for each technology in 2035, it uses these values as the starting capacities for the 2035 to 2050 expansion, with two exceptions.
First, lithium battery capacity is reset to zero at the start of the 2035 to 2050 expansion because their lifetimes are assumed to be around 15 years, so they must be entirely replaced in the 2050 expansion \cite{hesse_lithium-ion_2017}. Second, nuclear plants that are set to retire before 2050 (assuming a 60 year plant lifetime) have their capacities removed in order to model the need to replace them.
Finally, compared to the 2025 to 2035 expansion, we consider three new technologies for construction.
First, we include advanced nuclear plants, the focus of this study. Second, we model low-cost metal-air batteries suitable for multi-day energy storage \cite{wang_materials_2019}, and we source their operational and cost parameters from Baik et al.\ \cite{baik_what_2021}.
Finally, in decarbonized scenarios, we switch new natural gas plants built in 2025--2035 to burn a generic zero-carbon fuel (ZCF) as an alternative to building CCS, while retiring pre-existing coal and natural gas plants.
In practice, this ZCF could be hydrogen, biomethane (renewable natural gas), or synthetic methane, or it could continue being natural gas with fully offset emissions using direct air capture of CO$_2$.
For the purposes of this study, we consider the ZCF to be green hydrogen (with no embodied emissions), and we source future prices from Princeton University's Net-Zero America study \cite{larson_net-zero_2021}.
In the non-decarbonized policy scenarios, non-CCS natural gas plants can be built during this stage, as in 2025--2035.
Capacities of each technology in 2025 and in 2035, for both fully- and non-fully-decarbonized scenarios, are listed in~\ref{sec:caps}.

For reference, Table \ref{tab:inputmaps} maps each input to its location in the Appendix.
\begin{table}[h]
    \centering
    \caption{Inputs to the electricity system model are listed in different Appendices.}\label{tab:inputmaps}
    \begin{small}
    \begin{tabular}{>{\raggedright}p{0.60\linewidth}>{\raggedright\arraybackslash}p{0.20\linewidth}}
        \toprule
        \textbf{Input} & \textbf{Appendix} \\
        Geographical zones & \ref{sec:zones} \\
        Transmission & \ref{sec:transmission} \\
        Demand & \ref{sec:demand} \\
        VRE availability factors & \ref{sec:vre} \\
        Technology and fuel costs & \ref{sec:techcosts} \\
        2025 \& 2035 initial capacities & \ref{sec:caps} \\
        IRA policy credits & \ref{sec:ira} \\ 
        Capacity reserve margins & \ref{sec:reserves} \\
        \bottomrule
    \end{tabular}
    \end{small}
\end{table}

\subsection{Experimental design}\label{sec:scenarios}
To understand the value of advanced fission plants under different scenarios, we first define a reference case, calculate the break-even cost for plants in that reference case, and then analyze how that break-even cost changes in different scenarios.
We calculate the break-even cost of a plant by modeling the core with zero upfront capital costs and restricting the model to build no more than a certain capacity of advanced reactor core.
We then examine the dual value associated with this maximum nuclear capacity constraint (also referred to as the Lagrangian multiplier or ``shadow price'').
For linear programming problems (as a corollary of strong duality), the dual value of any constraint represents how much the objective function---in this case overall system cost---would \emph{lower} if we relax the right hand side of the constraint by one unit---in this case allowing the system to build one more megawatt of the free advanced reactor core.
We then add this dual value to the annualized cost of the same capacity of turbines to find the annualized value of a `base plant'.
So long as the annualized cost of a base plant (not including the cost of thermal storage or turbine capacity in excess of that matching the core capacity) is less than this value, deployment of additional capacity at the level of the constraint would lower total system costs, annualized costs equal to the dual value would break even, and costs greater than the dual value would increase total system costs at this level of deployment. 

Thus, we can use the dual value of the constraint on total reactor capacity to derive the break-even cost (or marginal value) of advanced fission plant deployment at a given capacity penetration level.
This same approach has been used in prior studies to represent the marginal value or break-even cost of different resources and is useful when the total cost of a technology is speculative \cite{sepulveda_design_2021, schwartz_value_2023, mallapragada_long-run_2020, mai_setting_2019}.
In effect, we `solve' backwards for the break-even cost required to reach a given penetration level rather than assume a speculative cost and then identify the optimal deployment level.
We repeat this at different capacity penetrations to see how reactor valuation erodes as the system becomes more saturated with nuclear.
More specifically, we run our model at 5, 10, 20, 40, 80, 150, 225, and 300 GWe nuclear capacity constraints, loosely following a geometric distribution.
For reference, 300 GW represents 55\% of the system's average load, and 31\% of its peak load.

Note that throughout the figures in this paper, we communicate this break-even cost both as an annual cost for the lifetime of a `base plant' (typically, left axes of our plots), which is the actual value derived from the dual and matching turbine costs, and an equivalent total upfront capital cost (typically, right axes of our plots)\footnote{Even when they are optimal to include in the plants, we do not include the costs of excess turbine capacity nor thermal storage in our plots or reporting in the text.
This would complicate the interpretation, because (a) the excess turbine capacity ratio and thermal storage duration vary (and generally decrease; see Fig.~\ref{fig:ratios}) with increasing fission capacity, so the marginal quantities of turbine or storage are different from their system-wide average values, and also because (b) fission plant costs in the literature rarely include these features.}.
The annual cost includes the construction investment annuity of the plant as well as its fixed operations and maintenance costs.
The total capital cost represents the justifiable upfront capital cost for the nuclear plant assuming fixed O\&M costs comprise 2.5\% of the total capital cost, and also assuming an asset life of 30 years and a real weighted average cost of capital of 4.5\% \cite{national_renewable_energy_laboratory_annual_2023}.
Given uncertainty about future nuclear plant financing costs, we also report in the text, where relevant, a range of capital cost targets corresponding to real WACC values of \qtyrange{3.5}{6.5}{\percent}. Table \ref{tab:convert} provides a means for easily converting our reported total capital costs (at 4.5\% real WACC and 30 year asset life) to any other WACC value in the range of \qtyrange{1}{10}{\percent} and any asset life from \numrange{20}{60} years. 

Our reference case consists of an advanced reactor with ramp rates typical of a standard reactor, but also coupled with thermal storage.
We assume a non-water-cooled reactor design capable of reaching 40\% conversion efficiency.
We assume a decarbonized electrical system by 2050.
We also assume IRA tax credits are not available for marginal projects coming online in the \numrange{2035}{2050} planning stage (e.g. assuming these credits expire circa 2035). Finally, for fuel and technology costs, we assume ``moderate'' projected values sourced from the NREL ATB 2023.
Advanced reactor costs and operational assumptions, which could not be obtained from PowerGenome, are included in Tables \ref{tab:coreinputs}--\ref{tab:turbinputs}.

\begin{table}
    \centering
    \caption[Assumptions for Nuclear Core]{The values assumed for all costs and operational parameters for the reference case nuclear core.}\label{tab:coreinputs}
    \begin{small}
    \begin{tabular}{>{\raggedright}p{0.4\linewidth}>{\raggedright}p{0.1\linewidth}>{\raggedright\arraybackslash}p{0.25\linewidth}}
        \toprule
        \textbf{Parameter} & \textbf{Value} & \textbf{Unit}\\
        \midrule
        \multicolumn{3}{l}{\textbf{Operational:}} \\
        Min stable power output & 50 & \% of rated capacity\\
        Ramp rate per hour & 25 & \% of rated capacity\\
        Min up commitment time & 12 & hours \\
        Min down commitment time & 12 & hours \\
        Heat rate & 3.45 & MMBtu / MWh$_\mathrm{th}$ \\
        \multicolumn{3}{l}{\textbf{Costs:}} \\
        Uranium Fuel Cost & 0.73 & \$ / MMBtu \\
        Variable O\&M & 0.16 & \$ / MWh$_{\textrm{th}}$ \\
        Startup cost & 178 & \$ / MW$_{\textrm{th}}$\\
        \bottomrule
        \vspace{1cm}
    \end{tabular}
    \end{small}
\end{table}
\medskip
\begin{table}
    \centering
    \caption[Assumptions for Thermal Storage]{The values assumed for all costs and operational parameters for the thermal storage and resistive heating.}\label{tab:storinputs}
    \begin{small}
    \begin{tabular}{>{\raggedright}p{0.45\linewidth}>{\raggedright}p{0.1\linewidth}>{\raggedright\arraybackslash}p{0.25\linewidth}}
        \toprule
        \textbf{Parameter} & \textbf{Value} & \textbf{Unit}\\
        \midrule
        \multicolumn{3}{l}{\textbf{Operational:}} \\
        Heat decay rate & 0.1 & \% lost per hour \\
        Resistive heating efficiency & 1 & MWh$_{\textrm{th}}$/MWh$_e$\\
        \multicolumn{3}{l}{\textbf{Costs:}} \\
        Thermal storage investment annuity & 1,845 & \$ / (MWh$_{\textrm{th}} \cdot$ yr)\\
        Resistive heating investment annuity & 2,935 & \$ / (MW$_{\textrm{e}} \cdot$ yr)\\
        \bottomrule
    \end{tabular}
    \end{small}
\end{table}
\begin{table}
    \centering
    \caption[Assumptions for Turbine]{The values assumed for all costs and operational parameters for the reference case turbine. FO\&M represents Fixed Operation and Maintenance Costs.}\label{tab:turbinputs}
    \begin{small}
    \begin{tabular}{>{\raggedright}p{0.4\linewidth}>{\raggedright}p{0.1\linewidth}>{\raggedright\arraybackslash}p{0.25\linewidth}}
        \toprule
        \textbf{Parameter} & \textbf{Value} & \textbf{Unit}\\
        \midrule
        \multicolumn{3}{l}{\textbf{Operational:}} \\
        Min stable power output & 20 & \% of rated capacity\\
        Hourly ramp rate & 64 & \% of rated capacity\\
        Min up commitment time & 6 & hours \\
        Min down commitment time & 6 & hours \\
        Efficiency & 0.4 & MWh$_e$ / MWh$_{\textrm{th}}$ \\ 
        \multicolumn{3}{l}{\textbf{Costs:}} \\
        Investment annuity & 83,752 & \$ / (MW$_{e} \cdot$ yr)\\
        FO\&M annuity & 34,106 & \$ / (MW$_{e} \cdot$ yr) \\
        VO\&M & 1.74 & \$ / MWh$_{e}$ \\
        Startup cost & 100 & \$ / MW$_{e}$\\
        \bottomrule
    \end{tabular}
    \end{small}
\end{table}

From this reference case, we consider variations in the following scenarios:
\begin{enumerate}[a)]
    \item Flexibility:
    We evaluate the added value from coupled thermal storage by considering reactors without storage.
    We analyze the potential for resistive heating to increase the value of reactors with storage.
    Finally, we consider the effect of having unlimited ramping capabilities on reactor value compared to typical ramp rates.
    \item Efficiency: We vary from the reference salt-cooled reactor efficiency of 40\% by considering traditional Light Water Reactor (LWR) efficiencies of 33\% \cite{department_of_energy_quadrennial_2015}, as well as highly efficient gas-cooled reactor efficiencies of 50\% \cite{guidez_fast_2023}.
    \item Policy: We consider three different policy scenarios in addition to the reference case.
    One considers the continuation of IRA policy (clean energy tax credits) for the 2035 to 2050 expansion.
    The other two cases consider partial decarbonization: a 90\% or 95\% reduction in emissions compared to 2005 levels, the year in which U.S.\ CO$_2$ emissions peaked \cite{congressional_budget_office_emissions_2022}.

    \item Fuel costs: We model the advanced reactor value sensitivity to fuel costs of both natural gas as well as nuclear fuel. For natural gas, we model both high and low cost scenarios by scaling reference costs up by 75\% or down by 25\% respectively (equivalent to EIA's 2021 Annual Energy Outlook cost predictions for low and high oil and gas supply scenarios \cite{energy_information_administration_annual_2021}). For nuclear fuel, only higher price scenarios are considered to represent the price of HALEU fuel if widespread domestic HALEU fuel production capacity is not established. We implement a ``high uranium cost'' scenario of 2.5x higher than base costs which we source from HALEU cost predictions in a Centrus report~\cite{centrus_status_2021}. Because HALEU costs are highly speculative, we model an additional ``very high uranium cost'' scenario with costs scaled 5 times higher.
    \item Technology costs: We analyze the advanced reactor value sensitivity to variations in the investment costs of other technologies. For all other technologies considered in the model, we vary their capital costs up and down by 10\% and 20\%, noting the resulting change in reactor value. We also consider the cost of molten salt thermal energy storage, and vary it in the same manner, to analyze its effect on the valuation of reactors with thermal storage. Finally, we consider different real WACC rates, from 3.5\% to 6.5\%.
\end{enumerate}
For each experiment we comment on how fission's competition with other resources is altered; and particularly on those resources for which the capacities or operations change most compared with the reference scenario.

\section{Results}
\subsection{Substitution}\label{sec:subst}

\begin{figure}[p]
\begin{center}
\includegraphics[width=\textwidth]{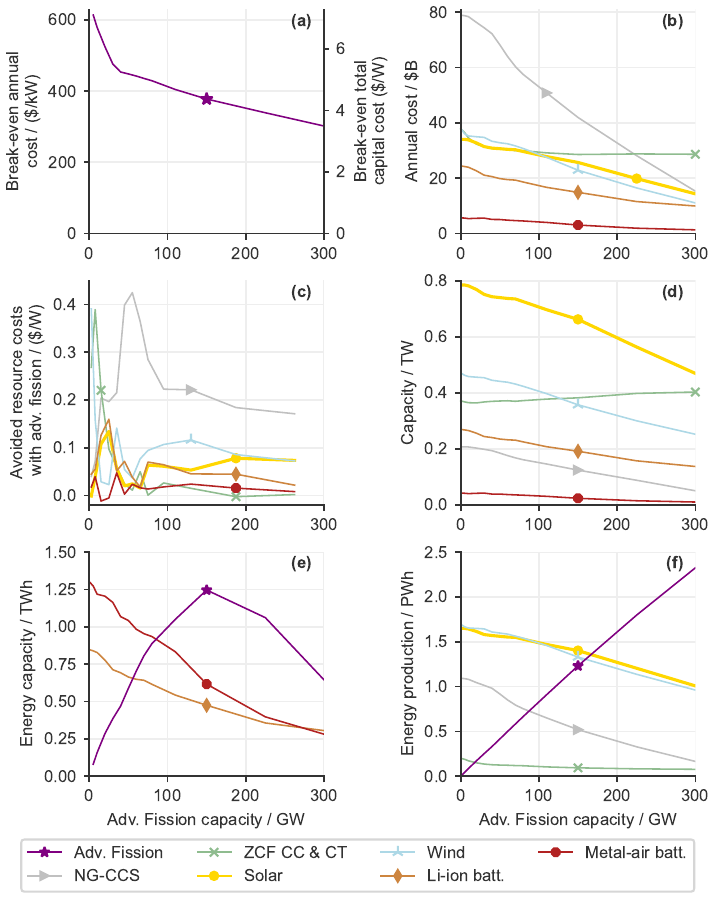}
\caption[Value of adv fission]{Part (a): marginal value of an advanced fission `base plant'.
The left axis expresses the break-even cost as an annuity, while right axis expresses it as a total capital cost (assuming 30 year asset life, 4.5\% WACC).
Part (b): Annual costs (investment, FO\&M, and production) of selected resource types.
Part (c): Changes in these costs with increasing advanced fission capacity. The marginal value of advanced fission (a) is equal to the sum of these avoided costs (plus avoided costs from expansion of transmission lines, which are small).
Parts (d), (e), (f): power capacity, energy storage capacity, and energy production, respectively of these resources.}\label{fig:displacement}
\end{center}
\end{figure}

The value of an advanced fission plant derives from its ability to displace capacity and energy costs associated with other resources in the system and therefore reduce total system cost.
Figure~\ref{fig:displacement}a shows the marginal value of an advanced fission plant as a function of fission's capacity penetration, while part~(b) shows the annual costs associated with selected resources, and part (c) shows the gradient of these costs with respect to the capacity penetration of advanced fission;  this directly shows which resources' displacement contributes to the marginal value of advanced fission at different capacity penetrations.
The marginal value declines steeply below \qty{30}{GW} as advanced fission displaces high-cost wind, solar, batteries, and the use of zero-carbon fuels.
At higher capacities, displacement of the latter supports most of advanced fission's value; displacement of wind, solar, and batteries are also significant contributors.
Parts~(d), (e) and (f) show how the presence of advanced fission changes the optimal power capacity, energy storage capacity, and overall energy production by the various resources, respectively.
From \qty{50}{GW} to \qty{300}{GW}, each GW of advanced fission displaces \qty{1.1}{GW} of solar,
\qty{0.75}{GW} of wind, \qty{0.5}{GW} of NG-CCS, as well as \qty{1.3}{GWh} of Li-ion battery storage and \qty{3}{GWh} of metal-air battery storage.
Part~(e) shows that advanced fission's thermal storage capacity peaks at \qty{150}{GW} of power capacity; past this point the need for storage would be displaced by the additional plants themselves.
Parts~(d) and~(f) show that while the capacity of plants burning zero-carbon fuels slowly rises (at about \qty{0.1}{GW} for each GW of advanced fission), the power generated by them falls; this indicates that they are being used to satisfy the system's capacity reserve margin constraints.
Part~(f) shows that advanced fission's largest competitor in energy production is NG-CCS followed by solar and wind.
Advanced fission becomes the largest producer at \qty{160}{GW} of capacity penetration.

Figures~\ref{fig:capmapThermalRTS20}--\ref{fig:capmapThermalRTS300} show the zones where advanced fission plants are constructed.
They are first built in the PJM-MACC zone.
This zone has a high population density so sites for onshore wind or utility-scale solar are highly constrained, and the geology does not permit using NG-CCS.
Beyond PJM-MACC at low capacity penetrations they are generally built along the eastern seaboard, especially in the Northeast.
At higher capacities, advanced fission is built further to the west and north, but generally not where wind dominates in the Great Plains.
Advanced fission also decreases the need for new transmission in 2035--2050, especially along east-west links.

\subsection{Flexibility}\label{sec:devflex}
\subsubsection{Thermal Storage}

\begin{figure}
\begin{center}
\includegraphics[width=\textwidth]{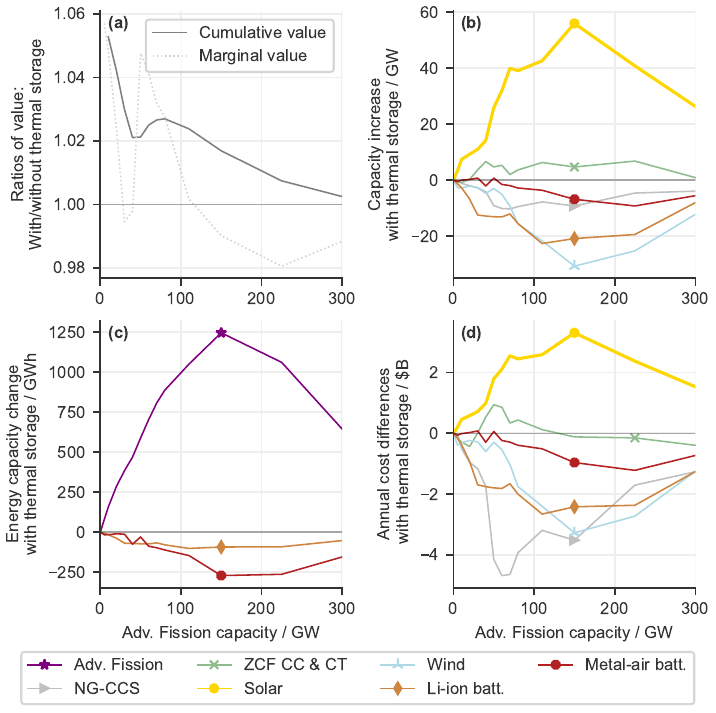}
\caption[Changes without thermal storage]{(a) Ratios of the break-even cost (marginal value) and the cumulative value---the integral of the marginal value---of the advanced fission plant with and without the option to build thermal storage as a function of the advanced fission capacity penetration.
Parts (b), (c), and (d): changes in the power capacity, energy storage capacity, and costs of selected resources with the option to build thermal storage.
}\label{fig:stor}
\end{center}
\end{figure}

In our reference configuration, advanced fission plants are equipped with thermal storage.
Here we explore the value provided by thermal storage by comparing cases without it.
We find that thermal storage adds value to advanced nuclear reactors, especially at low capacity penetrations.
Figure~\ref{fig:displacement}a shows the marginal value of plants with thermal storage, and Fig.~\ref{fig:stor}a shows the ratio of the marginal values with thermal storage to those without it.

At a low capacity penetration (\qty{5}{GW}) the break-even cost of an advanced reactor with thermal storage is \$7.1/W (\$6.0/W--\$7.7/W over 3.5\%--6.5\% WACC), while that for a reactor with no thermal storage is \$6.7/W (\$5.7/W--\$7.3/W). The former is about 5.5\% higher than the latter.
Fig.~\ref{fig:stor}a shows that for some capacity penetrations the marginal value of additional capacity in a fleet of plants with thermal storage is \textit{lower} than if thermal storage was not permitted.
This is not a paradox; the option to build storage always increases the cumulative value (the integral of the marginal value from 0 to $x$ GWe) provided by the fleet at a given capacity.
One can consider some of the value of an advanced fission fleet to be shifted from being captured at a higher capacity to a lower capacity; new value is created as well.

\begin{figure} \begin{center}
\includegraphics[width=\textwidth]{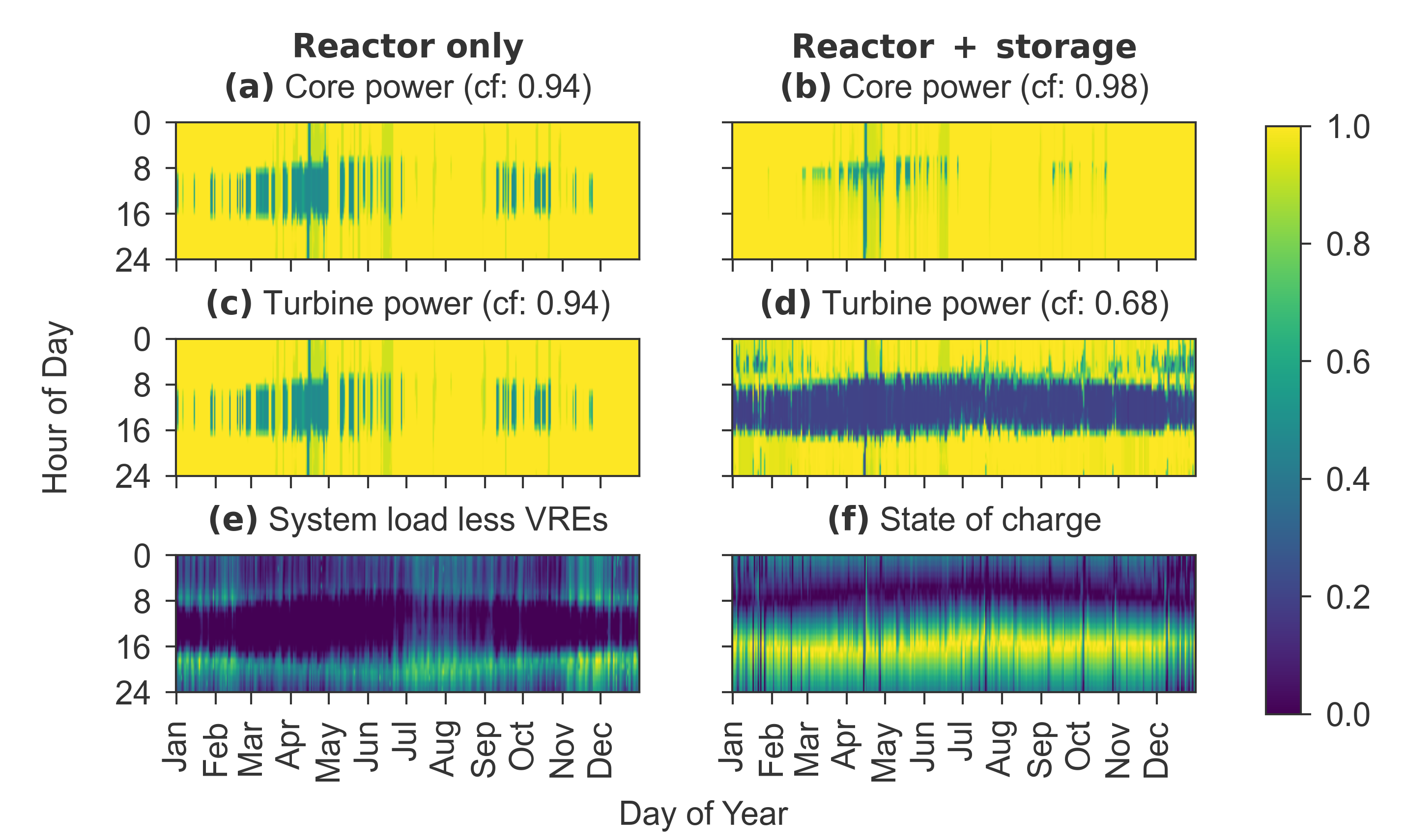}
	\caption[Advanced Reactor Operational Profile (20 GW)]{Heat maps illustrating advanced reactor operational profiles at 20 GW of nuclear capacity penetration. (a--b) Normalized nuclear core power output for all hours of the year, as well as capacity factors, for reactors with and without thermal storage. (c--d) Normalized turbine electrical power output for all hours of the year, as well as capacity factors, for reactors with and without thermal storage. (e) Normalized system load at all hours of the year, subtracting VRE generation. (f) Normalized state of charge for the thermal tanks at all hours of the year. For all graphs, we use an advanced nuclear capacity penetration of 20 GW, but similar patterns are observed at other capacities (Figures \protect{\ref{fig:op5}} - \protect{\ref{fig:op225}}).}\label{fig:ops}
\end{center} \end{figure}

The additional value for advanced reactors with thermal storage derives from their ability to store thermal energy during periods of low electricity prices and dispatch greater power generation during periods of higher prices, all while operating the reactor at a higher and more consistent utilization rate (Fig.~\ref{fig:ops}).

Fig.~\ref{fig:stor}b shows that this flexibility enables advanced fission plants with thermal storage to displace more wind, Li-ion batteries, metal-air batteries, and NG-CCS than plants without thermal storage; more solar and ZCF-burning plants are built instead.
Part (c) shows that the energy capacity of thermal storage in the advanced fission plants is much larger than the additional displaced energy storage of Li-ion and metal-air batteries; note that the efficiency of conversion to power on the grid is 40\% rather than 92\% and 65\% for Li-ion and metal-air batteries, respectively.
Part (d) shows that the additional cumulative value of plants with thermal storage is driven by displacement of NG-CCS, wind, and batteries, but more investment is made in solar.
Comparing these differences in capacities and costs with their absolute values (shown in Fig.~\ref{fig:stor}) we see that the changes can be larger than the change in cumulative value of the advanced fission plants themselves: for example, a 10\% lower total capacity of wind and 8\% higher solar capacity at \qty{150}{GW} of advanced fission, even though the cumulative value of advanced fission plants is less than 2\% larger with the option to build thermal storage.

To analyze why thermal storage displaces more batteries and enables additional solar, we consider the operational behavior of the nuclear plant.
Figure~\ref{fig:ops} demonstrates the difference in nuclear power plant operational behavior for plants with (right) and without storage (left).
For power plants without storage, the turbine power output (Fig.~\ref{fig:ops}c) is forced to match the core power output (Fig.~\ref{fig:ops}a), leading to high capacity factors of 94\% for both.
Despite the high capacity factor, the availability of zero marginal cost VRE regularly pushes net system load to zero during daytime solar production peaks (Fig.~\ref{fig:ops}e), requiring the nuclear core to ramp its power output significantly, while unit commitment constraints on reactor start-up and shut-down also force the turbine to continue producing even at times when system load is low (July and August, for example).
This production when demand is low causes VRE generation to be curtailed and does not contribute to the revenue or marginal value of reactors as marginal electricity prices are zero during periods of renewables curtailment.
On the other hand, nuclear power plants with thermal storage can better match the system demand profile.
At \qty{20}{\giga\watt} of core capacity, the fleet-average turbine is over-sized by a factor 1.42 and the plant has storage to hold 5.7 hours of heat production from the core.
The cores are able to operate at a higher capacity factor of 98\% and ramp less throughout the year.
Since they are over-sized, the turbines have a lower capacity factor of 68\%, (despite a higher average energy throughput).
Their power production resembles the system demand profile more closely (Figure~\ref{fig:ops}d), improving reactor economics.
Most of the plants' electrical energy production is shifted from periods with low or zero electricity prices (e.g. daytime hours with high solar production) to periods of higher value (when renewable energy production is insufficient to meet demand), increasing reactor revenues.
The (fleet-wide average) state of charge of the thermal tanks explains how the plants are able to achieve this (Fig.~\ref{fig:ops}f).
In the morning hours, thermal storage is fully discharged.
Throughout the day when demand is low, the thermal storage charges with heat from the nuclear cores, reaching a peak in the afternoon hours.
Then, thermal storage is discharged to meet the evening rise in demand.
This ability to shift power generation throughout the day to match net system demand serves the same function as lithium batteries and allows the plants to operate in synergy with solar; this increases the value of solar and decreases the value of batteries.

We find that the optimal amount of thermal storage built begins at \qty{6.4}{MWh} per MW of core thermal power for designs at low capacities, and decreases to \qty{1}{MWh} per MW of core power at high penetrations (Fig.~\ref{fig:ratios}).
Optimal turbine capacity sizing follows a similar trend, from 1.5 times larger than the core thermal power capacity at initial penetrations to 1.1 times larger at high penetrations (Fig.~\ref{fig:ratios}).
These are important design parameters that advanced nuclear reactor developers can consider when sizing the different components of their power plant.

\subsubsection{Resistive Heating}
We study whether advanced reactors with integrated thermal storage might additionally employ resistive heating during periods of low electricity prices to `charge' the thermal storage tank.
We find that the plant break-even costs do not increase by more than 0.1\% with the option to build resistive heaters at \$35/kW$_\mathrm{e}$.
If the three components (resistors, thermal storage, and excess turbine capacity) were significantly less expensive it could be economically feasible to form a cluster which would operate semi-independently from the fission reactor.

\subsubsection{Increased operational flexibility}
\begin{figure}
\begin{center}
\includegraphics[width=\textwidth]{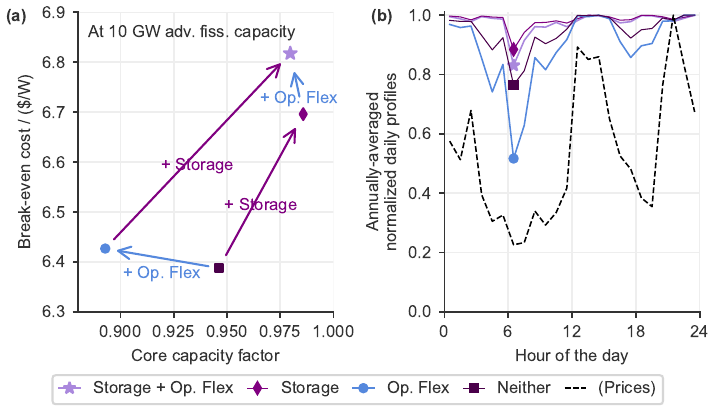}
\caption[Changes with flexible operation]{Part (a): operational flexibility (faster ramping, no minimum power, no minimum commitment duration) increases the break-even cost while decreasing the core capacity factor. The increase in marginal value is up to 2\% for plants with thermal storage, but \textless 1\% for plants without thermal storage. Part (b): Normalized daily average generation from advanced fission as well as the system-wide average price. Operational flexibility allows plants to avoid operation during hours when prices are lowest.
}\label{fig:opflex}
\end{center}
\end{figure}
The final element of flexibility we consider is an improved ramp rate, no minimum power level, and reduced commitment times for the core and power conversion system.
For designs with thermal storage, having a plant that can operate freely in this manner increases economic value by up to 2\%, about half as much as thermal storage, as shown in Fig.~\ref{fig:opflex}.
For designs not incorporating thermal storage, increasing operational flexibility increases the value by \textless 1\%.
The increased flexibility is used to avoid producing power when the cost of operation exceeds the price of electricity; it therefore decreases the plant's capacity factor while increasing its value.
Thermal storage alone increases value more than operational flexibility alone, but the two features work together: flexibility in the power conversion system enhances the value of plants with thermal storage because the generation avoided during the lowest-priced periods can be better shifted to higher-priced periods.
Note that this study does not consider operating reserve provision.
Prior work has estimated that reactors ramping down during periods of low electricity prices could use their spare capacity to provide ancillary services (frequency regulation and operating reserves) that could modestly increase annual revenues (on the order of 2--5\%), although ancillary service markets are relatively small and thus only a handful of reactors at most could benefit from these additional revenues \cite{jenkins_benefits_2018}.

\subsection{Efficiency}\label{sec:deveff}
Advanced reactors using non-water coolants can achieve higher input turbine temperatures and thus increased power generation efficiency.
We find that improved reactor efficiency modestly increases the marginal value of advanced reactors at each level of penetration.
Across capacity penetration levels, gas-cooled reactors with 50\% conversion efficiency consistently exhibit about 3\% higher marginal value than salt-cooled reactors with 40\% efficiency and 6\% greater value than traditional water-cooled reactors at 33\% efficiency.
As shown in Fig.~\ref{fig:eff}, the increased marginal value at higher efficiency comes mostly (roughly 3/4) from lowering the plant's fuel and variable operations \& maintenence costs; an increase in the thermal storage capacity (due the lower effective cost of storage) and subsequent displacement of other resources from the grid makes up the balance.
Thus, efficient reactors lower their variable costs while keeping their revenues equal, making them more lucrative investments, even if they provide little to no additional operational value to the grid.

\subsection{Policy}
\begin{figure}
\begin{center}
\includegraphics[width=\textwidth]{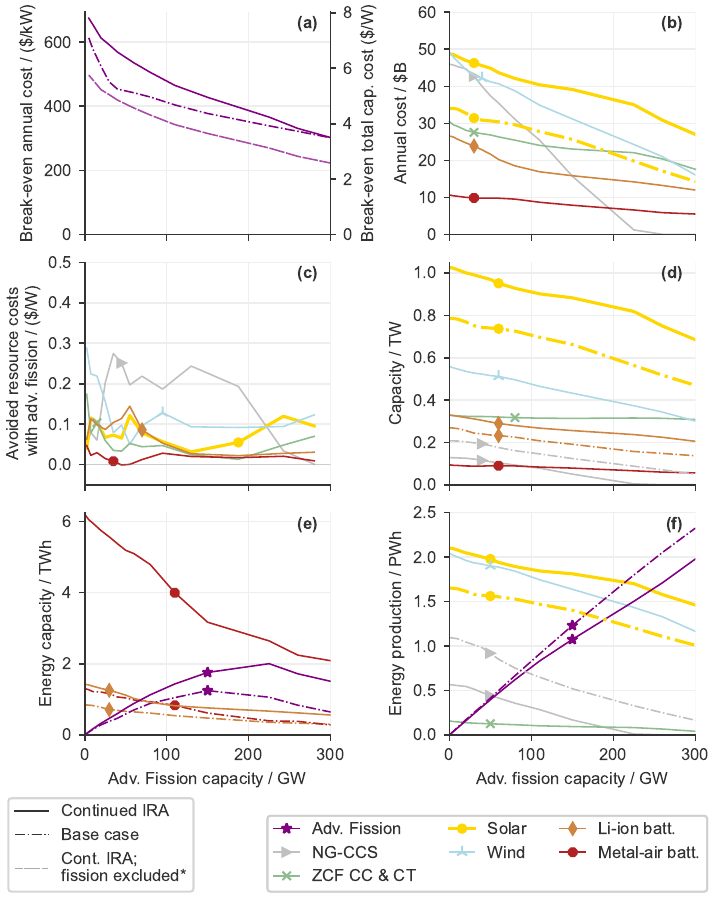}
	\caption[System under IRA Policy]{The break-even cost of advanced fission and the wider energy system under continued IRA policies. Dot-dashed lines are copied from the reference case (Fig.~\protect{\ref{fig:displacement}}) for comparison of selected quantities. The long-dashed line in Part (a) is an approximation to the value of fission in a continued-IRA scenario where fission was excluded from subsidies. }\label{fig:policyira}
\end{center}
\end{figure} 
We consider the effects of continuing the tax credits established by the Inflation Reduction Act of 2022 (IRA) from 2035 to 2050 under a fully decarbonized scenario, and alternately, different levels of incomplete decarbonization by 2050.
The IRA extends the production tax credit (PTC, worth \$27.5/MWh in 2022 USD) and investment tax credit (ITC, worth 30\% of eligible capital costs) for all carbon-free generation sources and offers generators the choice of whichever credit is more lucrative.
Additionally, bonus incentives increase the value of the PTC by 10\% and the value of the ITC by 10 percentage points if projects are built in qualifying `energy communities' or employ sufficient domestic content (these bonuses can be combined).
We assume that nuclear reactors qualify for one of the two bonus incentives and elect an ITC worth 40\% of capital costs (as this delivers greater net-present value than the PTC).
We calculate a break-even cost under IRA
with the ITC applied to the investment costs of the plant and including a transferability cost of 7.5\% of the tax credit value; see  \ref{sec:ira}.
We find that the maximum break-even cost of our reference case reactor under this extended tax credits scenario increases by 10\% at low penetrations to \$7.8/W, by up to 25\% (at \qty{40}{GW}) while this increase in value declines to about 1\% at \qty{300}{GW} (Figure~\ref{fig:policyira}a).

The increase in the break-even cost under IRA is much smaller than would be expected from the size of the ITC because it applies only to investment costs, not operations and maintenance costs, and because IRA also subsidizes competing clean energy resources.
The relative advantage for those resources is larger than that of advanced nuclear, which has significant non-subsidized operations, maintenance and fuel costs.

IRA shifts capacity and spending away from the non-subsidized but firm resources (NG-CCS and ZCF-burning plants) and toward the subsidized variable and storage resources (solar, wind, Li-ion and metal-air batteries).
For example, at zero advanced fission capacity, the annual cost of solar increases from \$33B/year to (post-subsidy) \$49B/year  (Fig.~\ref{fig:policyira}b) and solar capacity increases from \qty{0.8}{TW} to \qty{1.0}{TW}, while the capacity of NG-CCS plants decreases from \qty{0.21}{TW} to \qty{0.13}{TW} at zero advanced fission (Fig.~\ref{fig:policyira}d).
It is displaced by advanced fission at nearly the same rate as in the scenario without IRA, and it is totally displaced at \qty{225}{GW} of advanced fission capacity.
The value of advanced fission still derives from displacement of the same resources; the largest contributor is still the substitution of NG-CCS (Fig.~\ref{fig:policyira}c); it is followed by wind (particularly at low advanced fission capacity), solar, and batteries.

Under IRA the energy storage capacity in metal-air batteries increases by a factor of 4.8, from \qty{1.3}{TWh} to \qty{6.2}{TWh} at zero advanced fission capacity (Fig.~\ref{fig:policyira}e).
It also exhibits more displacement by advanced fission.
Displacement of metal-air energy storage capacity by advanced fission with thermal storage is minor in monetary terms, but significant in terms of the composition of storage on the grid.
Under IRA, the typical advanced fission thermal storage duration increases by 0.5 hours to 1 hour, to a maximum of 7 hours, and the fleet-average duration decreases more slowly with increasing reactor capacity, so the peak quantity of thermal storage energy capacity increases from \qty{1.25}{TWh} to \qty{2}{TWh}.
For the first \qty{150}{GW} of advanced fission, each \unit{GW} of capacity displaces \qty{20}{GWh} of metal-air energy capacity, four times more than without IRA.
Finally, under IRA the capacity factor of advanced fission lowers by 10\%--15\% in favor of additional production by renewables (Fig.~\ref{fig:policyira}f).

Note that the curves in Fig.~\ref{fig:policyira}c have a smaller sum than those in Fig.~\ref{fig:displacement}c.
They sum to the the \textit{unsubsidized} value of the fission cores, which is lower than the break-even cost with IRA subsidies. An upper-bound approximation to this value (where only the power conversion system and thermal storage system are subsidized, but not the core or the remainder of the plant) is shown inFig.~\ref{fig:policyira}a.

\begin{figure}
\begin{center}
\includegraphics[width=\textwidth]{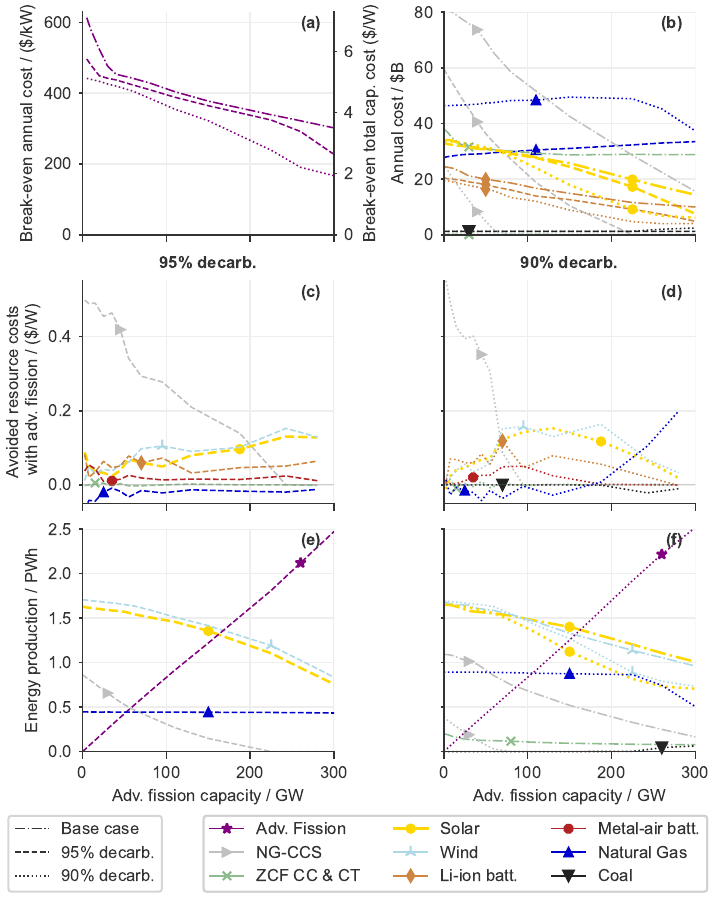}
\caption[System under non-fully-decarbonized policies]{The break-even cost of advanced fission and the wider energy system under 95\% and 90\% decarbonization policies. In these scenarios advanced fission derives its value from displacement of NG-CCS, wind, solar, and batteries.
Parts~(b), (e), and~(f) use the line styles as in~(a) to denote different scenarios; parts~(c) and~(d) use solid lines as all the traces are from the 95\% and 90\% decarbonized scenarios, respectively.}\label{fig:policynondec}
\end{center}
\end{figure} 

In scenarios where the U.S.\ mandates 90\% or 95\% decarbonization of the electricity supply by 2050 (expressed in CO$_2$-equivalent \unit{kg/kWh}) rather than 100\% decarbonization, the marginal value of nuclear declines modestly (Figure \ref{fig:policynondec}a).
The higher value below \qty{40}{GW} and under is diminished or lost entirely as the PJM-MACC zone can use natural gas instead of requiring costly ZCF fuels and high-priced wind.
From \qty{40}{GW} to \qty{100}{GW} the marginal value drops by 5\% and 10\% for the 95\% and 90\% decarbonization scenarios respectively, relative to the fully-decarbonized case.
In essence, for every 1\% reduction in an enforced decarbonization mandate, advanced fission reactors will lose 1\% of their economic value at low-to-moderate penetration levels.
However, at the largest penetrations studied (e.g. \qtyrange{225}{300}{\giga\watt}), reactors exhibit significantly lower value under incomplete decarbonization scenarios, with the marginal value/break-even cost of advanced fission declining 25\% and 45\% at 95\% and 90\% decarbonization, respectively.
To frame this another way, an advanced reactor design capable of achieving installed capital costs of \$4/W would have an addressable market of \qty{220}{GW} under a zero emissions scenario, but this would shrink to \qty{180}{GW} under a 95\% emissions reduction cap and to $\sim$\qty{120}{GW} with a 90\% emissions reduction cap.

We can understand the decline in fission's value through the changes in the resource mix, especially the firm resources.
Figure~\ref{fig:policynondec}b shows that under partial decarbonization spending on NG-CCS declines substantially, and ZCF-burning plants are totally eliminated (spending of \$0.1B/year or less); standard natural gas plants are built instead.
Spending on batteries declines at all penetrations and spending on solar declines at higher capacity penetrations; wind (not shown) behaves similarly.
Figure~\ref{fig:policynondec}e and (f) show the changes in production by various resources.
With 95\% decarbonization (e), natural gas production is flat across advanced fission capacity penetrations; the increase in spending (b) at higher capacity is because building (idle) natural gas plants is the lowest-cost way to satisfy capacity reserve requirements.
At 90\% decarbonization natural gas production is also flat (insensitive to the fission capacity) until it starts to be displaced by fission at \qty{225}{GW}; the carbon budget is then used to run coal plants which are less expensive than natural gas but have higher emissions.
Other than there, advanced fission does not compete against natural gas.
Parts (c) and (d) show that fission's value derives from the displacement of NG-CSS as a firm resource, and especially at 90\% decarbonization, displacement of wind, solar, and batteries.
In fact, above \qty{150}{GW} in the 95\% decarbonized scenario and above \qty{60}{GW} in the 90\% decarbonized scenario, most of fission's value comes from displacement of renewables and storage rather than displacement of NG-CCS.

\subsection{Fuel Costs}

\begin{figure}
\begin{center}
\includegraphics[width=\textwidth]{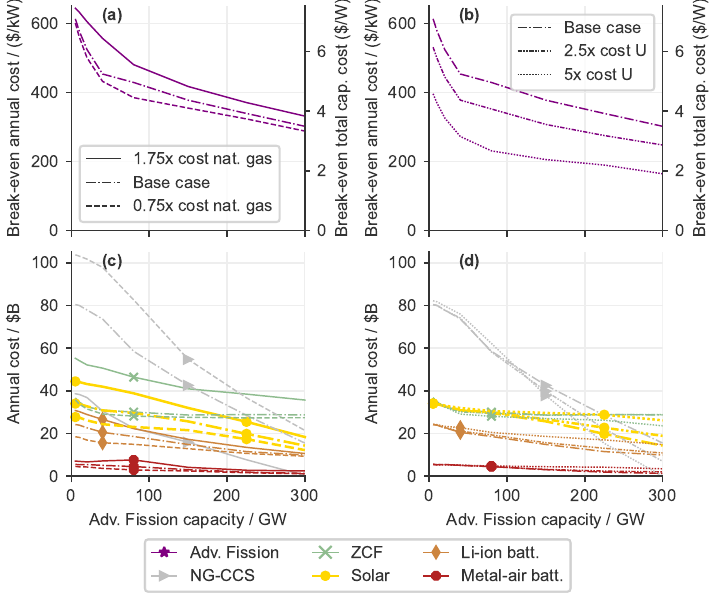}
\caption[The Impact of Fuel Costs on Reactor Value]{
The value of advanced fission changes with (left) the price of natural gas, used in NG-CCS plants, and (right) the price of uranium fuel.
The marginal value of the plants in these scenarios are compared with the reference case in parts (a) and (b).
High uranium costs are 2.5x the base costs (\$1.83/MMBtu), and very high uranium costs are 5x the base costs (\$3.65/MMBtu).
Parts (c) and (d) show the annual costs associated with solar, NG-CCS, and ZCF-burning plants.
Higher natural gas costs encourages the development of zero-carbon fuels and solar rather than NG-CCS, modestly raising the value of fission.
Higher uranium costs significantly reduce the value of advanced fission but have smaller changes on the composition of the wider electricity system.
}\label{fig:fuel}
\end{center}
\end{figure} 

Advanced reactor value is moderately sensitive to changes in natural gas costs but highly sensitive to variations in nuclear fuel costs (Figure~\ref{fig:fuel}).
A 75\% increase in natural gas cost from (\$3.79/MMBtu to \$6.64/MMBtu) adds about 10\% in value at low capacity penetrations, and that increase in value drops to roughly 3\% at high penetrations.
On the other hand, a 25\% decrease in natural gas fuel cost (to \$2.85/MMBtu) decreases reactor value by 5\%.
Only a portion of nuclear's marginal value comes from substitution of gas-fired generation at NG-CCS plants (see Figure~\ref{fig:displacement}).
With high-priced gas, fewer NG-CCS plants are built (Fig.~\ref{fig:fuel}c); ZCF becomes the largest resource category by cost and NG-CCS falls from first to fourth, behind solar and wind (not shown).
Nevertheless, displacement of NG-CCS is still the largest contributor to fission's value, as fission's variable costs are more similar to that of NG-CCS than ZCF. 
With low-priced gas, more NG-CCS plants are built and a larger portion of fission's value is from displacing it.

A ``high'' nuclear fuel cost future (2.5x higher than base) would reduce reactor valuation by 15\% (Figure~\ref{fig:fuel}b).
In a ``moderate'' uranium cost scenario, the fuel costs for producing one unit of electricity are \$6.3/MWh, while those in a high uranium costs scenario are \$15.7/MWh.
A ``very high'' nuclear fuel cost future (\$31.5/MWh, 5x higher than base) would have similar effects to the ``high'' nuclear future, but with a reduction of value of roughly 40\%.
In this scenario, the market-entry break-even cost of an advanced reactor would be \$4.6/W.

The step in fuel prices (from the base case to ``high'') by 1.5x of the base fuel price decreased marginal values by 15\% and the second step of 2.5x of the base price decreased marginal values by a further 25\%, indicating a linear relationship.
Thus as HALEU fuel becomes more expensive it will decrease reactor valuations proportionally.
Compared with the effects of a change in the price of natural gas, an increased uranium fuel cost has relatively small effects on the composition of the energy system (part (d)).
Only at very high fission capacity is there a difference: fission continues to displace NG-CCS but slows its displacement of solar (and wind, not shown); ZCF is unaffected.
At very high uranium costs, the variable costs of energy production are about half that of natural gas, so they are closer competitors than at standard uranium costs.
This higher price of energy production decreases the inframarginal rents for nuclear reactors and consequently the break-even cost.

\subsection{Technology Costs}\label{sec:techcostssensitivity}
\begin{figure}
\begin{center}
\includegraphics[width=\textwidth]{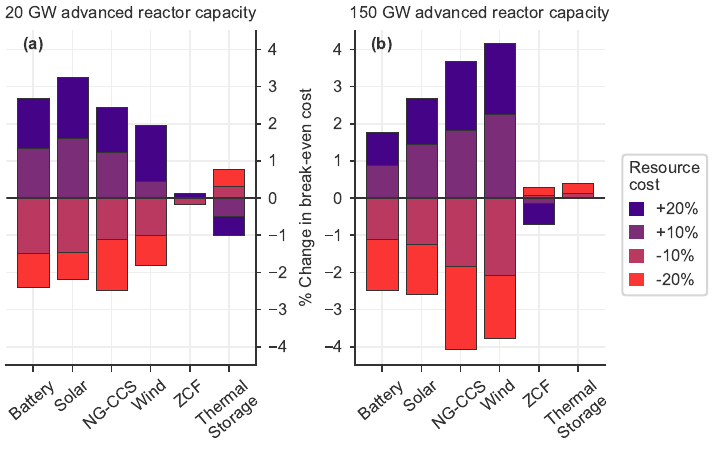}
\caption[The Impact of Technology Costs on Reactor Value]{The effect of variations in the costs of other technologies on the break-even cost of advanced reactors.
For each technology, 4 different price scenarios are considered: changes in the investment cost of $\pm10$\% and $\pm20\%$.
The considered technologies are lithium batteries, solar energy, natural gas with CCS, wind energy, and ZCF-burning generators.
The effect of cost variations for thermal storage is considered as well.}\label{fig:costs}
\end{center}
\end{figure} 
As the value of advanced fission plants derives from substitution of competing resources (see Section \ref{sec:subst}), we consider the sensitivity of our results to parametric uncertainties in the investment costs of these technologies.
At low capacity penetrations (\qty{20}{GW}), the marginal value of advanced fission changes by $\pm$2\% to 3\% when batteries, solar, NG-CCS, or wind have cost variations of $\pm 20\%$ (Fig.~\ref{fig:costs}a).
The value of fission is much less sensitive to cost variations for ZCF-burning plants (less than $\pm0.2\%$) as they are constructed largely to satisfy capacity reserve margin requirements and so are not displaced by fission as rapidly as the other technologies.
At larger capacities (150 GW), the sensitivity to battery costs decreases, while that to wind costs increases to $\pm4\%$ (Figure \ref{fig:costs}b), consistent with observed changes in marginal substitution rates at higher penetrations in Fig.~\ref{fig:displacement}c.

Figure \ref{fig:costs} also demonstrates that advanced reactor valuation is resilient against variations in thermal storage investment costs.
If the cost of storage rises by 20\%, the reactor will only lose about 1\% of its value at 20 GW.
The variation at 150 GW is smaller; as seen in Fig.~\ref{fig:stor}a this is near the marginal-value-neutral point where the increased integrated value of thermal storage has already been captured by the first 150 GW of fission capacity.
\begin{figure}
\begin{center}
\includegraphics[width=\textwidth]{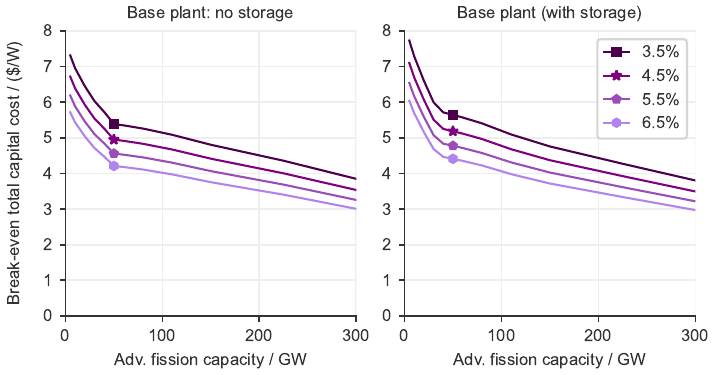}
	\caption{The effect of variations in WACC assumptions on break-even cost for (a) advanced fission plants and (b) advanced fission plants with coupled thermal storage. Break-even costs are expressed as a total capital cost assuming 30 year asset life (from 2023 NREL ATB \protect{\cite{national_renewable_energy_laboratory_annual_2023}}) and the WACC noted in the legend. Break-even costs are calculated for a `base plant' with a matched-capacity reactor and power conversion system, and excludes costs for coupled thermal storage (\protect{\ref{sec:cost}}). These results use Table \protect{\ref{tab:convert}} to calculate variations from our reference assumed WACC of 4.5\%.}\label{fig:rates}
\end{center}
\end{figure}

In addition to uncertainties in the capital costs of other technologies, we also analyze how sensitive our results are to uncertainties in our real weighted average cost of capital (WACC) assumptions.
Figure \ref{fig:rates} demonstrates that advanced nuclear reactors are strongly sensitive to variations in WACC. For both scenarios with and without storage, we note that value drops by 8\% for each 1\% increase in WACC.
Because total nuclear plant costs are dominated by their capital costs, it is reasonable to expect that their break-even costs are highly sensitive to the cost of capital to finance and build these plants.
Whereas at 20 GW penetration, advanced fission plants with storage would have a base-plant break-even cost of \$6.1/MW at a real WACC of 4.5\% (and 30 year asset life), the break-even cost would fall to \$5.2/MW at 6.5\% and increase to \$6.6/MW at 3.5\% WACC. Note that Table \ref{tab:convert} permits further conversion of reported break-even capital costs to equivalent break-even costs at any other WACC value in the range of 1-10\% and any asset life from 20-60 years.

\section{Discussion}
The initial economic value of an advanced reactor with no thermal storage is \$6.7/W (\$5.7/W--\$7.3/W over 3.5\%--6.5\% WACC), with marginal value declining steadily as capacity penetration increases to a value of \$3.5/W at \qty{300}{GW}. For the purposes of analyzing present-day investments in advanced nuclear, the low-penetration valuations are more useful, as they will indicate cost targets for the first tranche of new advanced reactors deployed at Nth-of-a-kind scale.
Additionally, to reach an eventual penetration of \qty{300}{GW} in 2050, the U.S.\ would need to add advanced nuclear reactor capacity at a pace of 15 GW per year, assuming construction begins around the year 2030.
This would be significantly higher than the record-setting 5.7 GW per year added in the U.S.\ between 1980 and 1987 \cite{world_nuclear_association_nuclear_2024}.
Thus, focusing on lower capacity penetrations between 0--150 GW is more realistic, even if the advent of SMRs and population growth in the U.S.\ could trigger record-setting growth in nuclear energy.

The estimated valuation for nuclear energy at low capacities is similar to or just below the total capital cost estimated in NREL's \textit{Annual Technology Baseline 2023} for small modular reactors in 2050, which is \$6.9/W \cite{national_renewable_energy_laboratory_annual_2023}.
Thus, based on the break-even costs established in this study, advanced reactors must achieve lower costs if they wish to be economically-competitive with other technologies in the future.
This study identifies a long-term extension of investment tax credits (ITC) established by the Inflation Reduction Act as an effective policy instrument to improve the economics of advanced reactors; we identify their break-even cost to be \$7.8/W, which is 10\% higher than it would be otherwise, if current ITC policy persists to 2050.
We do not, however, explore the impact that shifting this cost onto the federal government's budget may have on other energy or social programs, or on broader fiscal policy. Beyond tax credits, there are a few other avenues by which advanced reactors could increase their value or lower their construction costs in order to increase their economic competitiveness. For example, reactors that can co-generate heat, hydrogen, or other products may have two or more products to sell, which could increase the value of advanced fission plants and allow them to justify higher construction costs \cite{ingersoll_integration_2014}, although we do not explicitly model this possibility in this study. Furthermore, the advent of SMRs could facilitate factory fabrication of reactor components, leading to lower costs due to accelerated learning-by-doing, labor productivity, and lower construction risk \cite{glaser_small_2015}.

In practice, however, the promise of SMRs to lower construction costs and increase predictability may prove overly optimistic. For example, in 2023, NuScale Power increased construction cost estimates for their first commercial project --- a \qty{462}{MW} facility featuring six \qty{77}{MW} modular, factory-produced reactors --- by 75\%, and this cost increase would have been even more significant without the countervailing impact of IRA and DOE subsidies \cite{schlissel_eye-popping_2023}.
Based on this most recent cost estimate, we calculate that the full construction cost of NuScale's first SMR project is on the order of \$20/W, or \$\qty{11.4}{\per\watt} after IRA and DOE subsidies.
These costs significantly exceed the \$6.7/W marginal value identified in this study. NuScale cited construction material cost increases, as well as rising interest rates, as the primary reasons behind the cost increase \cite{schlissel_eye-popping_2023}.
Later in 2023, the Utah Associated Municipal Power Systems pulled out of the project due to the cost rises, leading to its cancellation \cite{bright_nuscale_2023}.
Thus, the NuScale project exemplifies how sensitive advanced reactors are to unfavorable investment environments and high interest rates, which is consistent with our results in Section \ref{sec:techcostssensitivity}.

Plants that incorporate coupled thermal storage into their designs can increase the initial marginal value of a reactor to \$7.1/W (\$6.0/W--\$7.7/W over 3.5\%--6.5\% WACC), justifying an up-to-5\% increase in break-even cost relative to the reactor-only design.
This estimated marginal value/break-even cost corresponds only to a `base' plant with a reactor coupled to a matched-capacity power conversion system. This excludes the costs of thermal storage and oversizing of the power conversion system, so the total cost of a plant could be higher.
Furthermore, this study has also demonstrated that reactor value is resilient against small variations in thermal storage capital costs.
Thus, integrating thermal storage with performance characteristics comparable to (or better than) the molten salt cost assumptions used in this study may be a profitable design choice for advanced reactor developers.

TerraPower is actively designing a salt-cooled fast reactor coupled with thermal storage, and their first generation cost estimates are around \$11.6/W upfront, or \$7.3/W after applying the IRA ITC \cite{terrapower_natrium_2023}. This subsidized cost estimate falls just above the \$7.1/W valuation for a reactor with thermal storage, implying the project would not be profitable if it were in this future modeled Eastern Interconnection, though it is closer to being so than the NuScale project.
Of course, because construction has not completed, it is impossible to know whether these estimates will be accurate, or whether TerraPower will experience the same cost overruns as NuScale. TerraPower's construction timeline has already been delayed after Russia's 2022 invasion of Ukraine halted the import of HALEU fuel used in the reactor \cite{white_catalyzing_2022}.
Furthermore, molten salt thermal storage is not yet a fully mature technology, and it could present challenges to the operation of the plant as a whole.
For example, a leak in the Crescent Dunes solar plant's molten salt storage halted operations in the project and eventually led to its bankruptcy \cite{boschee_becoming_2020}.
Even still, if engineering challenges are met and assumed costs are achieved, thermal storage gives advanced reactors a more competitive economic starting point.

Increasing turbine efficiency to the highest level achievable by current proposed designs provides a 3\% increase in marginal value for advanced reactors.
When considering that non-water coolants such as molten salt tend to be more corrosive and thus require special construction materials, this modest increase in value may be offset by higher construction costs \cite{glaser_small_2015}.
High-temperature gas-cooled reactors can avoid these corrosion issues and may therefore be better suited to take advantage of this increase in marginal value.

Furthermore, improving the efficiency of advanced reactors is likely to require a switch from LEU fuel to HALEU fuel.
Currently, there is no commercial source of HALEU fuel in the U.S., and HALEU fuel costs may be significantly higher than LEU fuel.
We find that the marginal value of advanced reactors will decrease by \qty{15}{\percent} if fuel costs are 2.5 times our base fuel assumptions (\$1.83/MMBtu vs \$0.73/MMBtu), and \qty{40}{\percent} if fuel costs are 5 times our base fuel assumptions (\$3.65/MMBtu vs \$0.73/MMBtu).
Therefore, the risk of developing advanced reactors without first establishing a HALEU supply chain could easily outweigh the added value that comes from greater efficiency.
Although the DOE has funded a HALEU enrichment plant in Ohio beginning to demonstrate HALEU production at the end of 2023, it is restricting its use for fueling advanced reactor prototypes only \cite{centrus_energy_centrus_2023}.
Thus, the U.S.\ is still a long ways away from establishing a domestic, commercial supply of HALEU fuel for future generations of reactors.
Nevertheless, the 2022 Inflation Reduction Act committed an additional \$700 million in funding through 2026 to promote the development of a domestic supply chain of HALEU, indicating a promising step in the right direction \cite{office_of_nuclear_energy_haleu_2024}.

\subsection{Limitations}
We note that GenX is an electricity-only energy systems model, which inherently limits the types of analyses that can be performed.
For example, we separately model heat and electricity production, but all heat is eventually converted back to electricity.
We are thus unable to model reactors capable of cogeneration, such as district heating, desalination, and hydrogen production.
If the entire energy system were to be decarbonized, not just electricity, this implementation would fail to capture nuclear energy's value in full.
Future work could analyze advanced reactors in other models that integrate multiple energy systems beyond electricity.
Similarly, our study focuses only on the value of nuclear energy for the Eastern Interconnection of the United States, but the findings could vary for countries or regions with different available resources and demand. 
Furthermore, we ignore broader real and perceived safety concerns related to nuclear energy, which could slow or constrain its commercial deployment.
Our model reflects an idealized electricity market, without for example separate day-ahead and real-time markets, profit-seeking behavior, or other types of non-optimal dispatch, so the revenue of the fission plants could differ if these complexities were included.
Note that we refer to a fleet of plants with a particular capacity as being ``profitable'' even if their cost equals the break-even cost for that capacity.
At that point there is zero \textit{excess} profit beyond the risk-adjusted return on equity included in the weighted average cost of capital. 
Finally, we run GenX with continuous capacity investment decisions and thus cannot capture potential economies of unit scale in nuclear costs, which may be relevant to analyze designs with different capacity size per reactor.

The one-at-a-time experimental approach in this study, where we vary parameters in specific categories while keeping all others at their reference values, is a good way to create initial indicators of value in the broad design space of advanced reactors. However, it misses capturing the full range of possible designs. For example, although we assess efficiency gains for reactors with thermal storage, we do not consider them for reactors without it. Similarly, although the effect of a 20\% increase in battery prices is assessed, the results of simultaneous increases in both battery and solar prices are not. A broader, combinatorial uncertainty analysis could yield additional insights.

\section{Conclusion}
Advanced nuclear reactors, which may offer greater flexibility, efficiency, and safety than traditional designs, have the potential to serve important roles in the United States' future energy systems.
This study aims to quantify the economic value that they could bring to a future fully- or partially-decarbonized power grid in order to establish cost targets for designs currently under development.
We achieve this by using the GenX electricity systems model to analyze how different advanced reactor design and policy scenarios affect their value in the Eastern Interconnection of the U.S.\ circa 2050. 

We find that the first tranche of advanced reactor capacity deployed will exhibit a marginal value (and thus maximum break-even capital cost) in the \$5.7/W to \$7.3/W range dependent on several factors, which is lower than recently reported costs of several designs under development today. Advanced reactor designs should thus focus centrally on achieving low capital costs to ensure economic viability.
To further improve their economics, advanced reactor developers can also consider coupling their reactors with thermal storage, which increases break-even reactor value by about 3--5\%.
This is the best way, under the options considered in this study, for reactors to achieve higher levels of operational flexibility, which can increase their value in a power grid with high amounts of solar and wind energy capacity.
Greater thermal efficiency offers a more modest opportunity for enhancing a design's value, but efficiency improvements should be considered in the context of potentially higher fuel costs for HALEU fuel as well, which reduce plant value.

Policymakers and private sector investors can use the results of this study to guide investment decisions related to advanced reactors and other technologies.
Without subsidies, recently reported costs of first generation advanced reactors would prohibit the technology from being  profitable.
However, continued access to investment tax credits, as implemented in the 2022 Inflation Reduction Act, has the potential to assist commercialization of advanced nuclear reactors, provided reactor vendors achieve sufficient capital costs.
Furthermore, any policy that lowers or raises fossil fuel prices will lower and raise nuclear plant value accordingly.
A large obstacle to widespread advanced reactor deployment, however, is the fact that most of these designs require HALEU nuclear fuel, none of which is commercially accessible in the U.S.\ today. Thus, policymakers advocating for a growth in advanced reactor investment must also provide the necessary incentives to establish a domestic HALEU economy, and HALEU funding in the 2022 Inflation Reduction Act serves as a strong first step.

\section*{Acknowledgements}
The authors thank Professor Denise M. Mauzerall for reading and providing feedback on the work. Also, the authors thank the referees for their timely and constructive feedback.
This work was funded by Princeton University’s Zero-Carbon Technology Consortium, which is funded by gifts from Google, GE, ClearPath, and Breakthrough Energy.
This work was supported by the U.S. Department of Energy under contract number DE-AC02-09CH11466. The United States Government retains a non-exclusive, paid-up, irrevocable, world-wide license to publish or reproduce the published form of this manuscript, or allow others to do so, for United States Government purposes.

\newpage
\appendix
\section{Calculating the Break-Even Cost}\label{sec:cost}
\subsection{Defining the Break-Even Cost}
The upfront capital costs of advanced nuclear reactors are not yet well-known, so instead of using them as inputs into the GenX model, we designed this study to calculate their break-even capital costs (the threshold at which they become economically competitive with other technologies) as outputs of the model. We do this by making the nuclear core free in the model, while constraining the model to build no more than a certain pre-determined amount of advanced nuclear core capacity.
We take the rate of change of the cost function with respect to nuclear core capacity (i.e. the dual of the capacity constraint) as indicative of the break-even cost of the reactor core at that given capacity penetration.
To evaluate how break-even cost varies at different capacity penetrations, we repeat this experiment at 5, 10, 20, 40, 80, 150, 225, and 300 GW$_{\textrm{e}}$ capacity constraints. 
\subsection{Adjusting for the cost of the power conversion system}
Because we only make the core of the reactor free, but not the power conversion system (PCS) (as steam generator and turbine construction costs are well-known), the model outputs the break-even cost of the nuclear plant without considering the cost of the turbine.
However, the literature usually refers to capital construction costs for a complete nuclear power plant, not just the nuclear core component.
Thus, to contextualize our results, we add the cost of the PCS, assumed to be \$1,364,230 per MW$_{\textrm{e}}$, to the break-even cost of the nuclear core (on a per-MW$_{\textrm{e}}$ basis).
This inherently assumes that there is a 1:1 ratio between the amount of nuclear core built and the amount of PCS built. This is a necessary assumption in order for us to communicate results that are consistent with those in the literature.
However, this is often not the case for nuclear reactors with thermal storage, as the PCS capacity often exceeds the core capacity to accommodate periods when stored energy is dispatched in addition to produced energy; we found the core-to-turbine ratio for reactors with storage to be between 1:1.1 and 1:1.5.
Because the cost of the remainder of the plant is larger than that of the PCS, these small variations in ratio have a small effect on the overall break-even cost. 
\subsection{Converting to Total Capital Costs}
GenX models one year of power system operation, so it considers all construction costs for generators as annuities paid for an initial capital investment. Thus, the break-even cost calculated by the model is communicated as a break-even annual cost. This annual costs includes both the financing cost of paying back the initial investment, as well as the yearly operational costs. We assume these operational costs comprise 2.5\% of the total capital costs of the reactor which is equivalent to the ratio in the NREL ATB \cite{national_renewable_energy_laboratory_annual_2023}. We also assume a real Weighted Average Cost of Capital (WACC) of 4.5\% and an asset lifetime of 30 years, also consistent with the ATB. These assumptions yield a capital recovery factor of 6.1\% per year. If we add this value to the 2.5\% operational costs, we find that the annual costs are 8.6\% of the total capital cost. We use this value to convert the GenX annuities into total upfront capital costs. Table \ref{tab:convert} provides conversion factors to calculate our total capital costs under different asset life and WACC assumptions. 

\begin{table}[t]
    \centering
    \begin{tabular}[t]{c|c c c c c c c c c}
        \toprule
\multicolumn{9}{c}{Asset life / years} \\
\multicolumn{1}{c}{WACC} & \multicolumn{1}{c}{20} & \multicolumn{1}{c}{25} & \multicolumn{1}{c}{\textbf{30}} & \multicolumn{1}{c}{35} & \multicolumn{1}{c}{40} & \multicolumn{1}{c}{45} & \multicolumn{1}{c}{50} & \multicolumn{1}{c}{55} & \multicolumn{1}{c}{60} \\
\midrule
1.00\% & 1.07 & 1.23 & 1.36 & 1.46 & 1.56 & 1.64 & 1.71 & 1.77 & 1.83 \\
2.00\% & 1.00 & 1.13 & 1.24 & 1.33 & 1.40 & 1.47 & 1.52 & 1.57 & 1.61 \\
3.00\% & 0.94 & 1.05 & 1.14 & 1.21 & 1.27 & 1.31 & 1.35 & 1.39 & 1.41 \\
4.00\% & 0.88 & 0.97 & 1.04 & 1.10 & 1.14 & 1.18 & 1.21 & 1.23 & 1.25 \\
\textbf{4.50}\% & 0.85 & 0.93 & \textbf{1.00} & 1.05 & 1.09 & 1.12 & 1.14 & 1.16 & 1.18 \\
5.00\% & 0.82 & 0.90 & 0.96 & 1.00 & 1.04 & 1.06 & 1.08 & 1.10 & 1.11 \\
6.00\% & 0.77 & 0.84 & 0.88 & 0.92 & 0.94 & 0.96 & 0.98 & 0.99 & 0.99 \\
7.00\% & 0.72 & 0.78 & 0.82 & 0.85 & 0.86 & 0.88 & 0.89 & 0.89 & 0.90 \\
8.00\% & 0.68 & 0.73 & 0.76 & 0.78 & 0.79 & 0.80 & 0.81 & 0.81 & 0.82 \\
9.00\% & 0.64 & 0.68 & 0.71 & 0.72 & 0.73 & 0.74 & 0.74 & 0.75 & 0.75 \\
10.00\% & 0.61 & 0.64 & 0.66 & 0.67 & 0.68 & 0.68 & 0.69 & 0.69 & 0.69 \\
\bottomrule
    \end{tabular}
    \caption{Conversion table for different financial assumptions. Any total capital cost in our results can be multiplied by the relevant coefficient in this table to see what the result would look like under different asset lives and WACCs.}
    \label{tab:convert}
\end{table}

\section{Data Availability}
 
All scenarios presented in this paper ran on a branch of GenX v0.3.3.
The source code for this branch is available at \url{https://github.com/emiliocanor/GenX/tree/senior_thesis}.
The standard GenX model is available at \url{https://github.com/GenXProject/GenX}.
PowerGenome is available at \url{https://github.com/PowerGenome/PowerGenome}.
Data will be archived at \url{https://datacommons.princeton.edu/}.

\section{Extent of geographic zones}\label{sec:zones}

\begin{figure}[H]
\begin{center}
\includegraphics[width=\textwidth]{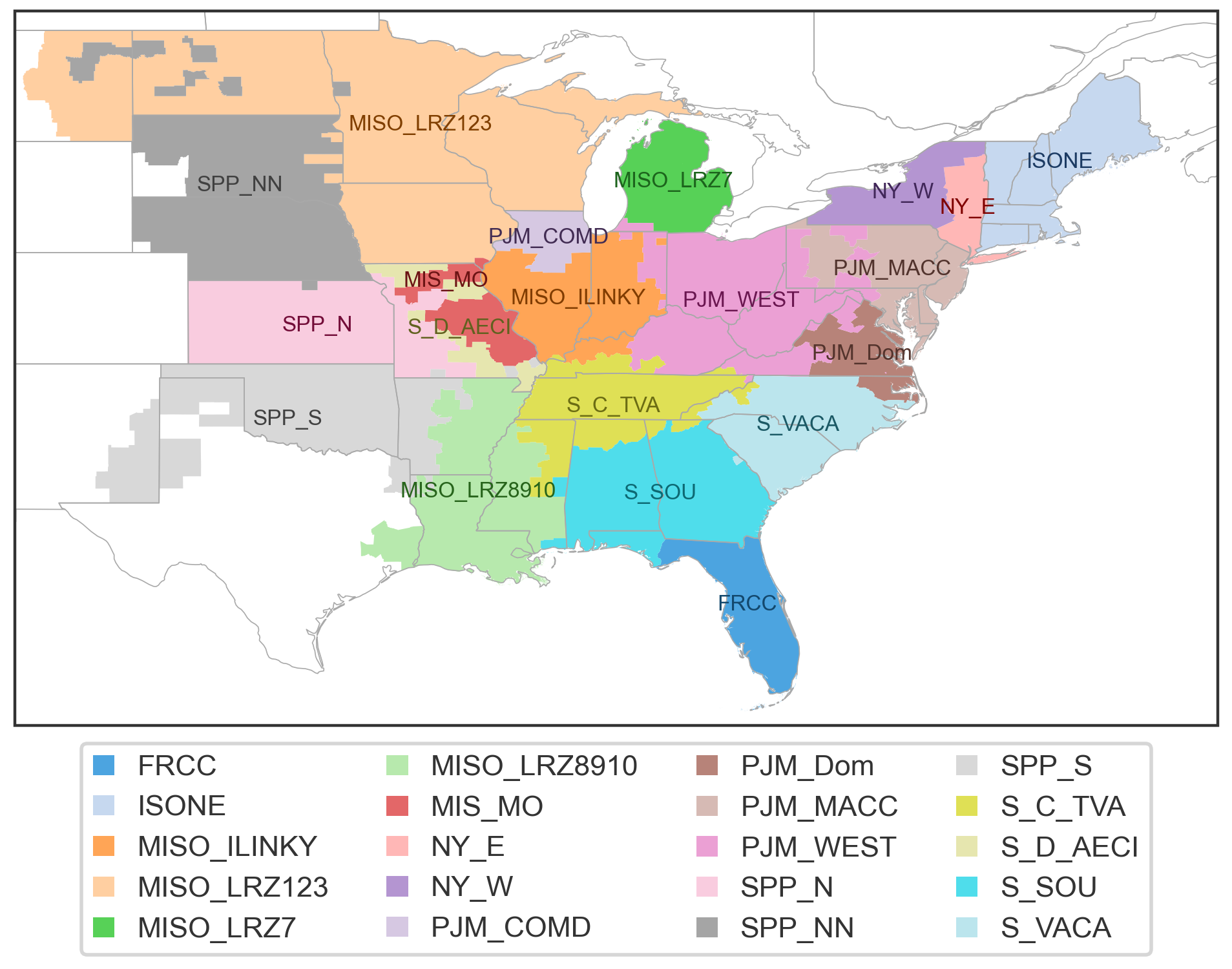}
	\caption[Eastern Interconnection of the US]{Eastern Interconnection of the United States divided into the 20 zones considered in this model. Figure adapted with permission from Schwartz et al.\ \protect{\cite{schwartz_value_2023}}.}\label{fig:map}
\end{center}
\end{figure}

Figure~\ref{fig:map} shows the zones, derived from those in the EPA's Integrated Planning Model (IPM), for the U.S.\ Eastern Interconnection power system~\cite{environmental_protection_agency_documentation_2021}.
The set of IPM regions making up each region in this study is found in Table~S1 of the Supplemental Information of Schwartz et al.\ \cite{schwartz_value_2023}.

\section{Transmission}\label{sec:transmission}
Table~\ref{tab:powerlines} lists the capacity of each transmission line at the start and end of the 2025--2035 run, and the maximum capacity expansion during the 2035--2050 runs.
The maximum capacity expansion for each line is 150\% of the existing capacity in 2025 or \qty{1.5}{GW}, whichever is greater.
Assumed costs are \$1M/GW$\cdot$mile, with a WACC of 0.069 and an economic lifetime of 60 years; this leads to an annual cost of 0.070283 of the capital cost. 
Late in the revision process we realized that these costs are low by up to a factor of 2--3; as a reference, doubling the transmission expansion cost of a reference case scenario with \qty{5}{GW} of fission capacity leads to an increase of 0.7\% in fission's marginal value.
As in \cite{schwartz_value_2023} losses along the line are proportional to its length at a rate of 1\% per hundred miles.

\begin{table}[H]
\caption{\label{tab:powerlines} Existing transmission lines and their potential for expansion in 2035--2050.}
\centering
\begin{footnotesize}
    
\begin{tabular}{@{}d{2.0}d{2.0}d{3.2}d{3.2}d{2.2}d{2.1}d{3.0}@{}} \toprule
\multicolumn{2}{c}{} & \multicolumn{1}{c}{2025 Capacity/}& \multicolumn{1}{c}{2035 Capacity/}& \multicolumn{1}{c}{Max. expansion/} & \multicolumn{1}{c}{Cost/} & \multicolumn{1}{c}{Length/} \\
\multicolumn{2}{c}{Zones} & 
\multicolumn{1}{c}{GW} & \multicolumn{1}{c}{GW} &  \multicolumn{1}{c}{GW} &
\multicolumn{1}{c}{(M\$/GW\,yr)} & 
\multicolumn{1}{c}{mile} \\ \midrule
 1 & 19 &  3.60 &  3.60 &  5.40 & 23.5 & 335 \\
 2 &  8 &  2.16 &  4.32 &  3.24 & 14.0 & 199 \\
 2 &  9 &  0.00 &  1.50 &  1.50 & 20.4 & 290 \\
 3 &  4 &  0.72 &  0.72 &  1.50 & 40.6 & 578 \\
 3 &  5 &  0.00 &  0.00 &  1.50 & 23.8 & 339 \\
 3 &  7 &  3.40 &  5.44 &  5.10 & 14.4 & 204 \\
 3 & 10 &  5.24 &  5.60 &  7.87 & 11.3 & 160 \\
 3 & 13 &  7.69 &  7.69 & 11.53 & 21.9 & 312 \\
 3 & 17 &  1.50 &  3.00 &  2.25 & 19.4 & 276 \\
 4 &  5 &  0.00 &  1.50 &  1.50 & 39.0 & 555 \\
 4 &  7 &  0.94 &  0.94 &  1.50 & 36.0 & 512 \\
 4 & 10 &  3.20 &  6.40 &  4.80 & 30.2 & 429 \\
 4 & 14 &  0.00 &  1.19 &  1.50 & 35.8 & 510 \\
 4 & 15 &  5.51 & 11.02 &  8.27 & 19.4 & 276 \\
 4 & 18 &  0.00 &  0.00 &  1.50 & 36.9 & 524 \\
 5 & 13 &  2.66 &  2.66 &  3.99 & 22.6 & 322 \\
 6 & 14 &  0.00 &  1.50 &  1.50 & 37.0 & 527 \\
 6 & 16 &  1.70 &  3.39 &  2.55 & 30.6 & 435 \\
 6 & 17 &  4.09 &  4.09 &  6.14 & 24.6 & 349 \\
 6 & 18 &  1.04 &  2.54 &  1.56 & 29.6 & 421 \\
 6 & 19 &  0.89 &  2.39 &  1.50 & 26.8 & 382 \\
 7 & 14 &  0.30 &  1.80 &  1.50 & 21.5 & 305 \\
 7 & 18 &  2.10 &  4.20 &  3.15 &  2.4 &  34 \\
 8 &  9 &  3.60 &  3.60 &  5.40 &  8.9 & 126 \\
 8 & 12 &  0.66 &  2.16 &  1.50 & 14.5 & 206 \\
 9 & 12 &  1.25 &  2.75 &  1.88 & 13.4 & 191 \\
10 & 13 &  0.98 &  0.98 &  1.50 & 27.8 & 396 \\
11 & 12 &  1.20 &  2.70 &  1.80 & 15.4 & 219 \\
11 & 13 &  6.93 &  6.93 & 10.39 & 19.1 & 271 \\
11 & 20 &  1.00 &  2.50 &  1.50 & 15.3 & 218 \\
12 & 13 &  3.88 &  3.89 &  5.83 & 21.1 & 300 \\
13 & 17 &  2.88 &  3.54 &  4.32 & 25.6 & 364 \\
13 & 20 &  1.22 &  2.48 &  1.83 & 23.0 & 327 \\
14 & 15 &  1.43 &  1.99 &  2.15 & 26.1 & 371 \\
14 & 16 &  3.37 &  3.40 &  5.06 & 16.8 & 239 \\
14 & 18 &  1.13 &  1.13 &  1.70 & 19.6 & 279 \\
15 & 16 &  0.00 &  0.58 &  1.50 & 40.8 & 581 \\
15 & 18 &  0.00 &  0.02 &  1.50 & 37.4 & 533 \\
16 & 18 &  1.17 &  1.18 &  1.76 & 29.4 & 418 \\
17 & 18 &  0.00 &  1.50 &  1.50 & 25.6 & 364 \\
17 & 19 &  3.20 &  6.39 &  4.79 & 16.7 & 237 \\
17 & 20 &  0.22 &  0.22 &  1.50 & 26.6 & 379 \\
19 & 20 &  1.40 &  1.40 &  2.10 & 23.3 & 331 \\
\bottomrule
\end{tabular}
\end{footnotesize}

\end{table}

\section{Demand}\label{sec:demand}
\begin{figure}[H]
\begin{center}
\includegraphics[width=2.5in]{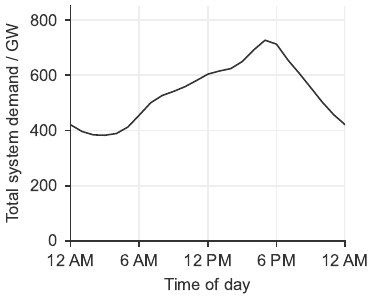}
\caption{Average total system demand (across all 20 zones and all 365 days) for each hour of the day.}\label{fig:demand_daily}
\end{center}
\end{figure}

\begin{figure}[H]
\begin{center}
\includegraphics[width=3.5in]{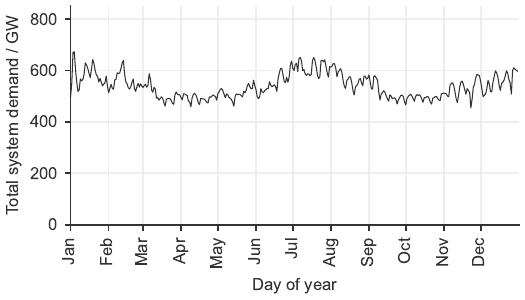}
\caption{Average total system demand (across all 20 zones and all 24 hours of the day) for each day of the year.}\label{fig:demand_yearly}
\end{center}
\end{figure}

Figure~\ref{fig:demand_daily} shows the average system demand by time of day, and Figure~\ref{fig:demand_yearly} shows the average system demand each day of the year.

\begin{figure}[H]
\begin{center}
\includegraphics[width=1.10\textwidth]{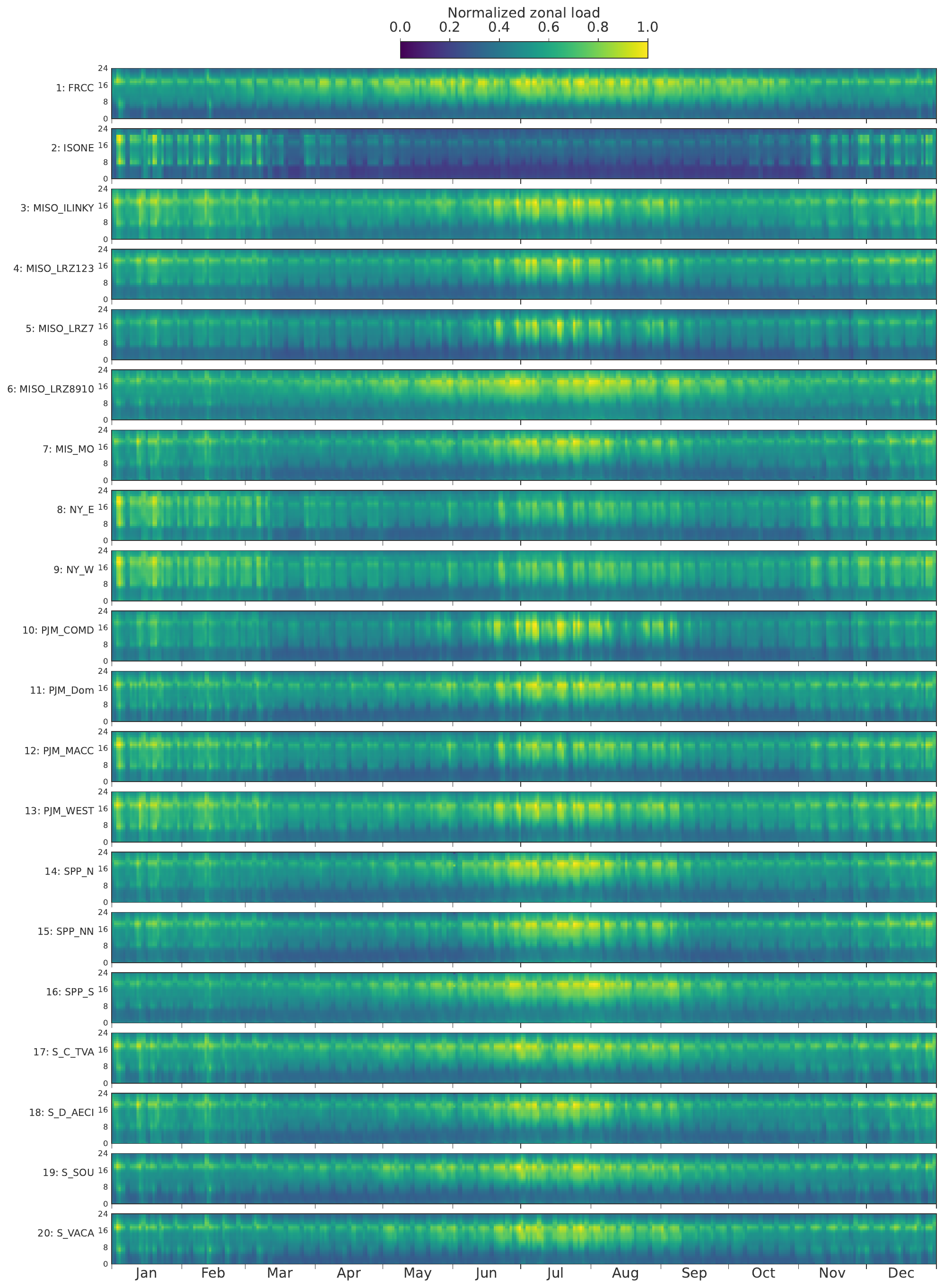}
\caption{Normalized hourly load profiles in each zone.}\label{fig:demand_profiles}
\end{center}
\end{figure}

Figure~\ref{fig:demand_profiles} shows the normalized hourly demand in each zone; one can see the daily, weekly, and seasonal variations.
Table~\ref{tab:demand} lists the peak demand and the peak quantity of shiftable demand (light duty vehicle charging and residential water heating) available in each zone. Note that the shiftable fraction changes hourly and has an average of about 25\%.

\begin{table}[H]
    \caption{\label{tab:tfloads}Peak hourly loads and shiftable loads in each zone.
    LDV is light duty vehicle charging and RWH is residential water heating. 
    LDV loads can be delayed by up to 5\,h and RWH loads can be advanced or delayed by 2\,h.
    All quantities are in MW.
    }
    \centering
    \begin{tabular}{@{}rrrr@{}}
    \toprule
    Zone & Peak load & LDV & RWH \\ \midrule
\verb|FRCC| & 85334 & 30936 & 1517 \\
\verb|ISONE| & 75732 & 15318 & 396 \\
\verb|MISO_ILINKY| & 43412 & 12498 & 472 \\
\verb|MISO_LRZ123| & 84904 & 25621 & 621 \\
\verb|MISO_LRZ7| & 38858 & 10661 & 207 \\
\verb|MISO_LRZ8910| & 57397 & 17886 & 461 \\
\verb|MIS_MO| & 16862 & 5831 & 152 \\
\verb|NY_E| & 36312 & 6170 & 318 \\
\verb|NY_W| & 12380 & 1968 & 101 \\
\verb|PJM_COMD| & 39905 & 7536 & 283 \\
\verb|PJM_Dom| & 37314 & 11415 & 346 \\
\verb|PJM_MACC| & 99227 & 27050 & 978 \\
\verb|PJM_WEST| & 108226 & 31297 & 1136 \\
\verb|SPP_N| & 27707 & 9275 & 210 \\
\verb|SPP_NN| & 20118 & 7314 & 173 \\
\verb|SPP_S| & 45523 & 15495 & 213 \\
\verb|S_C_TVA| & 58437 & 21225 & 758 \\
\verb|S_D_AECI| & 2779 & 960 & 25 \\
\verb|S_SOU| & 92609 & 35308 & 870 \\
\verb|S_VACA| & 78374 & 28075 & 976 \\ \bottomrule
    \end{tabular}
    \label{tab:demand}
\end{table}

\section{VRE Availability Factors}\label{sec:vre}
\begin{figure}[H]
\begin{center}
\includegraphics[width=0.95\textwidth]{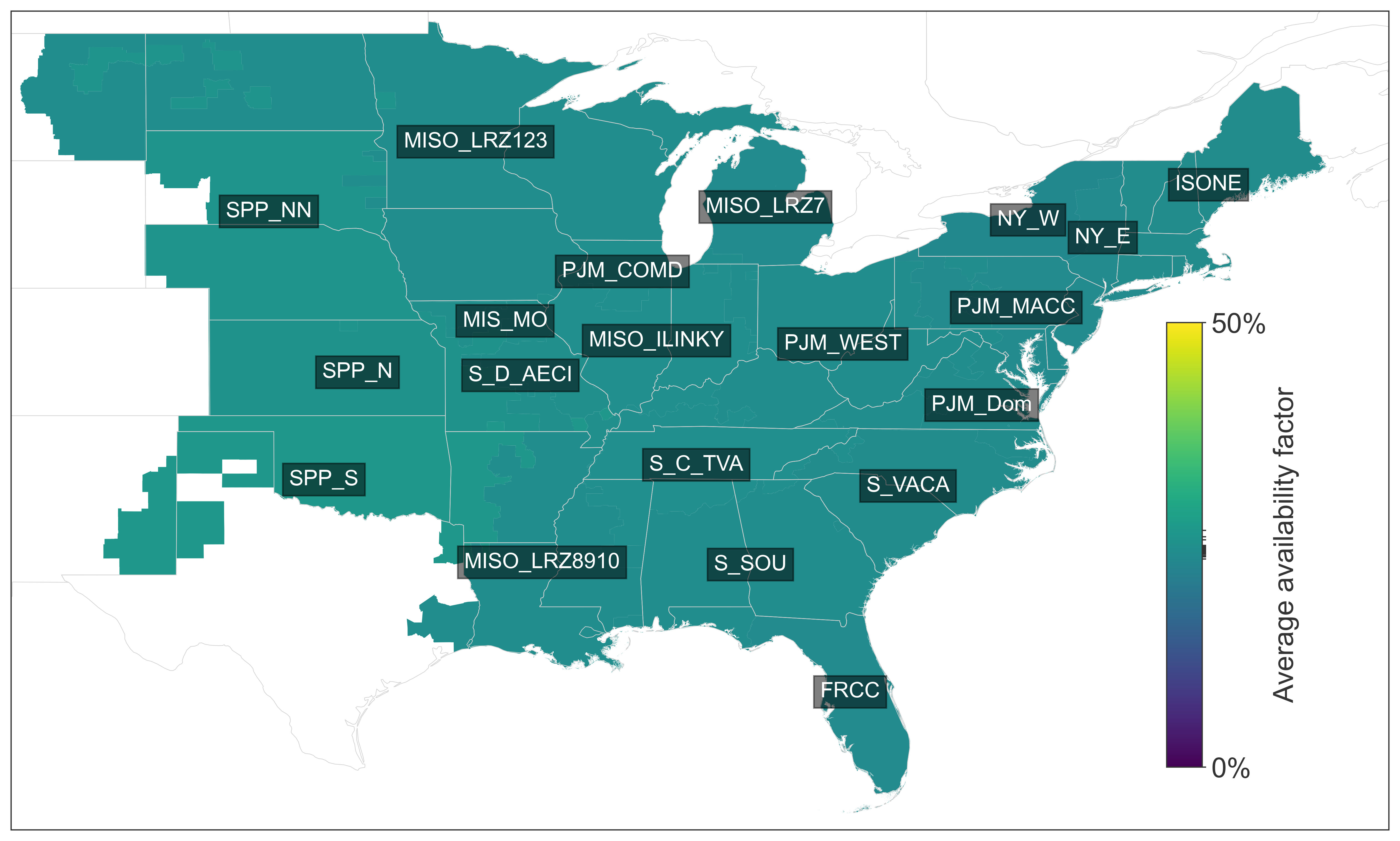}
\caption{Average availability factors (across all hours and all resource clusters within each zone) for solar resources in all 20 zones.}\label{fig:solar_map}
\end{center}
\end{figure}

\begin{figure}[H]
\begin{center}
\includegraphics[width=0.95\textwidth]{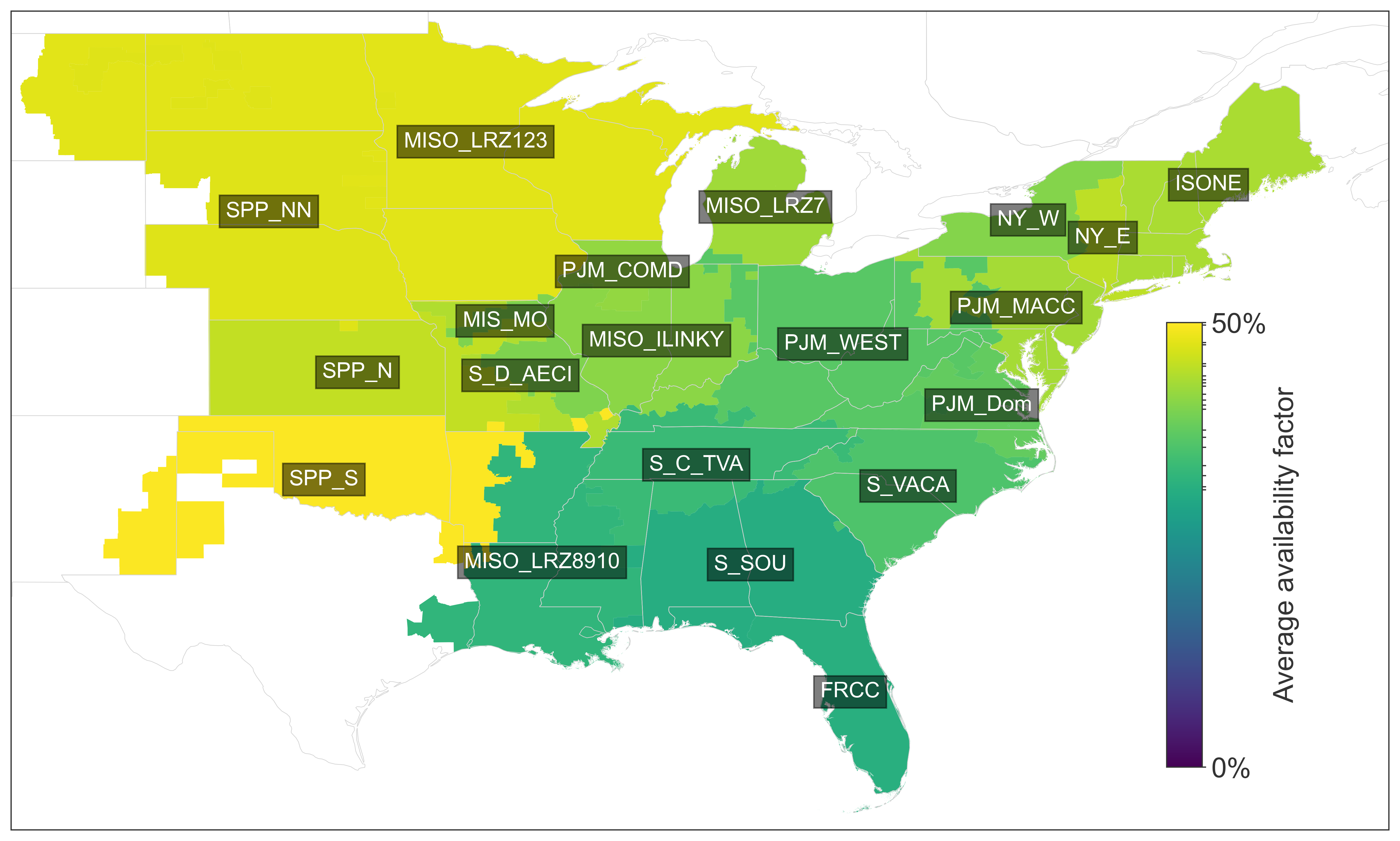}
\caption{Average availability factors (across all hours and all resource clusters within each zone) for wind resources in all 20 zones. Offshore wind resources are included as well.}\label{fig:wind_map}
\end{center}
\end{figure}

Figures~\ref{fig:solar_map} and~\ref{fig:wind_map} show the annual availability factors for solar and wind resources, respectively. The color bar is the same, for inter-comparison.

\section{Technology Cost Data}\label{sec:techcosts}

\begin{table}[ht]
    \centering
	\caption[Assumptions for all Technology Costs]{Investment costs and fixed costs for selected technologies. Those for advanced fission are included separately in Tables~\protect{\ref{tab:coreinputs}} through~\protect{\ref{tab:turbinputs}}. Storage technologies include two costs: for power capacity (\$/kW), then energy capacity (\$/kWh).}\label{tab:techcosts}
    \begin{small}
    \begin{tabular}{>{\raggedright}p{0.2\linewidth}
    >{\raggedright}p{0.15\linewidth}
    >{\raggedright}p{0.15\linewidth}
    >{\raggedright}p{0.10\linewidth}
    >{\raggedright\arraybackslash}p{0.15\linewidth}}
        \toprule
        \textbf{Technology} & \textbf{Inv cost} & \textbf{FO\&M} & \textbf{WACC} & \textbf{Lifetime}\\
        & \textbf{(\$/kW)} & \textbf{(\$/kW-yr)} & & \textbf{years}\\
        \midrule
Solar          &     683       & 13.62        & 0.044 & 30\\
Onshore wind   &     946       & 23.20        & 0.052 & 30\\
Offshore wind  &    2285       & 70.92        & 0.054 & 30\\
ZCF-CT         &     970       & 20.00        & 0.056 & 30\\
ZCF-CC         &    1090       & 23.87        & 0.056 & 30\\
NG-CCS         &    2052       & 55.87        & 0.056 & 30\\
Battery        &     251 / 164 &  6.28 / 4.11 & 0.044 & 15\\
Metalair       &    1200 /  12 & 30.00 / 0    & 0.044 & 25\\
Nat. gas CT    &     882       & 20.00        & 0.056 & 30\\
Nat. gas CC    &     990       & 23.87        & 0.056 & 30\\
Coal           &               & 67.42        &       &   \\
        \bottomrule
    \end{tabular}
    \end{small}
\end{table}

Table \ref{tab:techcosts} includes the investment and fixed operations and maintenance cost assumptions for all technologies other than advanced fission for which we considered construction, and coal.
For a few technologies, there are changes made to the NREL ATB data set from which PowerGenome sources its cost assumptions.
These changes align with those in the fusion study~\cite{schwartz_value_2023}.
\begin{enumerate}
    \item Metal-air batteries are not included in the ATB, so their cost and performance assumptions are sourced from Baik et al.\ \cite{baik_what_2021}.
    \item The ATB only includes cost assumptions for natural gas with 90\% CCS, but to have these technologies be considered in a decarbonized grid, it was necessary to model natural gas with 100\% CCS. Thus, costs are scaled up from the ATB assumptions, and increased CO$_2$ transportation costs are applied as well. For details see the fusion study\cite{schwartz_value_2023}.
    \item Resources have additional costs to account for spur lines; these are listed in Table~S4 of the fusion study\cite{schwartz_value_2023}.
\end{enumerate}

\begin{table}[ht]
    \centering
    \caption[Fuel and variable costs]{Average fuel and variable costs by resource type. Average fuel cost is included because costs vary geographically (per zone).}\label{tab:fuelcosts}
    \begin{small}
    \begin{tabular}{@{}ld{1.2}d{2.2}d{2.2}d{3.2}@{}}
        \toprule
        Technology &
        \multicolumn{1}{c}{VO\&M} &
        \multicolumn{1}{c}{Avg. fuel cost} &
        \multicolumn{1}{c}{Heat rate} &
        \multicolumn{1}{c}{Tot. var. cost} \\
        \multicolumn{1}{c}{} &
        \multicolumn{1}{c}{(\$/MWh)} &
        \multicolumn{1}{c}{(\$/MMBtu)} &
        \multicolumn{1}{c}{(MMBtu/MWh)} &
        \multicolumn{1}{c}{(\$/MWh)} \\ \midrule
ZCF-CT       & 6.08 & 14.41 &  9.72 & 146.15 \\
ZCF-CC       & 1.59 & 14.41 &  6.08 &  89.20 \\
NG-CCS       & 9.23 &  3.49 &  7.37 &  34.92 \\
Nat. gas CT  & 6.08 &  3.40 &  9.72 &  39.13 \\
Nat. gas CC  & 1.59 &  3.40 &  6.08 &  22.26 \\
Coal         & 1.86 &  1.69 & 11.86 &  21.89 \\
Li-ion batt. & 0.15 &       &       &   0.15 \\
        \bottomrule
    \end{tabular}
    \end{small}
\end{table}

Table~\ref{tab:fuelcosts} lists the variable operations and maintenance (VO\&M) cost, the average fuel costs the heat rate and the total variable cost for selected resources.

\begin{table}[ht]
    \centering
    \caption[Start cost]{Average unit-commitment start costs for applicable resources.
	Costs for the start fuel are listed in Table~\protect{\ref{tab:fuelcosts}}.}\label{tab:startcosts}
    \begin{small}
    \begin{tabular}{@{}ld{3.0}d{2.1}d{3.2}@{}}
        \toprule
        Technology &
        \multicolumn{1}{c}{Start cost} &
        \multicolumn{1}{c}{Start fuel} &
        \multicolumn{1}{c}{Tot. start cost} \\
        \multicolumn{1}{c}{} &
        \multicolumn{1}{c}{(\$/MW)} &
        \multicolumn{1}{c}{(MMBtu/MW)} &
        \multicolumn{1}{c}{(\$/MW)} \\ \midrule
ZCF-CT       &   0 &  3.5 &  50.44 \\
ZCF-CC       &   0 &  2.0 &  28.82 \\
NG-CCS       & 103 &  2.0 & 109.97 \\
Nat. gas CT  &   0 &  3.5 &  11.90 \\
Nat. gas CC  &   0 &  2.0 &   6.80 \\
Coal         & 122 & 16.5 & 149.87 \\
        \bottomrule
    \end{tabular}
    \end{small}
\end{table}

Table~\ref{tab:startcosts} lists the unit-commitment start costs for selected resources.
During the revision process we realized the ZCF and natural gas generators lacked start costs independent of the start fuel; correcting this in a reference-scenario case with \qty{5}{GW} of advanced fission led to a marginal value 0.8\% higher; the effect should be smaller elsewhere.

\section{Initial Capacities}\label{sec:caps}
\begin{table}[H]
    \centering
    \caption{Initial capacities (in GW) for each resource in each zone for the 2025 to 2035 expansion. In parentheses, percentage share of capacity that each resource holds in each zone. ``Hydro'' includes conventional hydroelectric and pumped storage. There is no initial battery capacity in 2025.}\label{tab:2025caps}
    \begin{footnotesize}
    \begin{tabular}{>{\raggedright}p{0.18\linewidth}>{\raggedright}p{0.09\linewidth}>{\raggedright}p{0.09\linewidth}>{\raggedright}p{0.09\linewidth}>{\raggedright}p{0.09\linewidth}>
    {\raggedright}p{0.09\linewidth}>{\raggedright\arraybackslash}p{0.09\linewidth}}
    \toprule
\textbf{Zone} & \textbf{Coal} & \textbf{N. Gas} & \textbf{Nuclear} & \textbf{Solar} & \textbf{Wind}& \textbf{Hydro} \\
 \midrule
FRCC & 3.1 (5.1\%) & 37.5 (61.2\%) & 2.2 (3.6\%) & 17.6 (28.7\%) & 0.0 (0.0\%) & 0.0 (0.0\%) \\
ISONE & 0.0 (0.0\%) & 13.4 (42.5\%) & 2.5 (7.9\%) & 9.0 (28.5\%) & 1.7 (5.4\%) & 3.7 (11.7\%) \\
MISO\_ILINKY & 8.5 (23.8\%) & 9.7 (27.2\%) & 1.1 (3.1\%) & 10.1 (28.3\%) & 5.5 (15.4\%) & 0.3 (0.8\%) \\
MISO\_LRZ123 & 12.2 (19.9\%) & 13.0 (21.2\%) & 0.0 (0.0\%) & 9.3 (15.2\%) & 24.0 (39.2\%) & 1.0 (1.6\%) \\
MISO\_LRZ7 & 0.1 (0.4\%) & 8.8 (39.3\%) & 1.2 (5.4\%) & 6.3 (28.1\%) & 3.7 (16.5\%) & 2.1 (9.4\%) \\
MISO\_LRZ8910 & 6.8 (14.3\%) & 25.9 (54.4\%) & 4.6 (9.7\%) & 9.0 (18.9\%) & 0.3 (0.6\%) & 0.7 (1.5\%) \\
MIS\_MO & 0.0 (0.0\%) & 2.0 (32.6\%) & 1.2 (19.6\%) & 1.6 (26.1\%) & 0.5 (8.2\%) & 0.7 (11.4\%) \\
NY\_E & 0.0 (0.0\%) & 10.0 (64.5\%) & 0.0 (0.0\%) & 3.5 (22.6\%) & 0.0 (0.0\%) & 1.6 (10.3\%) \\
NY\_W & 0.0 (0.0\%) & 0.1 (0.8\%) & 2.1 (16.6\%) & 2.8 (22.1\%) & 3.2 (25.2\%) & 4.3 (33.9\%) \\
PJM\_COMD & 0.0 (0.0\%) & 13.7 (48.2\%) & 7.2 (25.3\%) & 2.8 (9.8\%) & 4.4 (15.5\%) & 0.0 (0.0\%) \\
PJM\_Dom & 0.8 (3.1\%) & 9.5 (36.9\%) & 2.0 (7.8\%) & 9.5 (36.9\%) & 0.4 (1.6\%) & 3.5 (13.6\%) \\
PJM\_MACC & 1.4 (2.4\%) & 30.0 (50.5\%) & 9.4 (15.8\%) & 13.0 (21.9\%) & 1.3 (2.2\%) & 3.4 (5.7\%) \\
PJM\_WEST & 10.8 (15.5\%) & 34.6 (49.7\%) & 5.2 (7.5\%) & 12.4 (17.8\%) & 4.6 (6.6\%) & 1.8 (2.6\%) \\
SPP\_N & 5.7 (26.1\%) & 4.9 (22.5\%) & 1.3 (6.0\%) & 0.9 (4.1\%) & 8.5 (39.0\%) & 0.1 (0.5\%) \\
SPP\_NN & 4.3 (21.7\%) & 4.2 (21.2\%) & 0.0 (0.0\%) & 0.7 (3.5\%) & 7.3 (36.8\%) & 2.7 (13.6\%) \\
SPP\_S & 5.5 (12.1\%) & 13.2 (29.0\%) & 0.0 (0.0\%) & 3.6 (7.9\%) & 20.2 (44.3\%) & 2.5 (5.5\%) \\
S\_C\_TVA & 0.6 (1.7\%) & 16.7 (48.7\%) & 6.1 (17.8\%) & 4.1 (12.0\%) & 0.0 (0.0\%) & 6.7 (19.5\%) \\
S\_D\_AECI & 1.4 (22.7\%) & 3.0 (48.7\%) & 0.0 (0.0\%) & 0.1 (1.6\%) & 1.6 (26.0\%) & 0.1 (1.6\%) \\
S\_SOU & 5.5 (9.0\%) & 32.2 (52.7\%) & 7.4 (12.1\%) & 10.2 (16.7\%) & 0.0 (0.0\%) & 5.7 (9.3\%) \\
S\_VACA & 4.8 (9.3\%) & 19.4 (37.7\%) & 7.8 (15.2\%) & 13.9 (27.0\%) & 0.0 (0.0\%) & 5.3 (10.3\%) \\
\midrule
\textbf{Total} & \textbf{71.5 (10.0\%)} & \textbf{301.9 (42.1\%)} & \textbf{61.4 (8.6\%)} & \textbf{140.2 (19.5\%)} & \textbf{87.2 (12.2\%)} & \textbf{46.3 (6.5\%)}\\
 \bottomrule
    \end{tabular}
    \end{footnotesize}
\end{table}

\begin{table}[H]
    \centering
    \caption{Initial capacities (in GW) for each resource in each zone for the decarbonized 2035 to 2050 expansion. In parentheses, percentage share of capacity that each resource holds in each zone. ``Hydro'' includes conventional hydroelectric and pumped storage. There is no initial battery capacity in 2035. NG-CCS is not included; its total capacity $< 0.004$ GW.}\label{tab:2035caps}
    \begin{footnotesize}
    \begin{tabular}{>{\raggedright}p{0.18\linewidth}>{\raggedright}p{0.09\linewidth}>{\raggedright}p{0.09\linewidth}>{\raggedright}p{0.09\linewidth}>{\raggedright}p{0.09\linewidth}>{\raggedright\arraybackslash}p{0.09\linewidth}}
\toprule
\textbf{Zone} & \textbf{Nuclear} & \textbf{Solar} & \textbf{Wind}& \textbf{ZCF} & \textbf{Hydro} \\
 \midrule
FRCC & 0.0 (0.0\%) & 42.8 (65.0\%) & 0.2 (0.3\%) & 22.8 (34.6\%) & 0.0 (0.0\%) \\
ISONE & 0.0 (0.0\%) & 9.2 (14.6\%) & 10.4 (16.6\%) & 39.4 (62.7\%) & 3.7 (5.9\%) \\
MISO\_ILINKY & 0.0 (0.0\%) & 19.3 (43.3\%) & 14.5 (32.5\%) & 10.4 (23.3\%) & 0.3 (0.7\%) \\
MISO\_LRZ123 & 0.0 (0.0\%) & 30.0 (52.9\%) & 20.4 (36.0\%) & 5.3 (9.3\%) & 1.0 (1.8\%) \\
MISO\_LRZ7 & 0.0 (0.0\%) & 20.8 (58.8\%) & 5.1 (14.4\%) & 7.3 (20.6\%) & 2.1 (5.9\%) \\
MISO\_LRZ8910 & 0.0 (0.0\%) & 26.7 (53.8\%) & 4.8 (9.7\%) & 17.4 (35.0\%) & 0.7 (1.4\%) \\
MIS\_MO & 0.0 (0.0\%) & 4.0 (34.2\%) & 6.0 (51.3\%) & 0.9 (7.7\%) & 0.7 (6.0\%) \\
NY\_E & 0.0 (0.0\%) & 4.2 (15.3\%) & 0.3 (1.1\%) & 21.3 (77.6\%) & 1.6 (5.8\%) \\
NY\_W & 0.0 (0.0\%) & 3.0 (17.5\%) & 8.2 (47.8\%) & 1.6 (9.3\%) & 4.3 (25.1\%) \\
PJM\_COMD & 0.0 (0.0\%) & 2.8 (83.9\%) & 0.5 (15.0\%) & 0.0 (0.0\%) & 0.0 (0.0\%) \\
PJM\_Dom & 0.0 (0.0\%) & 12.0 (47.2\%) & 10.0 (39.3\%) & 0.0 (0.0\%) & 3.5 (13.8\%) \\
PJM\_MACC & 0.0 (0.0\%) & 13.7 (30.4\%) & 12.8 (28.4\%) & 15.2 (33.7\%) & 3.4 (7.5\%) \\
PJM\_WEST & 0.0 (0.0\%) & 45.2 (45.2\%) & 19.1 (19.1\%) & 33.9 (33.9\%) & 1.8 (1.8\%) \\
SPP\_N & 0.0 (0.0\%) & 8.2 (43.9\%) & 10.4 (55.6\%) & 0.0 (0.0\%) & 0.1 (0.5\%) \\
SPP\_NN & 0.0 (0.0\%) & 19.2 (56.1\%) & 12.3 (36.0\%) & 0.0 (0.0\%) & 2.7 (7.9\%) \\
SPP\_S & 0.0 (0.0\%) & 24.6 (50.8\%) & 21.3 (44.0\%) & 0.0 (0.0\%) & 2.5 (5.2\%) \\
S\_C\_TVA & 2.5 (4.0\%) & 27.3 (43.6\%) & 26.0 (41.6\%) & 0.0 (0.0\%) & 6.7 (10.7\%) \\
S\_D\_AECI & 0.0 (0.0\%) & 3.2 (29.9\%) & 7.5 (70.0\%) & 0.0 (0.0\%) & 0.1 (0.9\%) \\
S\_SOU & 2.2 (4.1\%) & 35.3 (65.5\%) & 0.0 (0.0\%) & 10.6 (19.7\%) & 5.7 (10.6\%) \\
S\_VACA & 0.0 (0.0\%) & 31.9 (61.3\%) & 5.6 (10.8\%) & 9.2 (17.7\%) & 5.3 (10.2\%) \\
\midrule
\textbf{Total} & \textbf{4.8 (0.6\%)} & \textbf{383.4 (46.4\%)} & \textbf{195.4 (23.7\%)} & \textbf{195.3 (23.7\%)} & \textbf{46.3 (5.6\%)}\\
 \bottomrule
    \end{tabular}
    \end{footnotesize}
\end{table}

\begin{table}[H]
    \centering
    \caption{Initial capacities (in GW) for each resource in each zone for the 90\% and 95\% decarbonized 2035 to 2050 expansion. In parentheses, percentage share of capacity that each resource holds in each zone. ``Hydro'' includes conventional hydroelectric and pumped storage. There is no initial battery or ZCF capacity in 2035. }\label{tab:2035caps}
    \begin{footnotesize}
    \begin{tabular}{>{\raggedright}p{0.18\linewidth}>{\raggedright}p{0.09\linewidth}>{\raggedright}p{0.09\linewidth}>{\raggedright}p{0.09\linewidth}>{\raggedright}p{0.09\linewidth}>{\raggedright}p{0.09\linewidth}>{\raggedright\arraybackslash}p{0.09\linewidth}}
\toprule
\textbf{Zone} & \textbf{Coal} & \textbf{N. Gas} & \textbf{Nuclear} & \textbf{Solar} & \textbf{Wind}& \textbf{Hydro} \\
 \midrule
FRCC & 0.5 (0.6\%) & 39.5 (47.4\%) & 0.0 (0.0\%) & 42.8 (51.3\%) & 0.2 (0.2\%) & 0.0 (0.0\%) \\
ISONE & 0.0 (0.0\%) & 43.2 (64.4\%) & 0.0 (0.0\%) & 9.2 (13.7\%) & 10.4 (15.5\%) & 3.7 (5.5\%) \\
MISO\_ILINKY & 2.2 (4.6\%) & 11.8 (24.5\%) & 0.0 (0.0\%) & 19.3 (40.0\%) & 14.5 (30.1\%) & 0.3 (0.6\%) \\
MISO\_LRZ123 & 3.2 (5.1\%) & 7.4 (11.8\%) & 0.0 (0.0\%) & 30.0 (47.8\%) & 20.4 (32.5\%) & 1.0 (1.6\%) \\
MISO\_LRZ7 & 0.0 (0.0\%) & 10.1 (26.4\%) & 0.0 (0.0\%) & 20.8 (54.4\%) & 5.1 (13.3\%) & 2.1 (5.5\%) \\
MISO\_LRZ8910 & 0.7 (1.2\%) & 25.0 (42.9\%) & 0.0 (0.0\%) & 26.7 (45.8\%) & 4.8 (8.2\%) & 0.7 (1.2\%) \\
MIS\_MO & 0.0 (0.0\%) & 1.0 (8.5\%) & 0.0 (0.0\%) & 4.0 (34.1\%) & 6.0 (51.2\%) & 0.7 (6.0\%) \\
NY\_E & 0.0 (0.0\%) & 24.8 (80.0\%) & 0.0 (0.0\%) & 4.2 (13.6\%) & 0.3 (1.0\%) & 1.6 (5.2\%) \\
NY\_W & 0.0 (0.0\%) & 1.6 (9.3\%) & 0.0 (0.0\%) & 3.0 (17.4\%) & 8.2 (47.7\%) & 4.3 (25.0\%) \\
PJM\_COMD & 0.0 (0.0\%) & 3.6 (51.6\%) & 0.0 (0.0\%) & 2.8 (40.1\%) & 0.5 (7.2\%) & 0.0 (0.0\%) \\
PJM\_Dom & 0.8 (2.6\%) & 5.0 (16.0\%) & 0.0 (0.0\%) & 12.0 (38.4\%) & 10.0 (32.0\%) & 3.5 (11.2\%) \\
PJM\_MACC & 1.0 (1.6\%) & 31.8 (50.5\%) & 0.0 (0.0\%) & 13.7 (21.7\%) & 12.8 (20.3\%) & 3.4 (5.4\%) \\
PJM\_WEST & 3.1 (2.6\%) & 49.7 (41.7\%) & 0.0 (0.0\%) & 45.2 (38.0\%) & 19.1 (16.0\%) & 1.8 (1.5\%) \\
SPP\_N & 1.3 (6.4\%) & 0.0 (0.0\%) & 0.0 (0.0\%) & 8.2 (40.4\%) & 10.4 (51.3\%) & 0.1 (0.5\%) \\
SPP\_NN & 1.1 (2.9\%) & 2.3 (6.0\%) & 0.0 (0.0\%) & 19.2 (50.4\%) & 12.3 (32.3\%) & 2.7 (7.1\%) \\
SPP\_S & 0.6 (1.1\%) & 2.8 (5.4\%) & 0.0 (0.0\%) & 24.6 (47.1\%) & 21.3 (40.8\%) & 2.5 (4.8\%) \\
S\_C\_TVA & 0.5 (0.7\%) & 5.5 (8.0\%) & 2.5 (3.6\%) & 27.3 (39.8\%) & 26.0 (37.9\%) & 6.7 (9.8\%) \\
S\_D\_AECI & 0.0 (0.0\%) & 0.0 (0.0\%) & 0.0 (0.0\%) & 3.2 (29.8\%) & 7.5 (69.9\%) & 0.1 (0.9\%) \\
S\_SOU & 0.7 (1.2\%) & 16.8 (27.6\%) & 2.2 (3.6\%) & 35.3 (58.1\%) & 0.0 (0.0\%) & 5.7 (9.4\%) \\
S\_VACA & 2.7 (4.3\%) & 16.7 (26.8\%) & 0.0 (0.0\%) & 31.9 (51.2\%) & 5.6 (9.0\%) & 5.3 (8.5\%) \\
\midrule
\textbf{Total} & \textbf{18.6 (2.0\%)} & \textbf{298.6 (31.4\%)} & \textbf{4.8 (0.5\%)} & \textbf{383.4 (40.3\%)} & \textbf{195.4 (20.5\%)} & \textbf{46.3 (4.9\%)}\\
 \bottomrule
    \end{tabular}
    \end{footnotesize}
\end{table}

\section{IRA Tax Credits}\label{sec:ira}
PTCs and ITCs are included to directly model the IRA provisions.
The base ITC from IRA of 30\% is used, as is the base PTC of \$\qty{27.5}{\per\MWh} \cite{environmental_protection_agency_inflation_2023}.
For each of these, we assumed they could benefit from at least one of the IRA's bonus credits. Bonus credits are applied to projects with a minimum percentage of manufactured products coming from the U.S., as well as those sited in an ``energy community'' --- a region of the U.S.\ that has historically relied on coal, oil, or gas \cite{environmental_protection_agency_inflation_2023}.
After applying these bonus credits, the ITC rises to 40\% and the PTC rises to \$\qty{28.6}{\per\MWh}. The tax credits are paid out as reductions in the amount of owed tax from a clean energy business over the duration of the tax credit, but these are often sold to other businesses that owe greater tax.
Thus, we apply a 7.5\% `haircut' to the IRA tax credits which represents the lost value from the resale of the tax credits as done by Ricks et al.\ \cite{ricks_minimizing_2023}.
Finally, we convert the tax credits into their net present value in 2019 dollars.
These were incorporated into GenX as reductions in variable operation costs (PTC) and investment costs (ITC) for each eligible technology. 

\section{Capacity Reserve Margins}\label{sec:reserves}
Table~\ref{tab:crm} describes the capacity reserve margin (CRM) constraints.
Each zone is a member of one of six CRM regions.
Each region requires a certain fraction of the hourly demand to be backed by unused capacity held in reserve.
Table~\ref{tab:crmCont} lists the maximum contribution of each technology's unused capacity to the reserve.
Data for both these tables are obtained from PowerGenome \cite{schivley_powergenomepowergenome_2022}.
For a description of how the CRM policy is implemented, refer to the GenX documentation.

\begin{table}[H]
\caption{\label{tab:crm} Capacity reserve margin constraint regions and included zones.}
\centering
\begin{tabular}{@{}lll@{}} \toprule
\textbf{CRM region} & \textbf{Margin} &  \textbf{Zones} \\ \midrule
1 & 0.183 & \verb|ISONE| \\
2 & 0.15 & \verb|NY_E|, \verb|NY_W| \\
3 & 0.155 & \verb|PJM_COMD|, \verb|PJM_Dom|, \\
  &       & \verb|PJM_MACC|, \verb|PJM_WEST| \\
4 & 0.12 & \verb|SPP_N|, \verb|SPP_NN|, \verb|SPP_S| \\
5 & 0.12 & \verb|MISO_ILINKY|, \verb|MISO_LRZ123|, \verb|MISO_LRZ7| \\
  &  & \verb|MISO_LRZ8910|, \verb|MIS_MO| \\
6 & 0.15 & \verb|FRCC|, \verb|S_C_TVA|, \verb|S_D_AECI| \\
  &      & \verb|S_SOU|, \verb|S_VACA| \\
\bottomrule \end{tabular}
\end{table}

\begin{table}[H]
    \centering
    \caption[Capacity Reserve Margins contributions]{Maximum capacity reserve margins contribution by technology type.}\label{tab:crmCont}
    \begin{small}
    \begin{tabular}[h]{@{}ll@{}}
        \toprule
        \textbf{Fraction of capacity} & \textbf{Technology}\\
        \midrule
        0.8 & Solar, Wind, Hydroelectric power \\
        0.9 & Nuclear, Coal, Natural gas, ZCF generator, Biomass \\
        0.95 & Lithium battery, Metal-air battery, Hydroelectric storage \\
        \bottomrule
    \end{tabular}
    \end{small}
\end{table}

\section{Advanced Reactor Costs and Parameters}\label{sec:adv_assumps}
This appendix defines assumptions made for the costs and operational parameters of advanced reactors. Here, we focus on the definitions for the reference cases, as the values taken in other scenarios are described in Section \ref{sec:scenarios}.
\subsection{Nuclear Core}
Table \ref{tab:coreinputs} summarizes the input values assumed for reference case nuclear cores. We began by defining the ramp rate to be 25\% of the core's capacity every hour, and the minimum stable output power to be 50\% of the rated capacity. These values are consistent with the NRC licensing documents for the Westinghouse AP1000 reactor \cite{westinghouse_ap1000_2011}, and also match those employed for traditional reactors in other electricity system models \cite{sepulveda_role_2018, schwartz_value_2023}. Although reactors (especially advanced ones) can theoretically ramp much faster than this, in practice, ramping rates are kept low to reduce stress on reactor components \cite{jenkins_benefits_2018}. Also, the minimum stable output of a reactor varies significantly depending on the amount of time the fuel has been in operation, and 50\% is close to the midpoint of this potential range \cite{jenkins_benefits_2018}. Although these parameters are fairly conservative, more flexible values are explored in Section \ref{sec:devflex}. Minimum up and down commitment times are set at 12 hours, a slight improvement from traditional times, to reflect a more flexible advanced core design; again, more and less optimistic values are analyzed in Section \ref{sec:devflex}. Finally, the heat rate of the reactor (heat per unit of fuel) is set to be the same as that for a nuclear reactor from the ATB. This is an area of potential improvement for advanced reactors, but the possible range of values is not yet well defined, so speculating potential values would produce results without justification. Even still, efficiency gains are considered in Section \ref{sec:deveff}.
\par Costs for advanced reactors follow those of reactors in the ATB closely, as there is no literature on their expected operational costs. For variable operation and maintenance (VO\&M) costs, we take the NREL ATB values and subtract the VO\&M costs for the turbine. We take the same approach for startup costs. Startup fuel use is set to zero, as it is zero for reactors in the ATB as well.
\subsection{Thermal Storage}
Table \ref{tab:storinputs} summarizes the input values assumed for the reference case thermal storage.
Thermal storage charge rate is limited by nuclear core and resistive heating power capacity, while discharge rate is limited by the power conversion system capacity, so no explicit rates are modeled. Only a heat decay rate is included to simulate heat loss, and it is derived from a molten salt thermal storage study~\cite{riahi_thermal_2020}.
Resistive heating is widely considered to have 100\% conversion efficiency \cite{stack_performance_2019}.
As for the costs, we assume \$22/kWh$_\mathrm{th}$ for thermal storage \cite{riahi_thermal_2020}, and \$35/kW$_{e}$ for resistive heating \cite{stack_performance_2019}.
\subsection{Power conversion system}
Table \ref{tab:turbinputs} summarizes the input values assumed for the reference case power conversion system, which we sometimes refer to as the ``turbine''.
We assumed ramp rates, minimum stable powers, and minimum up and down commitment times consistent with a stand-alone generator for a concentrated solar power plant \cite{turchi_csp_2019}.
We set the efficiency at 40\%, consistent with molten salt reactors, and a rough midpoint for the potential ranges achievable. We set the startup fuel to zero for the same reasons as the nuclear core.
The investment costs are from the DOE \cite{department_of_energy_combined_2016}, while the FO\&M, VO\&M, and startup costs are obtained from a concentrated solar power study with a stand-alone turbine \cite{turchi_csp_2019}.
Both of these costs are converted into annuities using the same financial assumptions as the nuclear core, under the assumption that the power plant (core + turbine) is financed as one lump sum. 

\newpage

\bibliographystyle{elsarticle-num}
\bibliography{my_bibliography}

\newpage

\setcounter{figure}{0}
\setcounter{table}{0}
\renewcommand{\appendixname}{} 
\section*{Supplemental Figures}\label{sec:supp}
\renewcommand{\thefigure}{S.\arabic{figure}}
\renewcommand{\thetable}{S.\arabic{table}}

\begin{figure}[H] \begin{center}
\includegraphics[width=\textwidth]{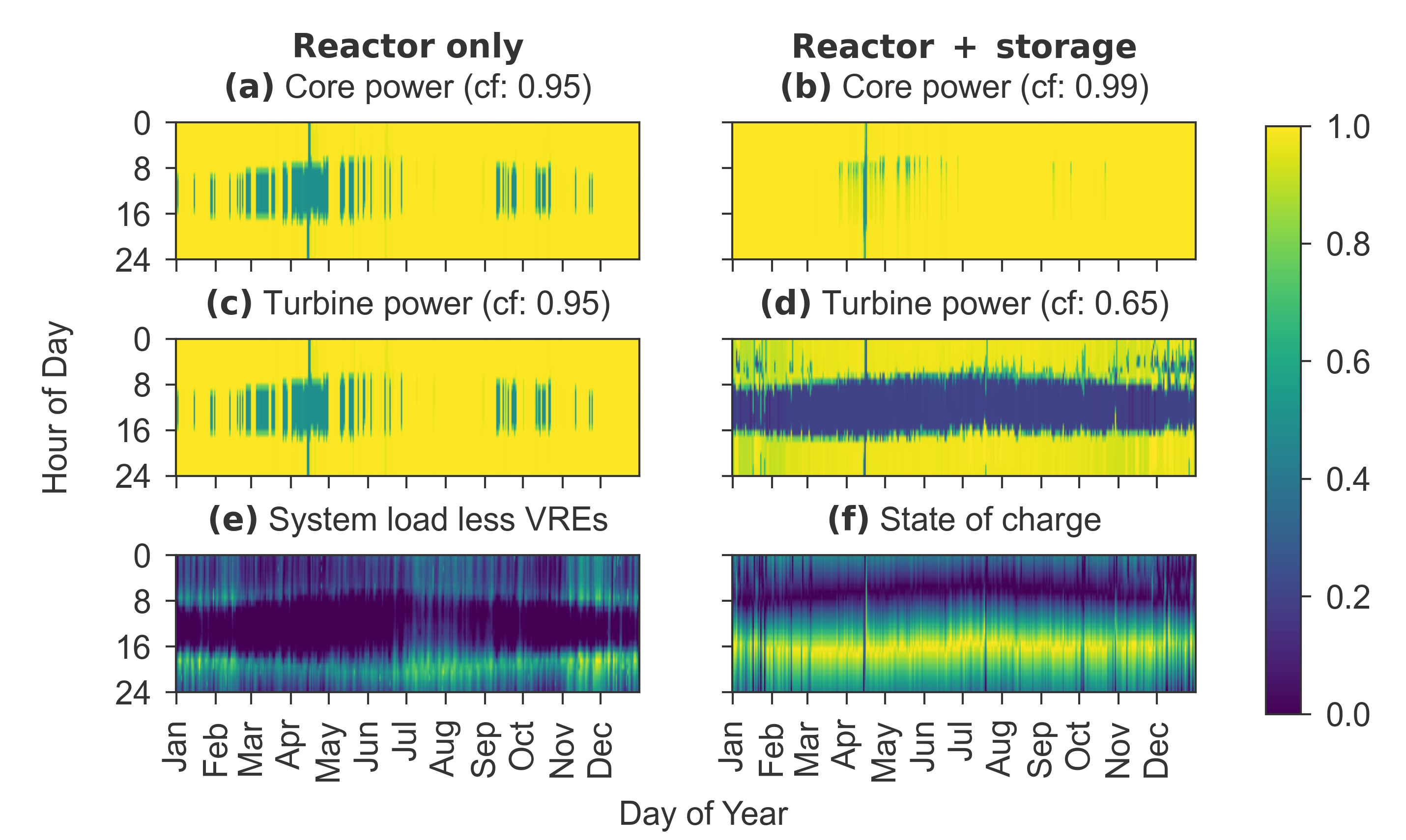}
\caption{Heat maps illustrating advanced reactor operational profiles at 5 GW of nuclear capacity penetration. (a--b) Normalized nuclear core power output for all hours of the year, as well as capacity factors, for reactors with and without thermal storage. (c--d) Normalized turbine electrical power output for all hours of the year, as well as capacity factors, for reactors with and without thermal storage. (e) Normalized system load at all hours of the year, subtracting VRE generation. (f) Normalized state of charge for the thermal tanks at all hours of the year.}\label{fig:op5}
\end{center} \end{figure} 

\begin{figure}[H] \begin{center}
\includegraphics[width=\textwidth]{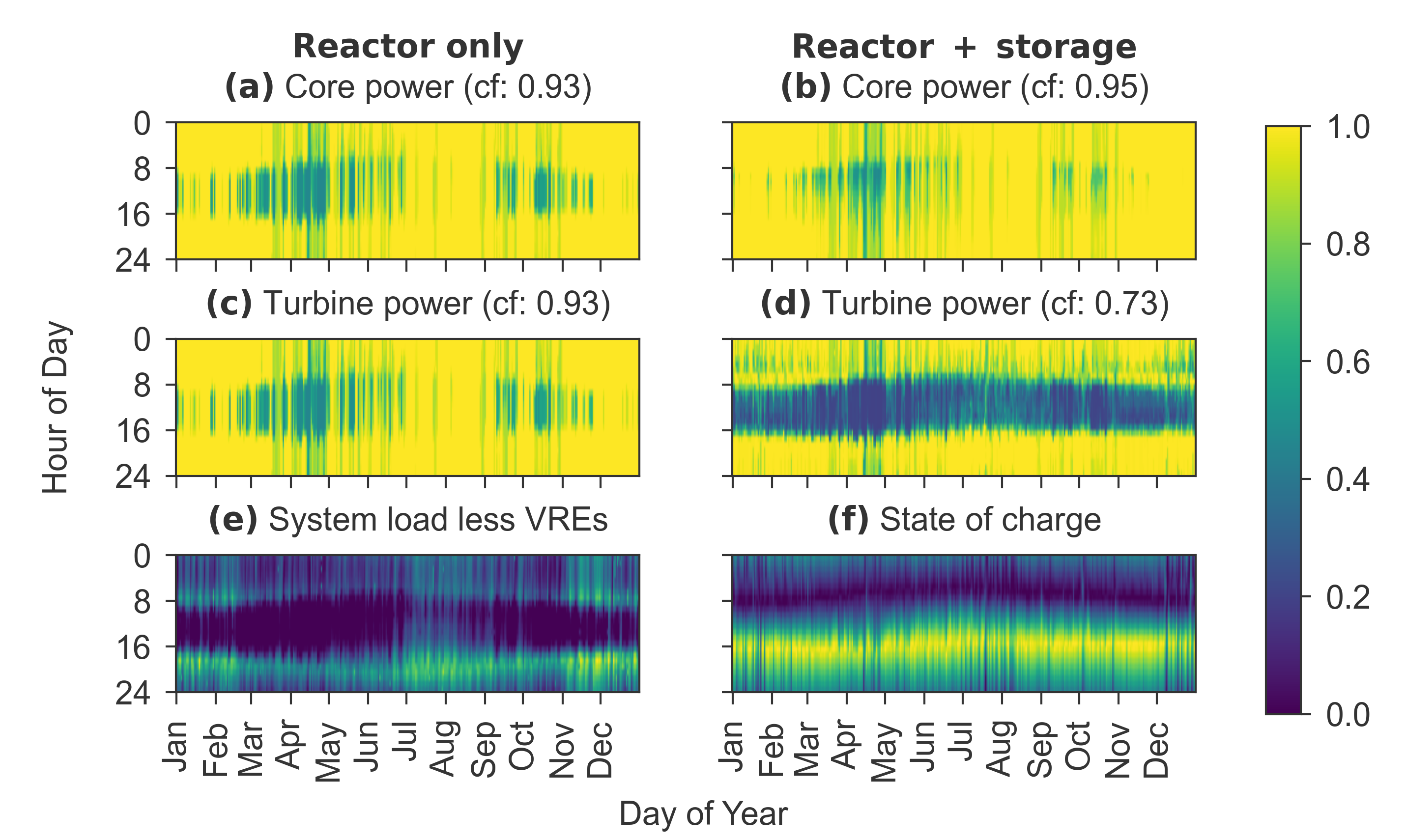}
\caption{Heat maps illustrating advanced reactor operational profiles at 80 GW of nuclear capacity penetration. (a--b) Normalized nuclear core power output for all hours of the year, as well as capacity factors, for reactors with and without thermal storage. (c--d) Normalized turbine electrical power output for all hours of the year, as well as capacity factors, for reactors with and without thermal storage. (e) Normalized system load at all hours of the year, subtracting VRE generation. (f) Normalized state of charge for the thermal tanks at all hours of the year.}\label{fig:op80}
\end{center} \end{figure} 

\begin{figure}[H] \begin{center}
\includegraphics[width=\textwidth]{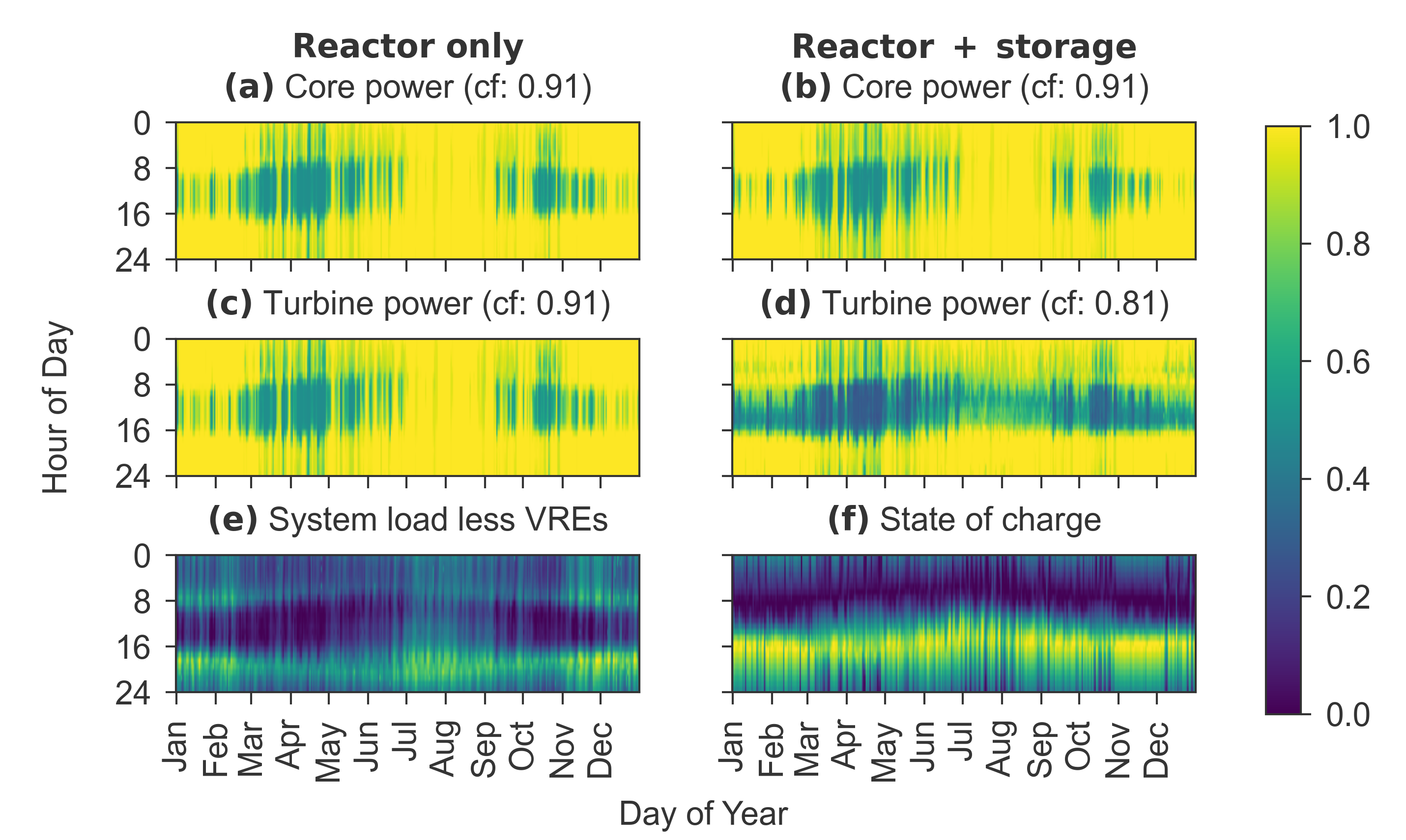}
\caption{Heat maps illustrating advanced reactor operational profiles at 225 GW of nuclear capacity penetration. (a--b) Normalized nuclear core power output for all hours of the year, as well as capacity factors, for reactors with and without thermal storage. (c--d) Normalized turbine electrical power output for all hours of the year, as well as capacity factors, for reactors with and without thermal storage. (e) Normalized system load at all hours of the year, subtracting VRE generation. (f) Normalized state of charge for the thermal tanks at all hours of the year.}\label{fig:op225}
\end{center} \end{figure}

\begin{figure}[H]
\begin{center}
\includegraphics[width=2.5in]{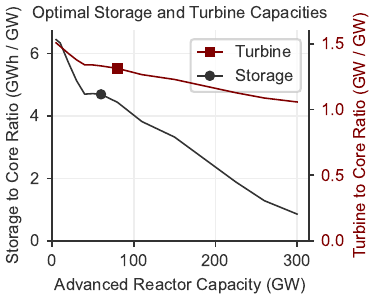}
\caption[Optimal Thermal Storage and Turbine Capacities]{(Left axis) Optimal amounts of thermal storage construction at varying amounts of thermal core construction. (Right axis) Optimal amounts of turbine construction at varying amounts of thermal core construction.}\label{fig:ratios}
\end{center}
\end{figure}

\begin{figure}[H]
\begin{center}
\includegraphics[width=\textwidth]{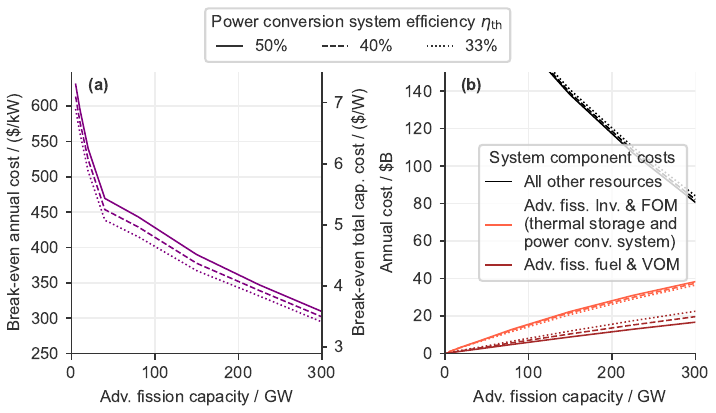}
\caption[The Value of Efficiency]{Efficiency mostly increases a plant's value by decreasing its fuel costs.
Part (a): Nuclear reactor break-even cost as a function of capacity penetration for designs with varying turbine conversion efficiencies (50\% for gas-cooled, 40\% for salt-cooled, and 33\% for water-cooled reactors). Left axis expresses break-even cost as an annuity, while right axis expresses it as a total capital cost (assuming 30 year asset life, 4.5\% WACC). Break-even costs are calculated for a ``base plant'' without excess turbine capacity or thermal storage, (\ref{sec:cost}). Part (b): costs of selected components of the electricity system.
The sum of the investment, fixed operations and maintenance (FO\&M), fuel and variable costs from all other resources (black) decreases slightly as advanced fission becomes more efficient; this is mostly balanced by an increased investment in thermal storage and power conversion systems at the advanced fission plants (light red) since the effective cost of thermal storage is lower.
Most of the increase in a the marginal value of a plant, seen in Part (a) is from reduction the cost of uranium fuel (dark red) with increased efficiency.
}\label{fig:eff}
\end{center}
\end{figure}


\newpage
\begin{figure}[H]
\begin{center}
\includegraphics[width=1.0\textwidth]{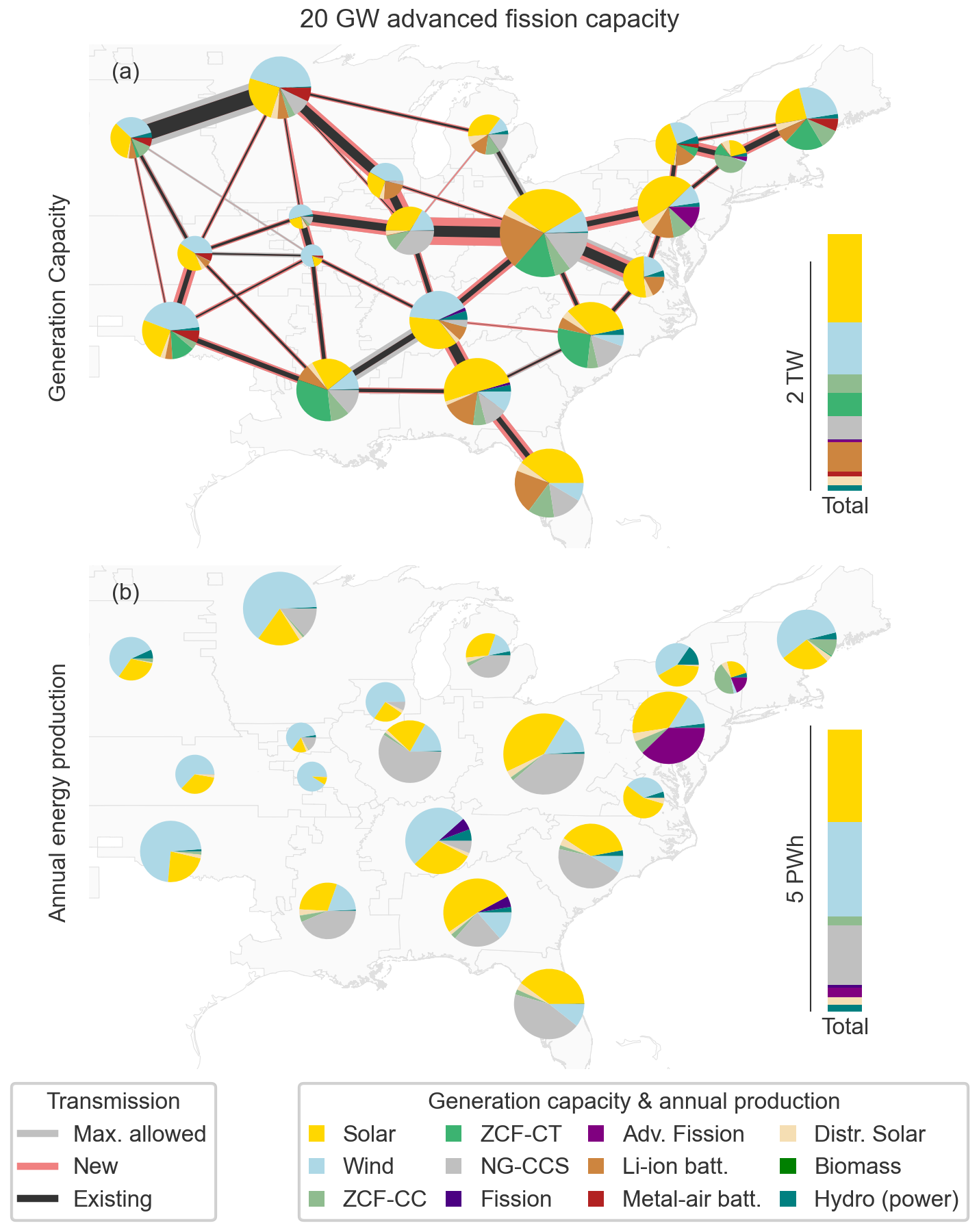}
\caption{Part (a): zonal generation capacity by resource type and transmission line capacity between zones. Part (b): energy production by resource type. Fission reactor with thermal storage. Quantities are proportional to the chart area. }\label{fig:capmapThermalRTS20}
\end{center}
\end{figure}

\newpage
\begin{figure}[H]
\begin{center}
\includegraphics[width=1.0\textwidth]{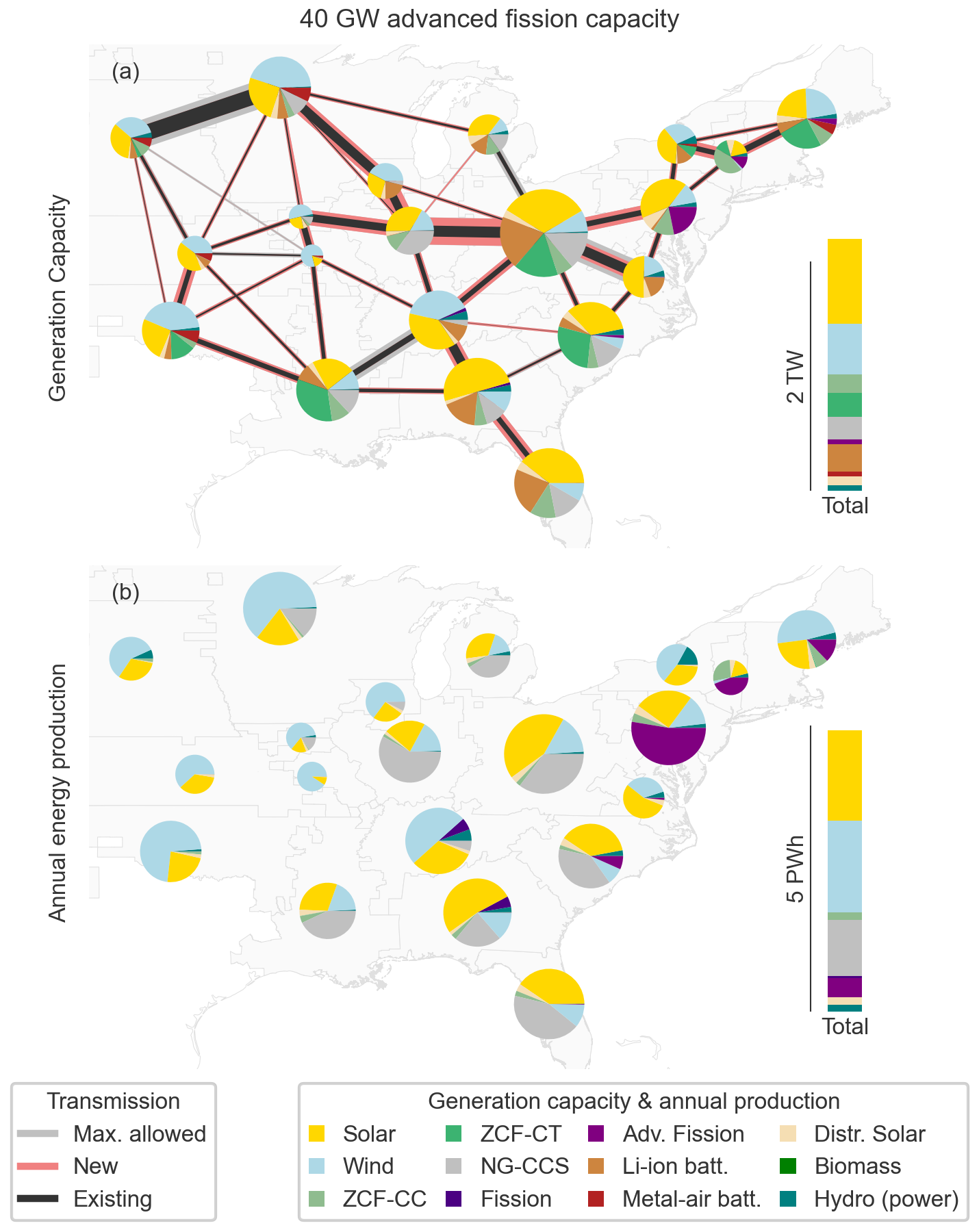}
\caption{Part (a): zonal generation capacity by resource type and transmission line capacity between zones. Part (b): energy production by resource type. Fission reactor with thermal storage. Quantities are proportional to the chart area. }\label{fig:capmapThermalRTS40}
\end{center}
\end{figure}

\newpage
\begin{figure}[H]
\begin{center}
\includegraphics[width=1.0\textwidth]{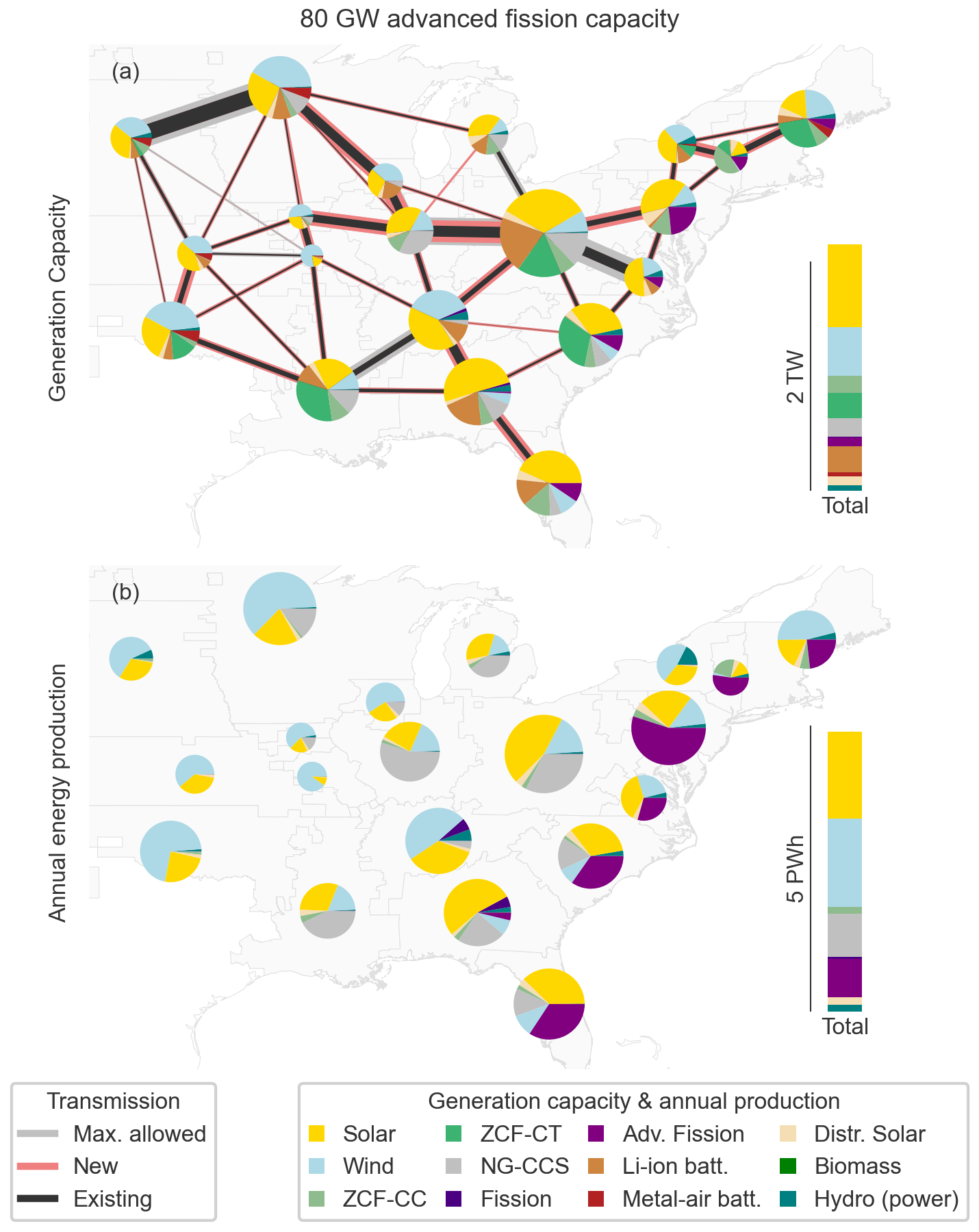}
\caption{Part (a): zonal generation capacity by resource type and transmission line capacity between zones. Part (b): energy production by resource type. Fission reactor with thermal storage. Quantities are proportional to the chart area. }\label{fig:capmapThermalRTS80}
\end{center}
\end{figure}

\newpage
\begin{figure}[H]
\begin{center}
\includegraphics[width=1.0\textwidth]{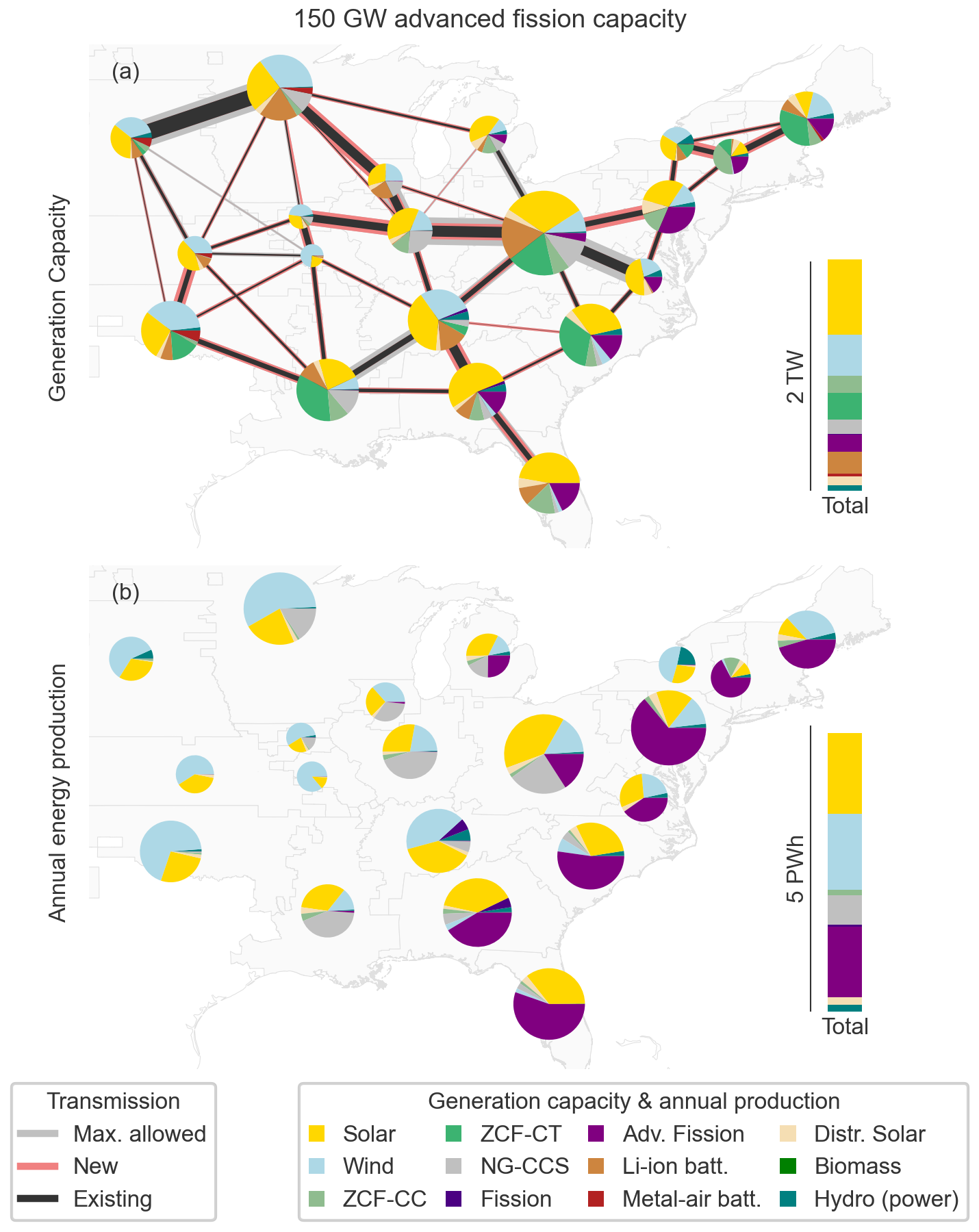}
\caption{Part (a): zonal generation capacity by resource type and transmission line capacity between zones. Part (b): energy production by resource type. Fission reactor with thermal storage. Quantities are proportional to the chart area. }\label{fig:capmapThermalRTS150}
\end{center}
\end{figure}

\newpage
\begin{figure}[H]
\begin{center}
\includegraphics[width=1.0\textwidth]{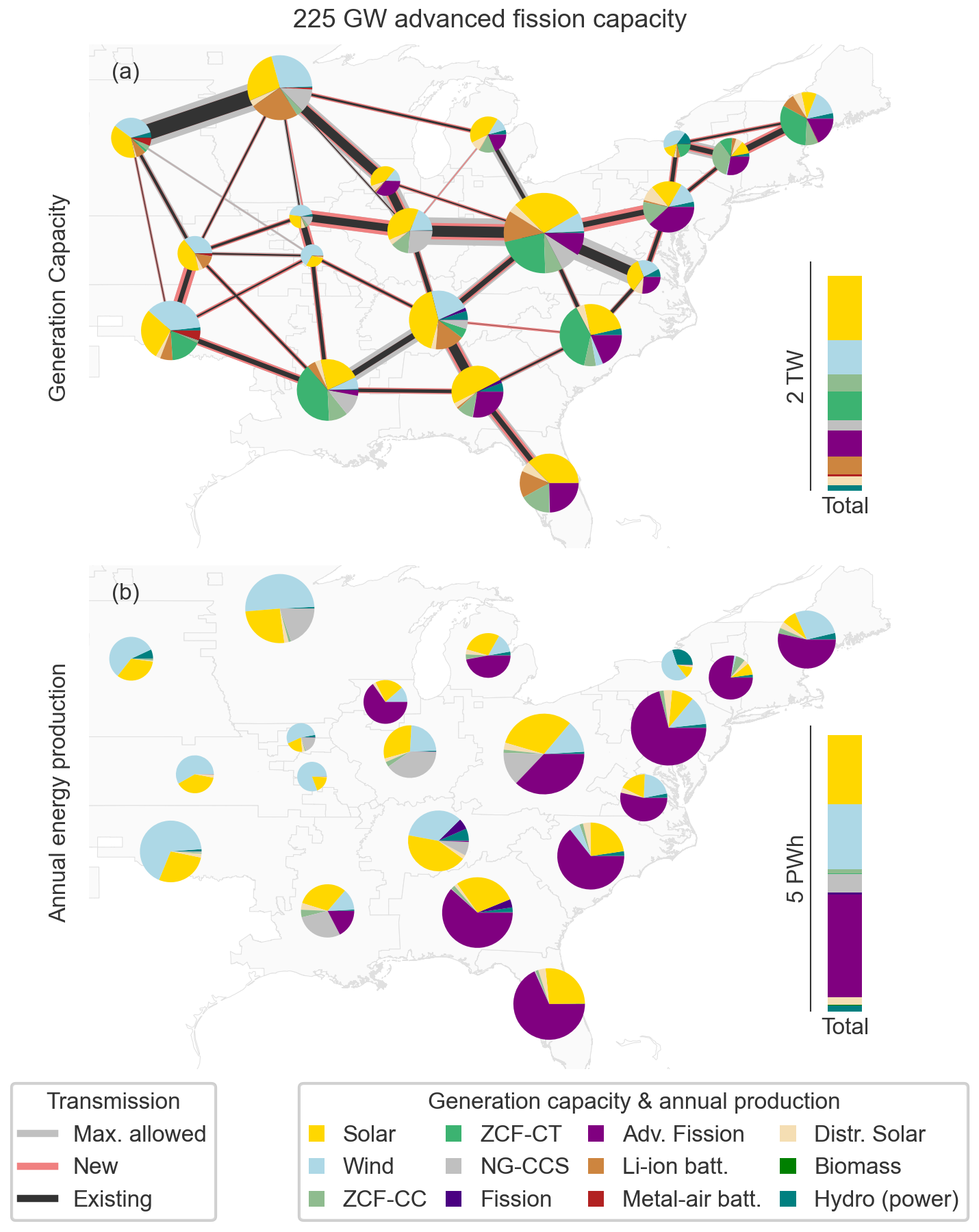}
\caption{Part (a): zonal generation capacity by resource type and transmission line capacity between zones. Part (b): energy production by resource type. Fission reactor with thermal storage. Quantities are proportional to the chart area. }\label{fig:capmapThermalRTS225}
\end{center}
\end{figure}

\newpage
\begin{figure}[H]
\begin{center}
\includegraphics[width=1.0\textwidth]{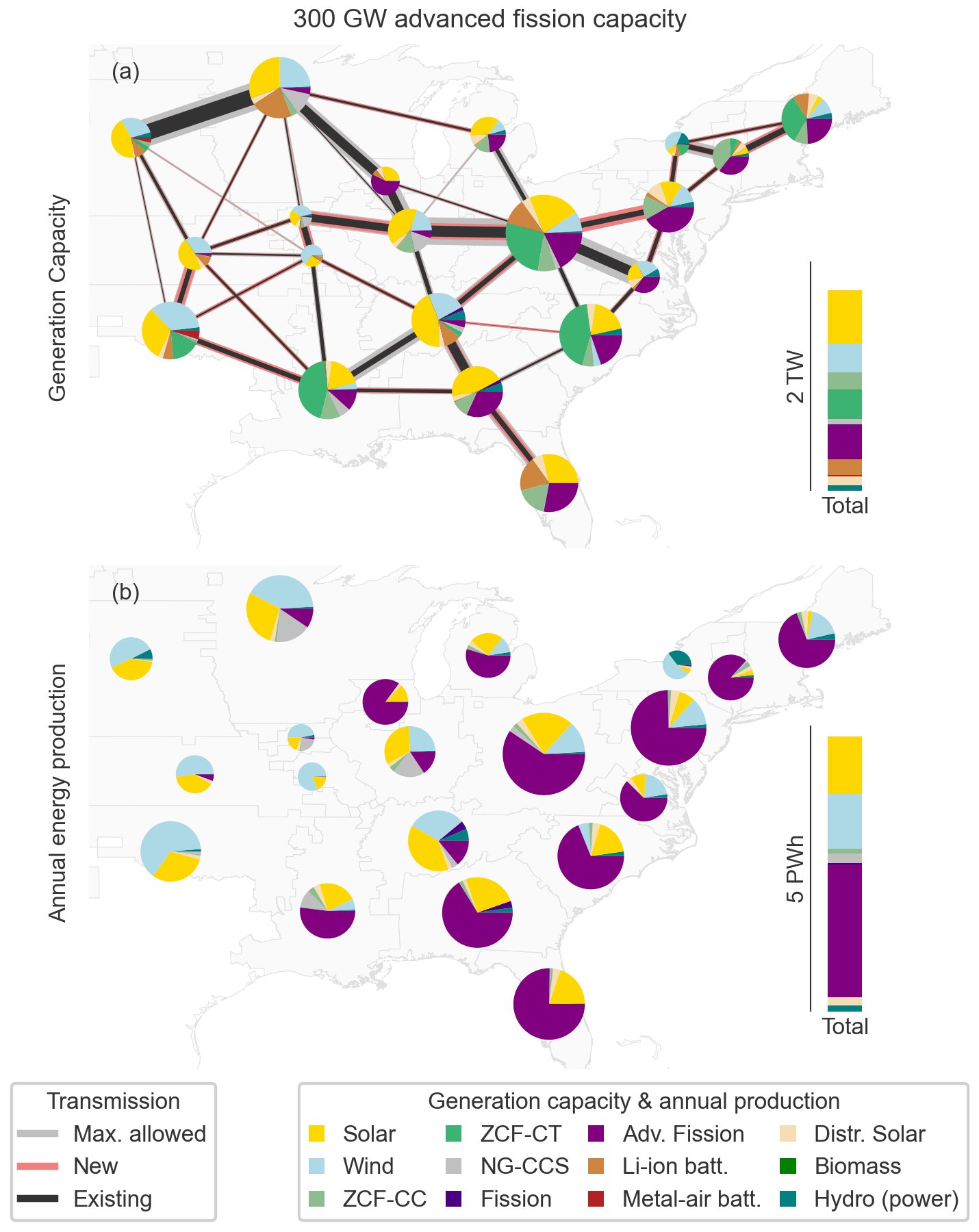}
\caption{Part (a): zonal generation capacity by resource type and transmission line capacity between zones. Part (b): energy production by resource type. Fission reactor with thermal storage. Quantities are proportional to the chart area. }\label{fig:capmapThermalRTS300}
\end{center}
\end{figure}


\newpage
\begin{figure}[H]
\begin{center}
\includegraphics[width=1.0\textwidth]{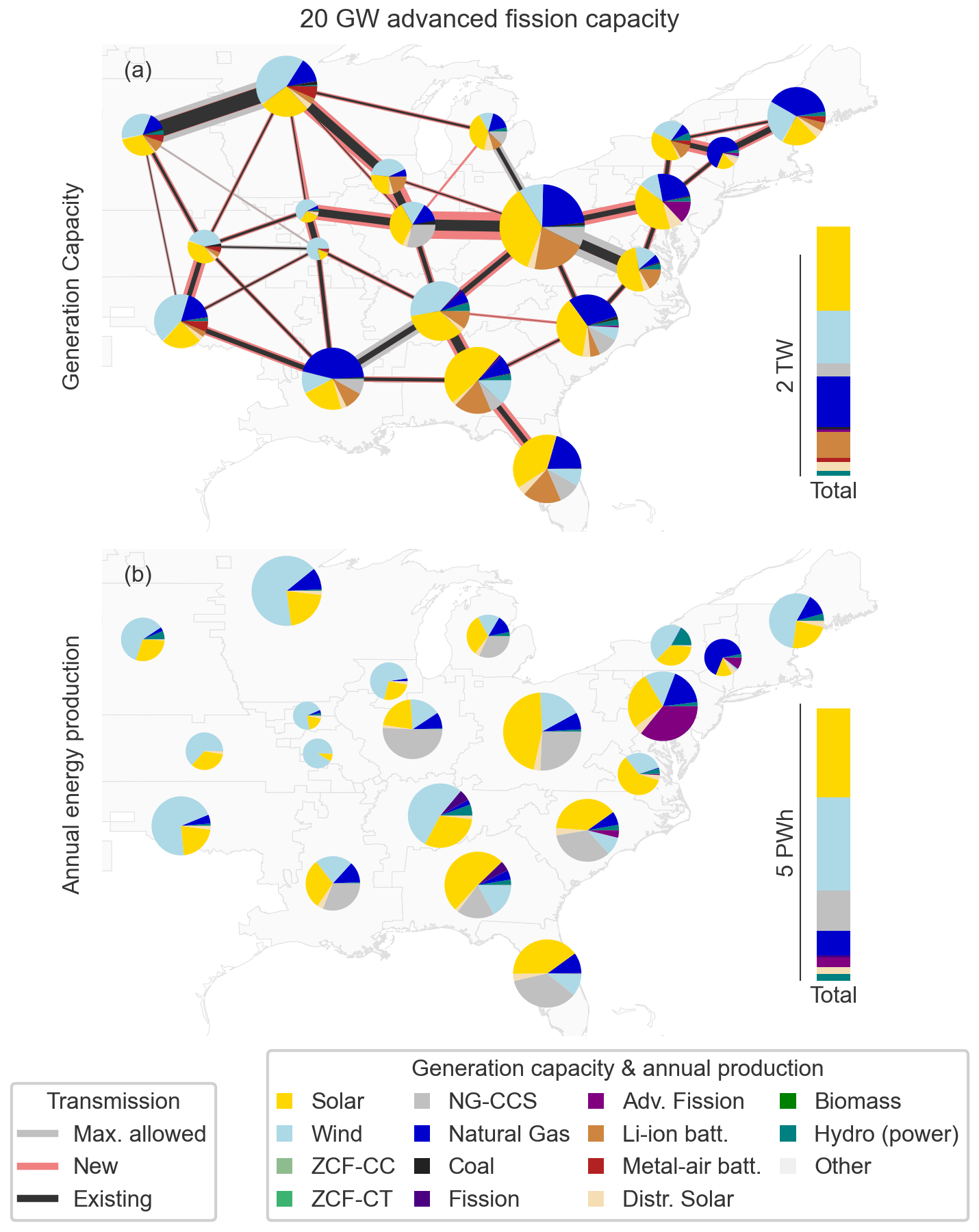}
\caption{Part (a): zonal generation capacity by resource type and transmission line capacity between zones. Part (b): energy production by resource type. Fission reactor with thermal storage, 95\% carbon emissions reduction case. Quantities are proportional to the chart area. }\label{fig:capmap_C9520}
\end{center}
\end{figure}

\newpage
\begin{figure}[H]
\begin{center}
\includegraphics[width=1.0\textwidth]{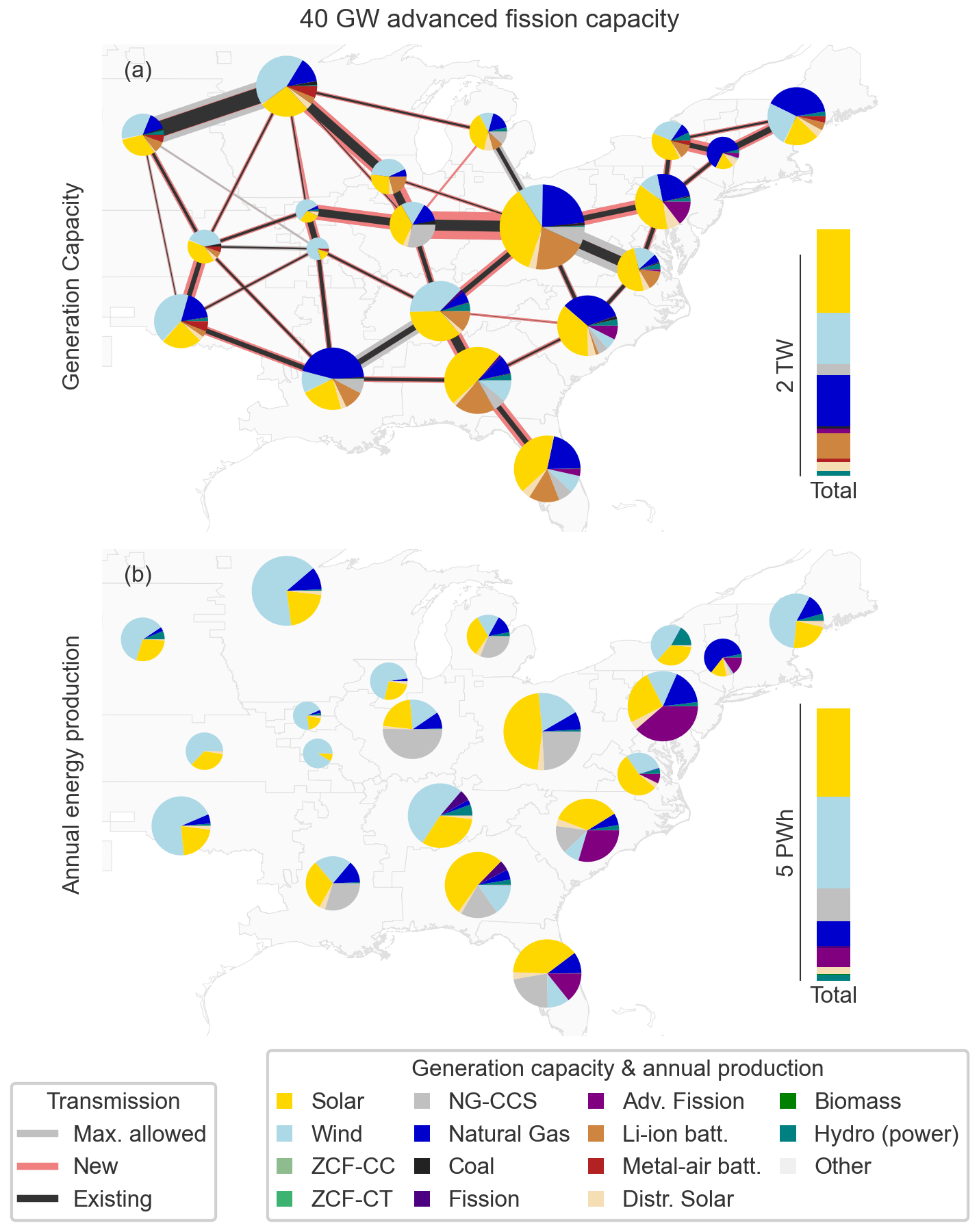}
\caption{Part (a): zonal generation capacity by resource type and transmission line capacity between zones. Part (b): energy production by resource type. Fission reactor with thermal storage, 95\% carbon emissions reduction case. Quantities are proportional to the chart area. }\label{fig:capmap_C9540}
\end{center}
\end{figure}

\newpage
\begin{figure}[H]
\begin{center}
\includegraphics[width=1.0\textwidth]{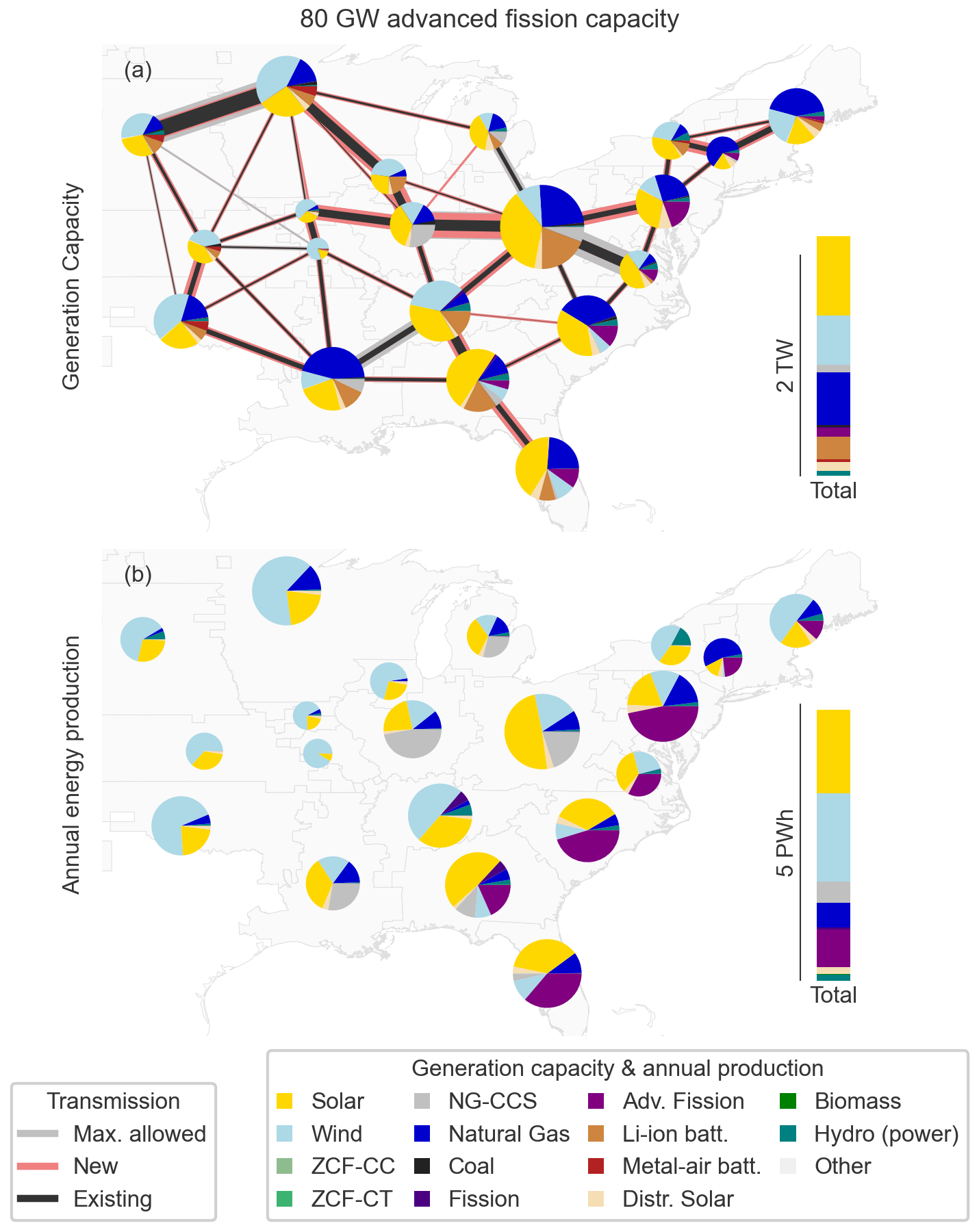}
\caption{Part (a): zonal generation capacity by resource type and transmission line capacity between zones. Part (b): energy production by resource type. Fission reactor with thermal storage, 95\% carbon emissions reduction case. Quantities are proportional to the chart area. }\label{fig:capmap_C9580}
\end{center}
\end{figure}

\newpage
\begin{figure}[H]
\begin{center}
\includegraphics[width=1.0\textwidth]{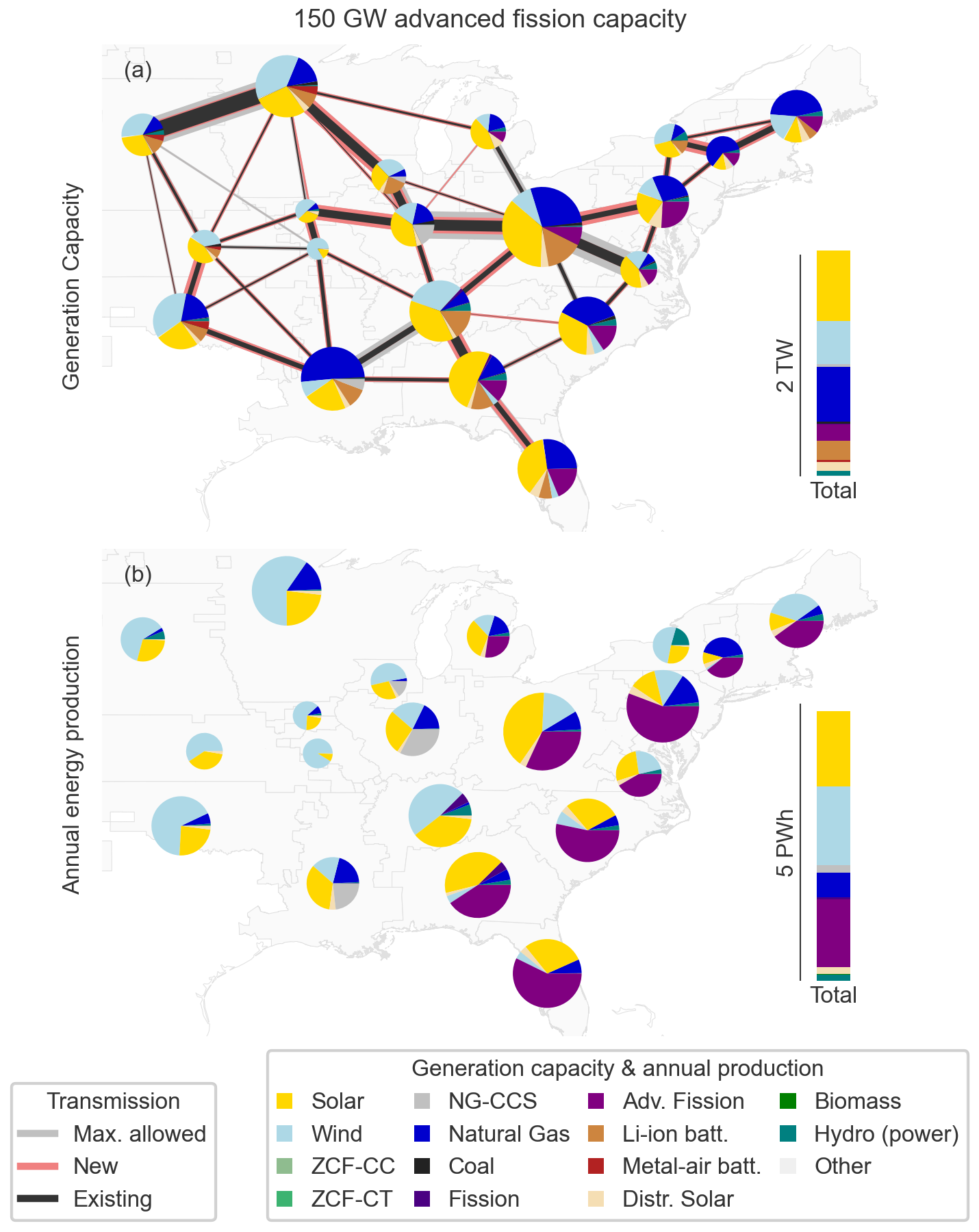}
\caption{Part (a): zonal generation capacity by resource type and transmission line capacity between zones. Part (b): energy production by resource type. Fission reactor with thermal storage, 95\% carbon emissions reduction case. Quantities are proportional to the chart area. }\label{fig:capmap_C95150}
\end{center}
\end{figure}

\newpage
\begin{figure}[H]
\begin{center}
\includegraphics[width=1.0\textwidth]{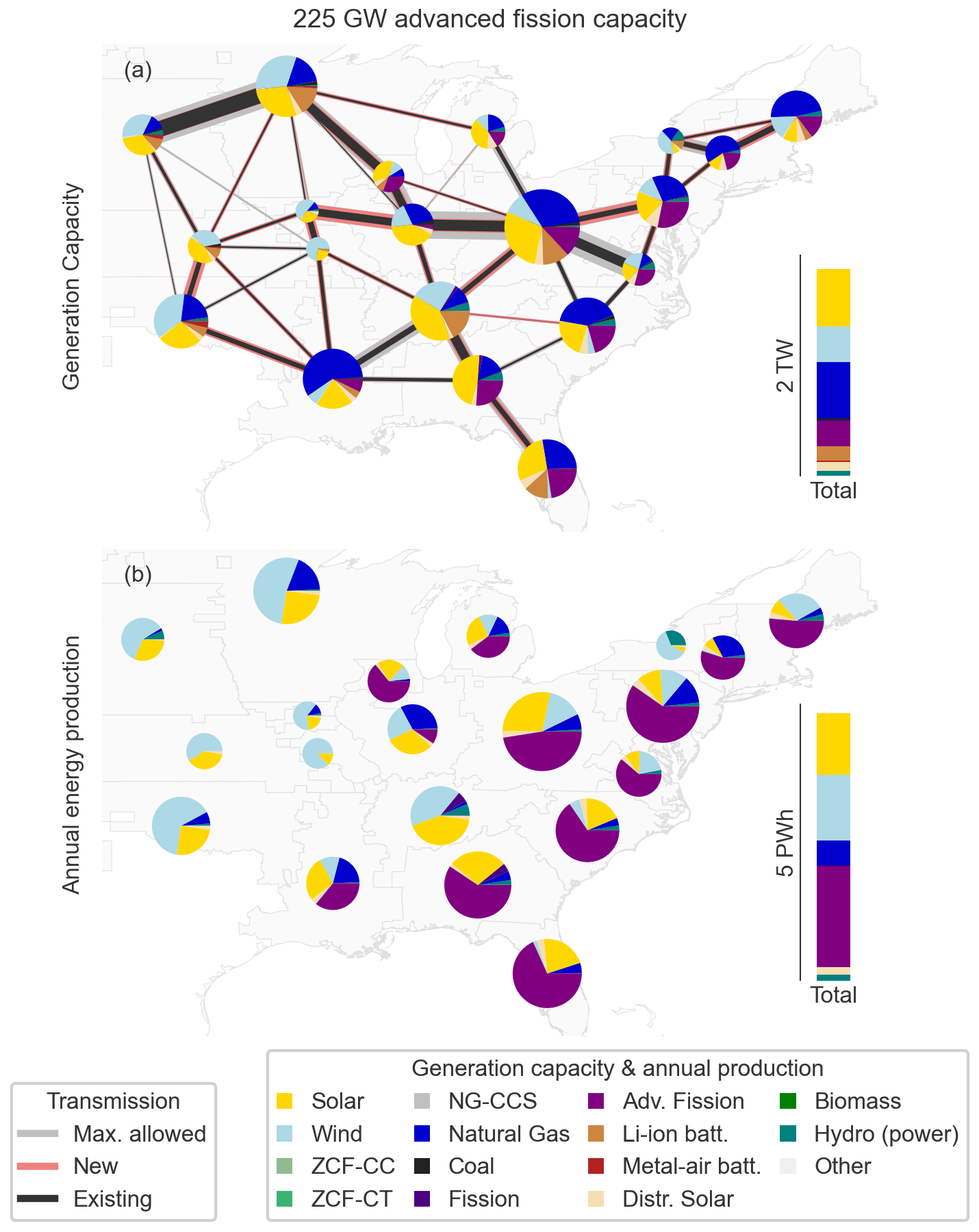}
\caption{Part (a): zonal generation capacity by resource type and transmission line capacity between zones. Part (b): energy production by resource type. Fission reactor with thermal storage, 95\% carbon emissions reduction case. Quantities are proportional to the chart area. }\label{fig:capmap_C95225}
\end{center}
\end{figure}

\newpage
\begin{figure}[H]
\begin{center}
\includegraphics[width=1.0\textwidth]{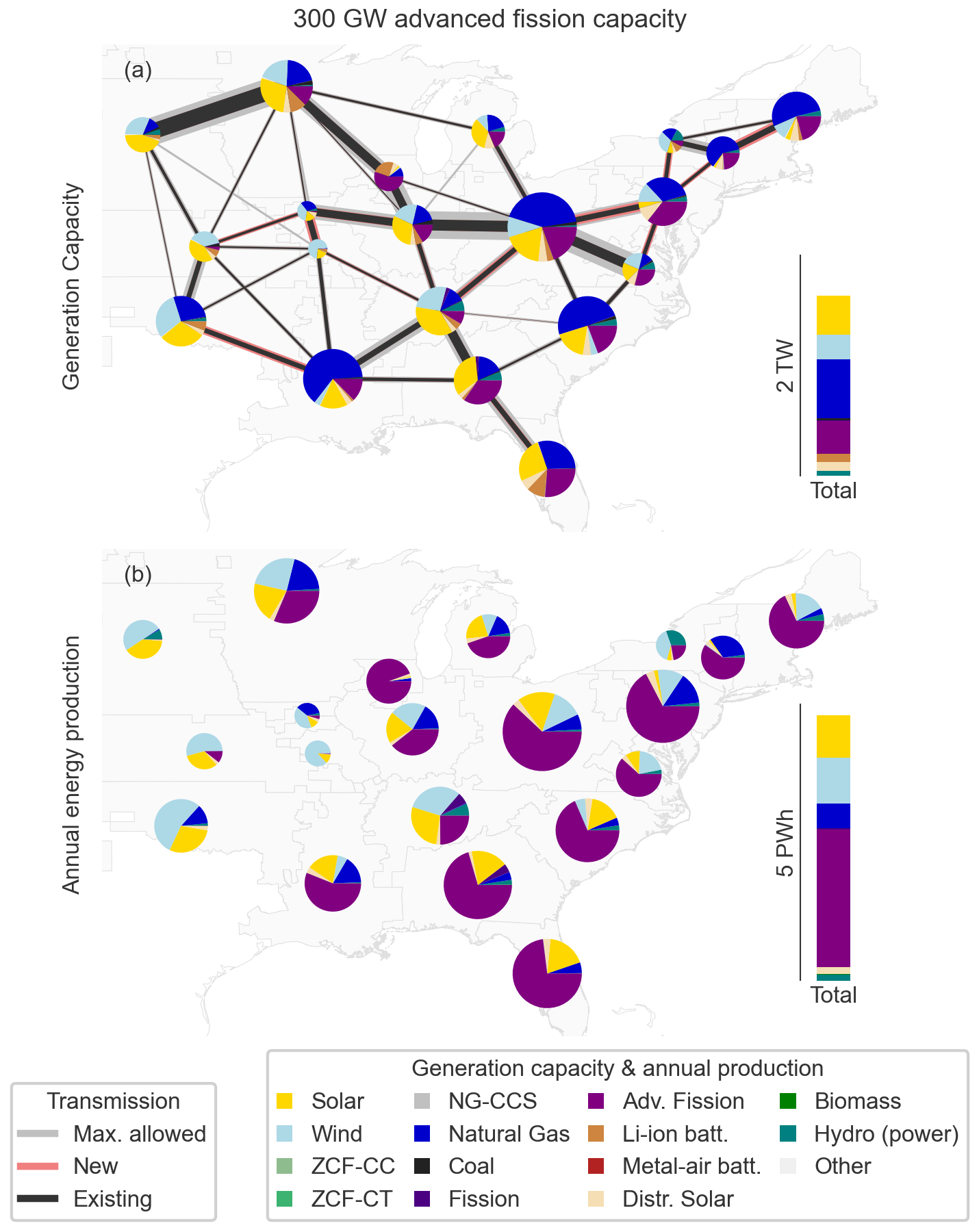}
\caption{Part (a): zonal generation capacity by resource type and transmission line capacity between zones. Part (b): energy production by resource type. Fission reactor with thermal storage, 95\% carbon emissions reduction case. Quantities are proportional to the chart area. }\label{fig:capmap_C95300}
\end{center}
\end{figure}


\newpage
\begin{figure}[H]
\begin{center}
\includegraphics[width=1.0\textwidth]{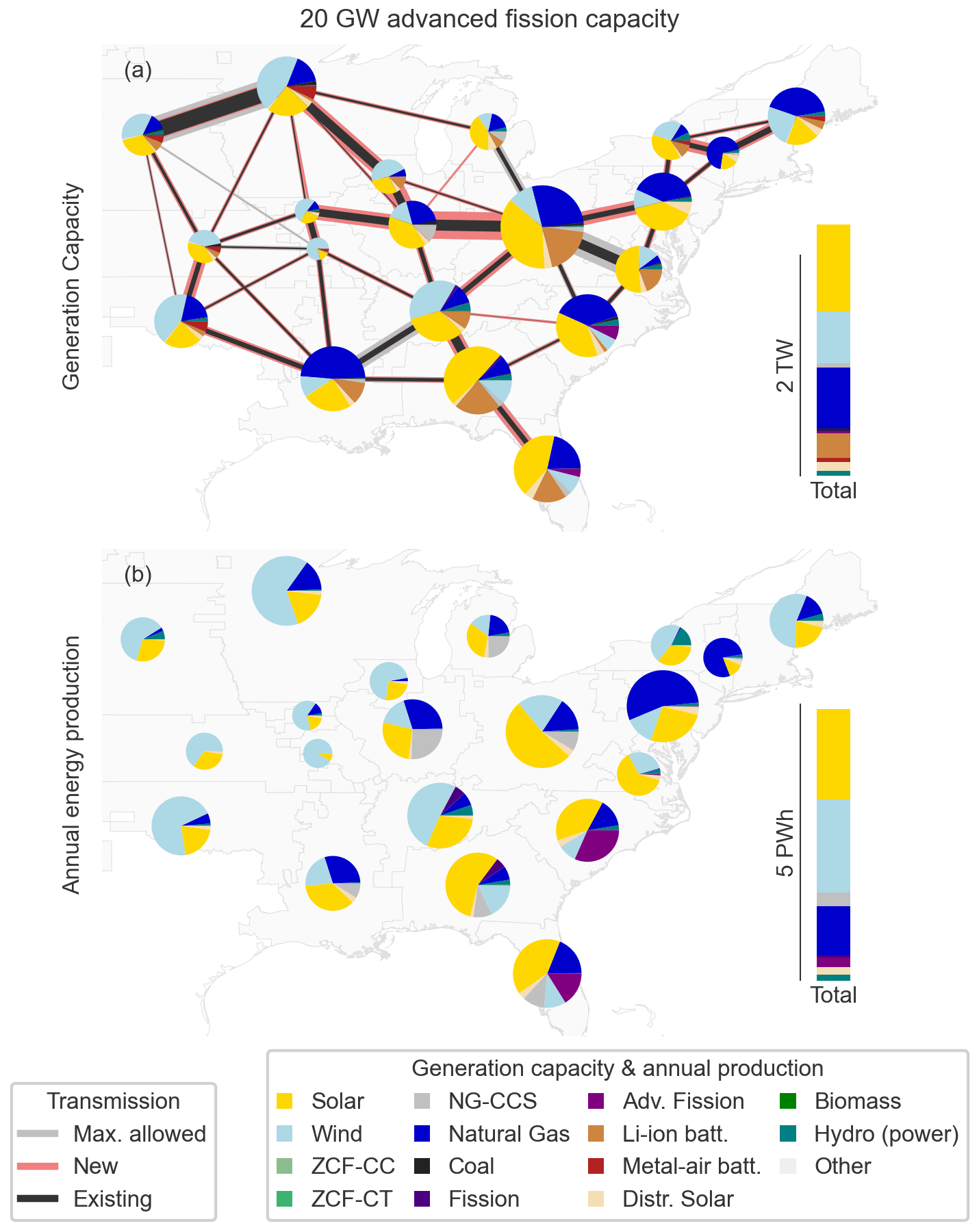}
\caption{Part (a): zonal generation capacity by resource type and transmission line capacity between zones. Part (b): energy production by resource type. Fission reactor with thermal storage, 90\% carbon emissions reduction case. Quantities are proportional to the chart area. }\label{fig:capmap_C9020}
\end{center}
\end{figure}

\newpage
\begin{figure}[H]
\begin{center}
\includegraphics[width=1.0\textwidth]{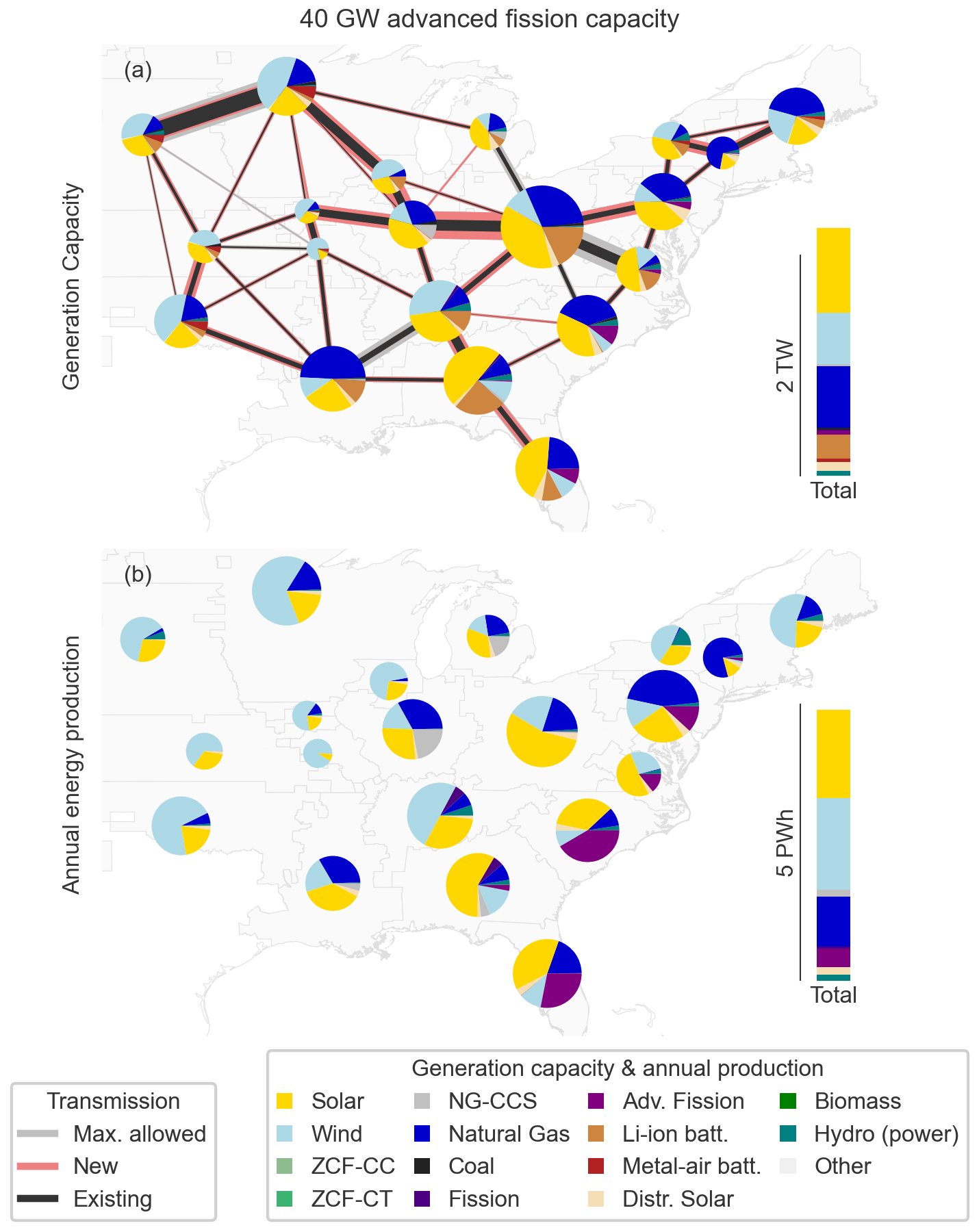}
\caption{Part (a): zonal generation capacity by resource type and transmission line capacity between zones. Part (b): energy production by resource type. Fission reactor with thermal storage, 90\% carbon emissions reduction case. Quantities are proportional to the chart area. }\label{fig:capmap_C9040}
\end{center}
\end{figure}

\newpage
\begin{figure}[H]
\begin{center}
\includegraphics[width=1.0\textwidth]{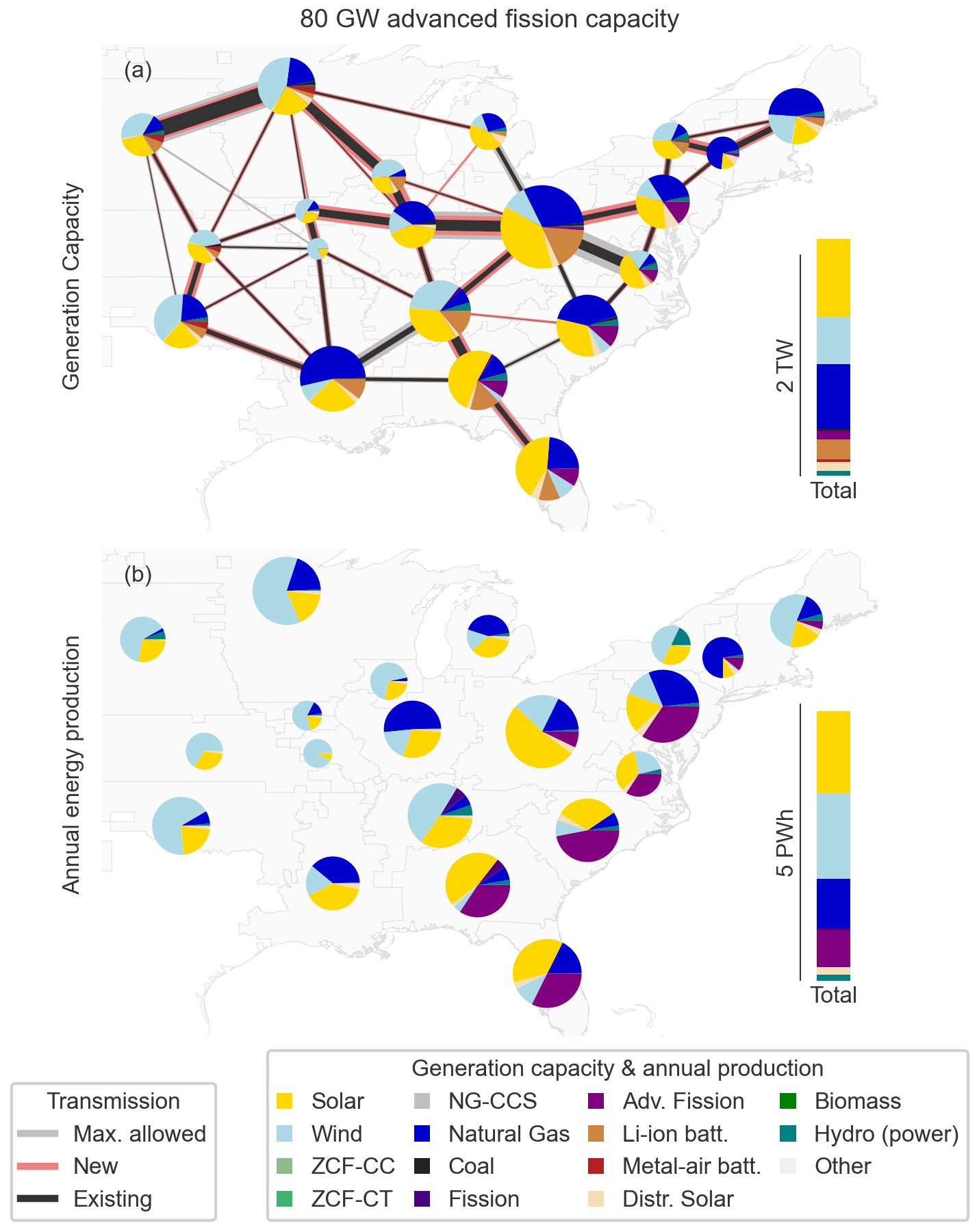}
\caption{Part (a): zonal generation capacity by resource type and transmission line capacity between zones. Part (b): energy production by resource type. Fission reactor with thermal storage, 90\% carbon emissions reduction case. Quantities are proportional to the chart area. }\label{fig:capmap_C9080}
\end{center}
\end{figure}

\newpage
\begin{figure}[H]
\begin{center}
\includegraphics[width=1.0\textwidth]{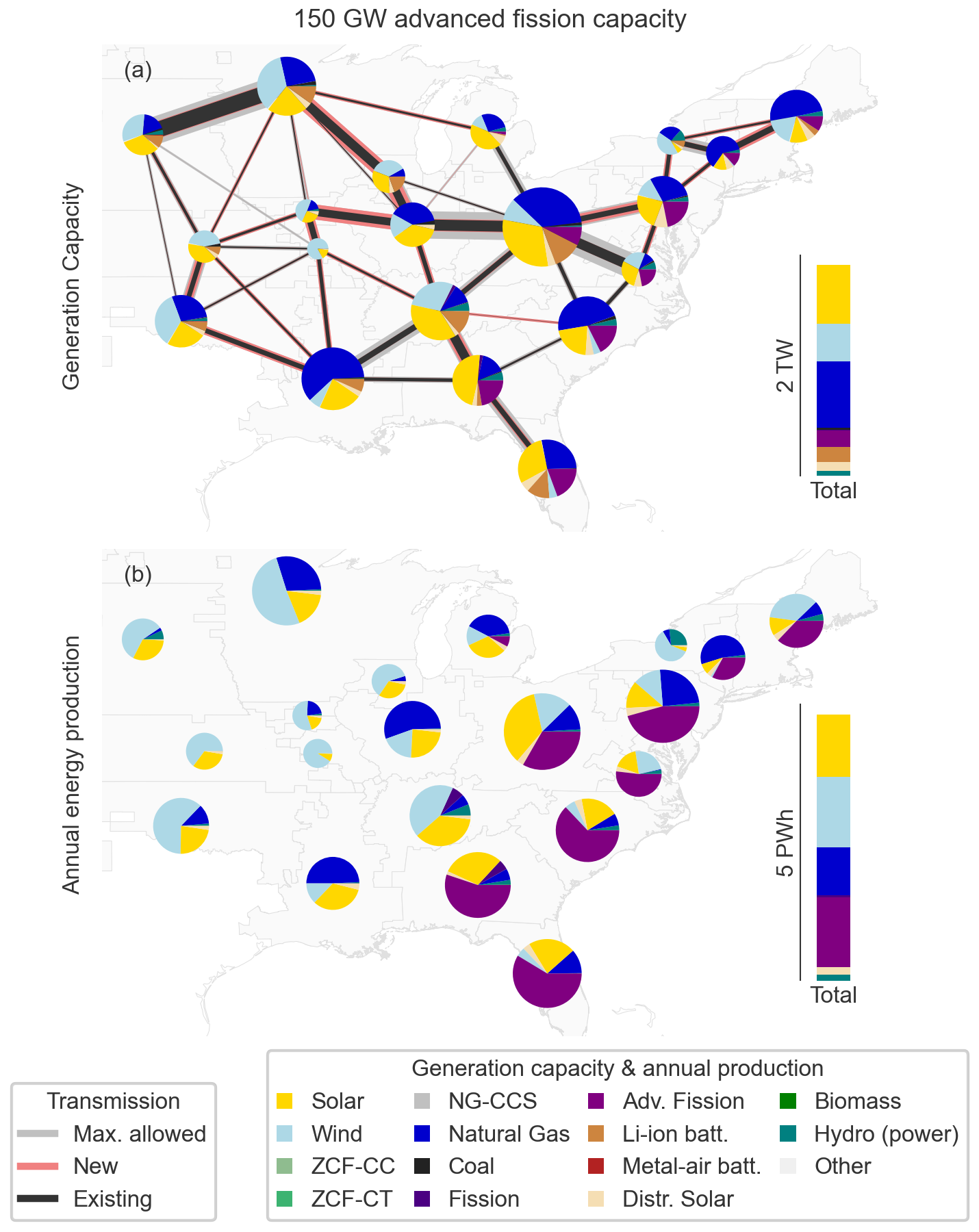}
\caption{Part (a): zonal generation capacity by resource type and transmission line capacity between zones. Part (b): energy production by resource type. Fission reactor with thermal storage, 90\% carbon emissions reduction case. Quantities are proportional to the chart area. }\label{fig:capmap_C90150}
\end{center}
\end{figure}

\newpage
\begin{figure}[H]
\begin{center}
\includegraphics[width=1.0\textwidth]{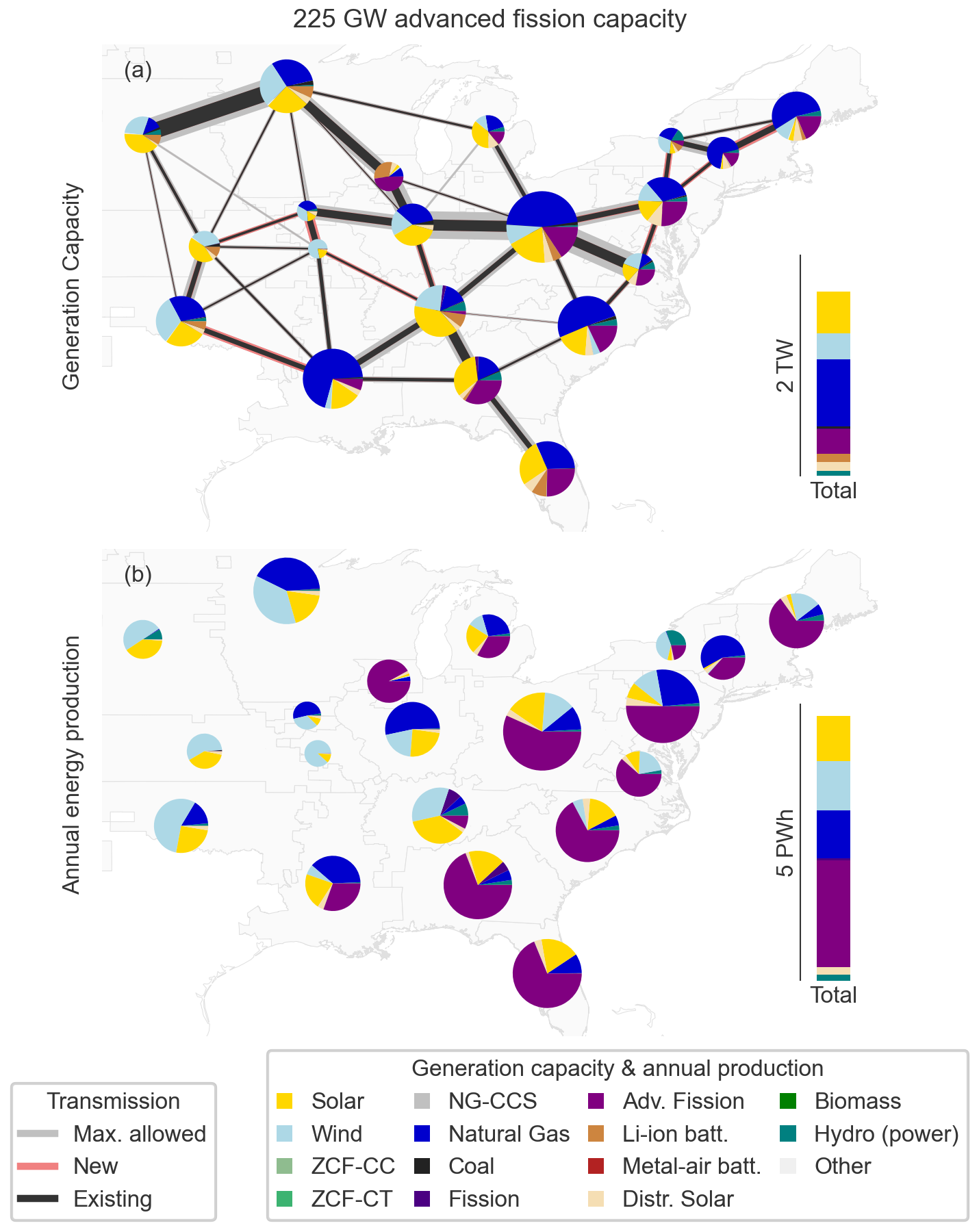}
\caption{Part (a): zonal generation capacity by resource type and transmission line capacity between zones. Part (b): energy production by resource type. Fission reactor with thermal storage, 90\% carbon emissions reduction case. Quantities are proportional to the chart area. }\label{fig:capmap_C90225}
\end{center}
\end{figure}

\newpage
\begin{figure}[H]
\begin{center}
\includegraphics[width=1.0\textwidth]{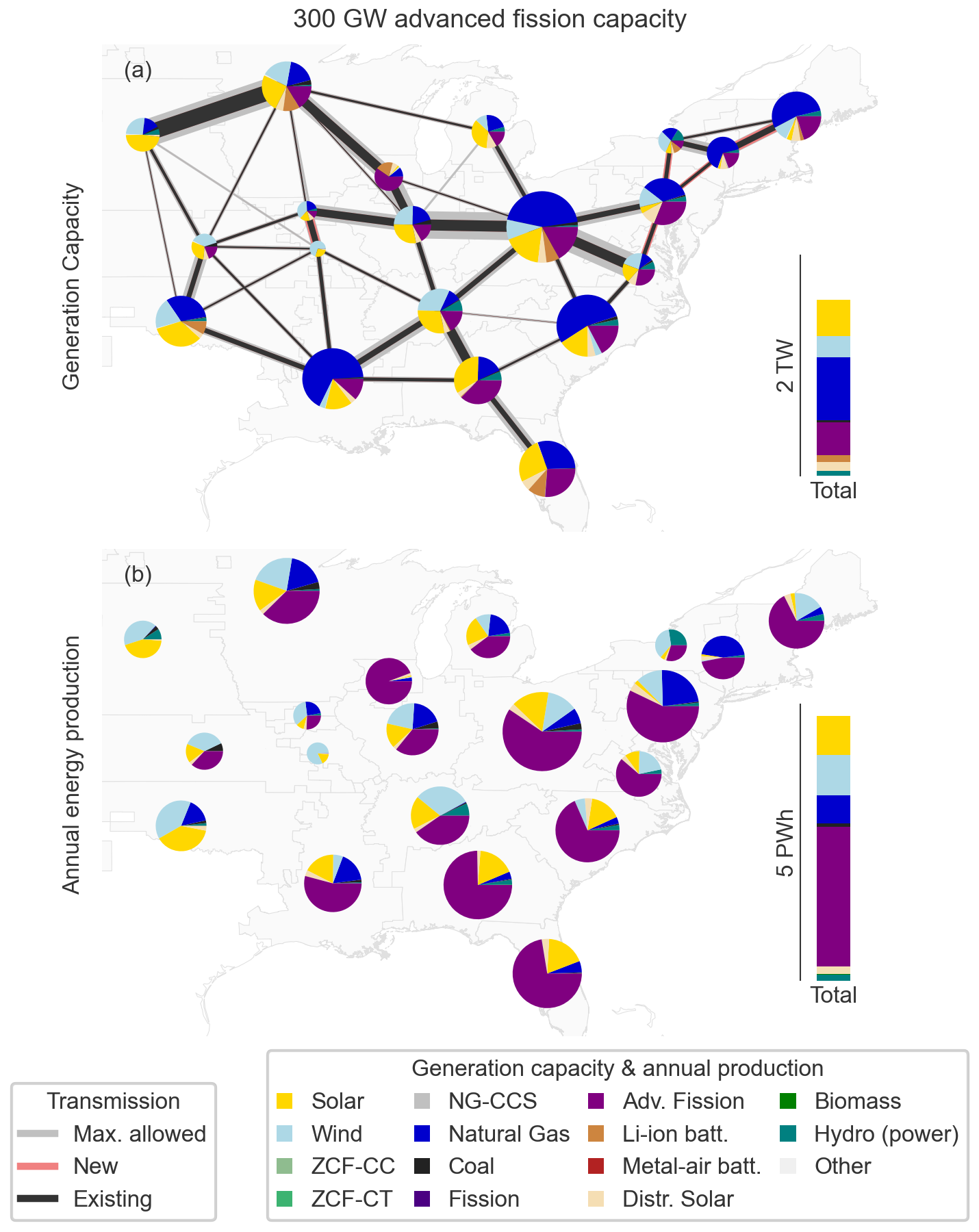}
\caption{Part (a): zonal generation capacity by resource type and transmission line capacity between zones. Part (b): energy production by resource type. Fission reactor with thermal storage, 90\% carbon emissions reduction case. Quantities are proportional to the chart area. }\label{fig:capmap_C90300}
\end{center}
\end{figure}


\newpage
\begin{figure}[H]
\begin{center}
\includegraphics[width=1.0\textwidth]{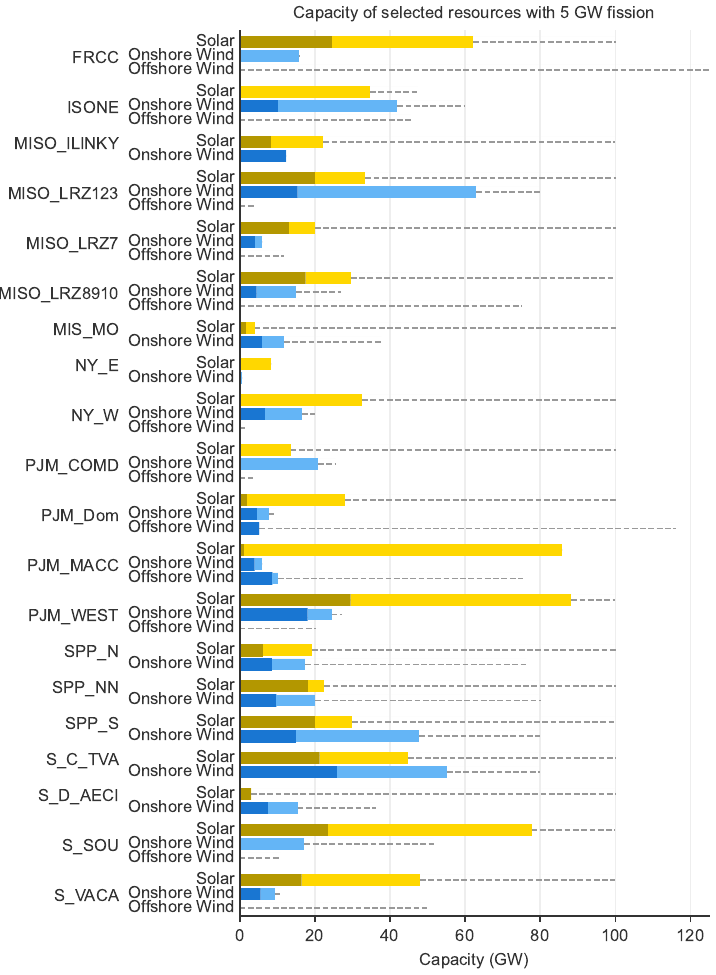}
\caption{Solar, onshore wind, and offshore wind capacity by zone in the reference scenario. Each bar shows the existing capacity in 2035 (darker, hashed), the capacity built 2035--2050 (solid), and the remaining capacity up to the maximum which could be built (dashed line).
 }\label{fig:buildout_clusters_thermal_r_5}
\end{center}
\end{figure}

\newpage
\begin{figure}[H]
\begin{center}
\includegraphics[width=1.0\textwidth]{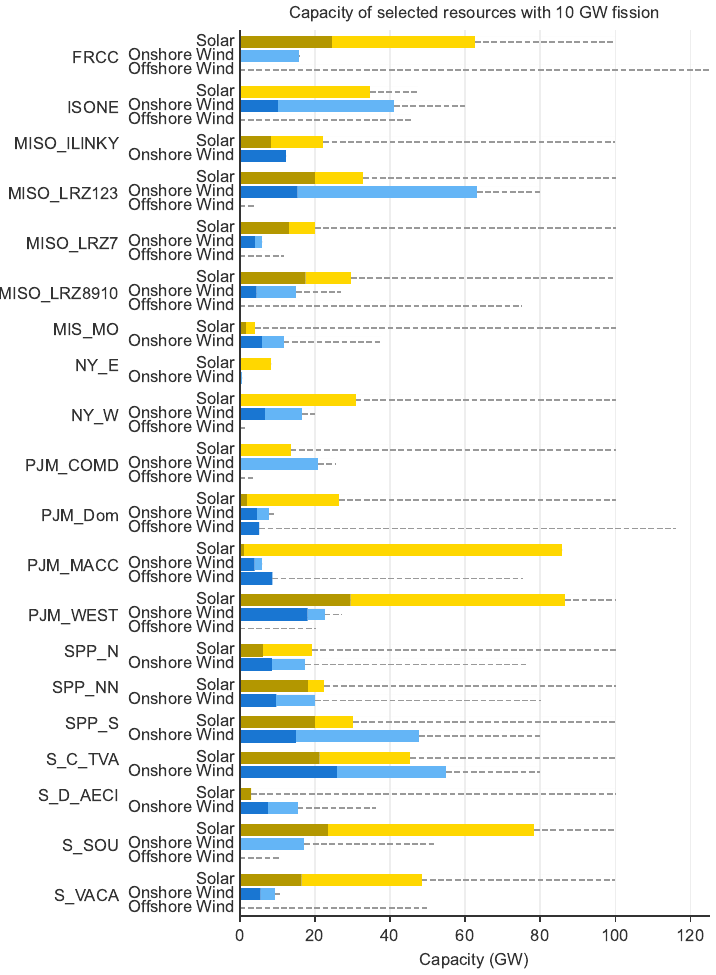}
\caption{Solar, onshore wind, and offshore wind capacity by zone in the reference scenario. Each bar shows the existing capacity in 2035 (darker, hashed), the capacity built 2035--2050 (solid), and the remaining capacity up to the maximum which could be built (dashed line).
 }\label{fig:buildout_clusters_thermal_r_10}
\end{center}
\end{figure}

\newpage
\begin{figure}[H]
\begin{center}
\includegraphics[width=1.0\textwidth]{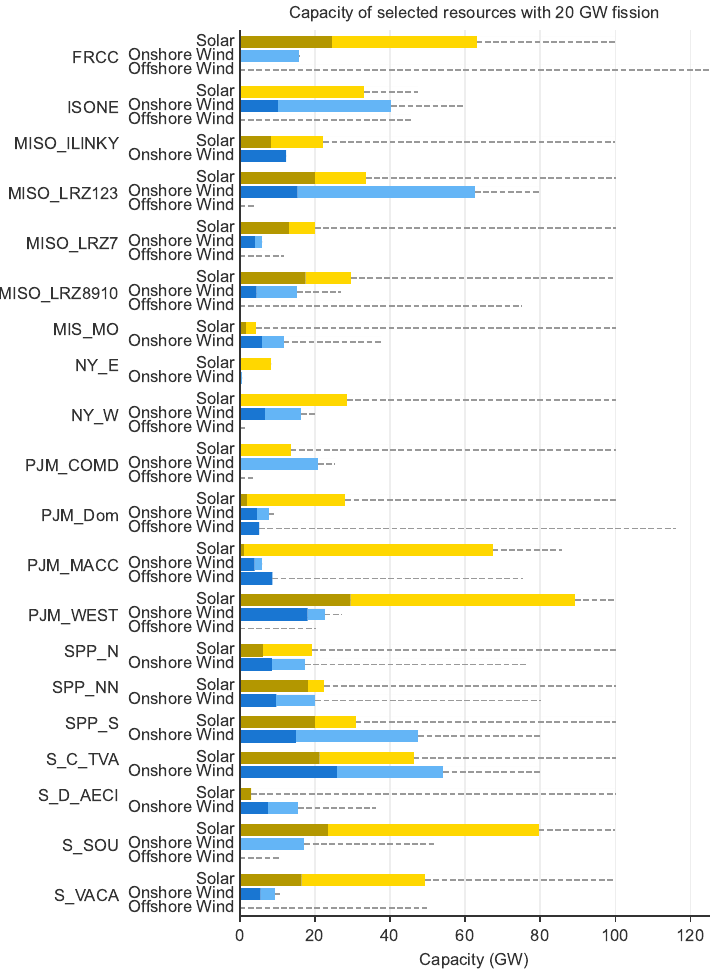}
\caption{Solar, onshore wind, and offshore wind capacity by zone in the reference scenario. Each bar shows the existing capacity in 2035 (darker, hashed), the capacity built 2035--2050 (solid), and the remaining capacity up to the maximum which could be built (dashed line).
 }\label{fig:buildout_clusters_thermal_r_20}
\end{center}
\end{figure}

\newpage
\begin{figure}[H]
\begin{center}
\includegraphics[width=1.0\textwidth]{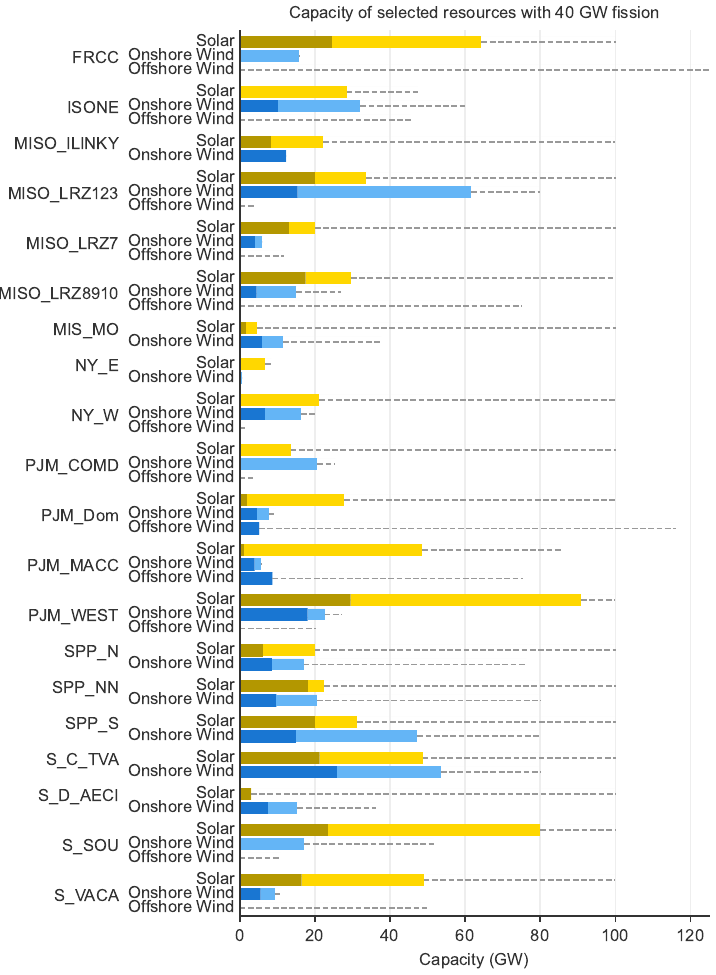}
\caption{Solar, onshore wind, and offshore wind capacity by zone in the reference scenario. Each bar shows the existing capacity in 2035 (darker, hashed), the capacity built 2035--2050 (solid), and the remaining capacity up to the maximum which could be built (dashed line).
 }\label{fig:buildout_clusters_thermal_r_40}
\end{center}
\end{figure}

\newpage
\begin{figure}[H]
\begin{center}
\includegraphics[width=1.0\textwidth]{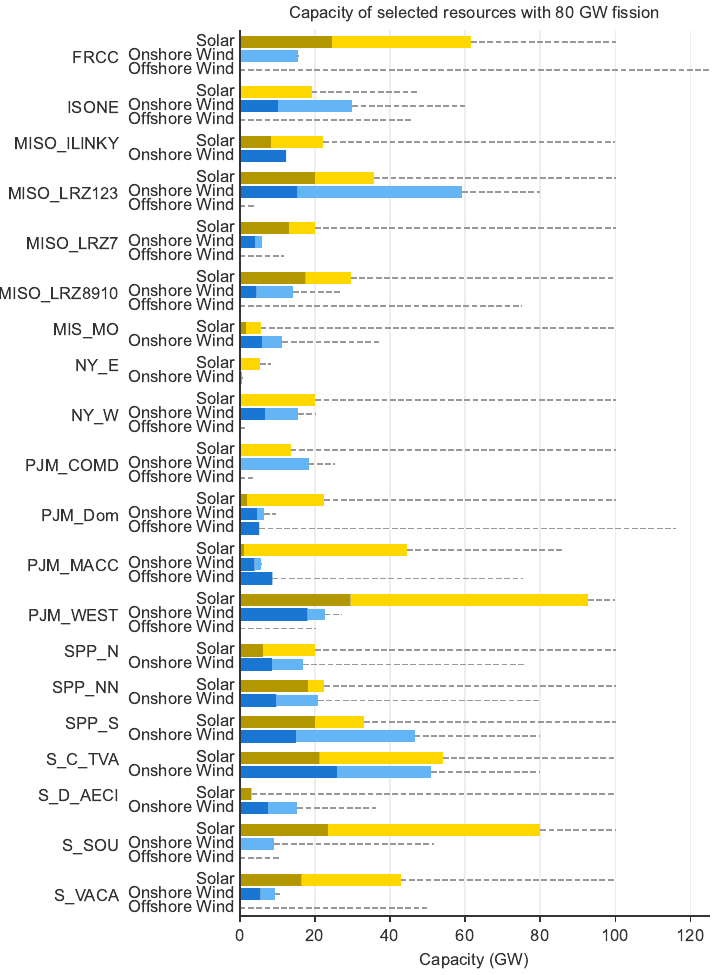}
\caption{Solar, onshore wind, and offshore wind capacity by zone in the reference scenario. Each bar shows the existing capacity in 2035 (darker, hashed), the capacity built 2035--2050 (solid), and the remaining capacity up to the maximum which could be built (dashed line).
 }\label{fig:buildout_clusters_thermal_r_80}
\end{center}
\end{figure}

\newpage
\begin{figure}[H]
\begin{center}
\includegraphics[width=1.0\textwidth]{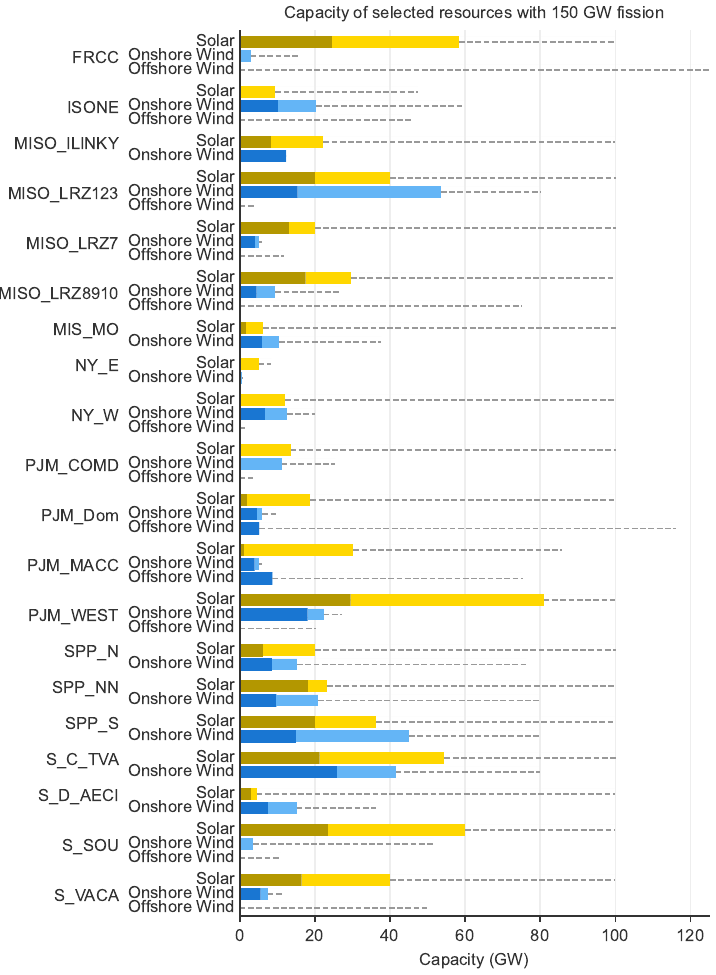}
\caption{Solar, onshore wind, and offshore wind capacity by zone in the reference scenario. Each bar shows the existing capacity in 2035 (darker, hashed), the capacity built 2035--2050 (solid), and the remaining capacity up to the maximum which could be built (dashed line).
 }\label{fig:buildout_clusters_thermal_r_150}
\end{center}
\end{figure}

\newpage
\begin{figure}[H]
\begin{center}
\includegraphics[width=1.0\textwidth]{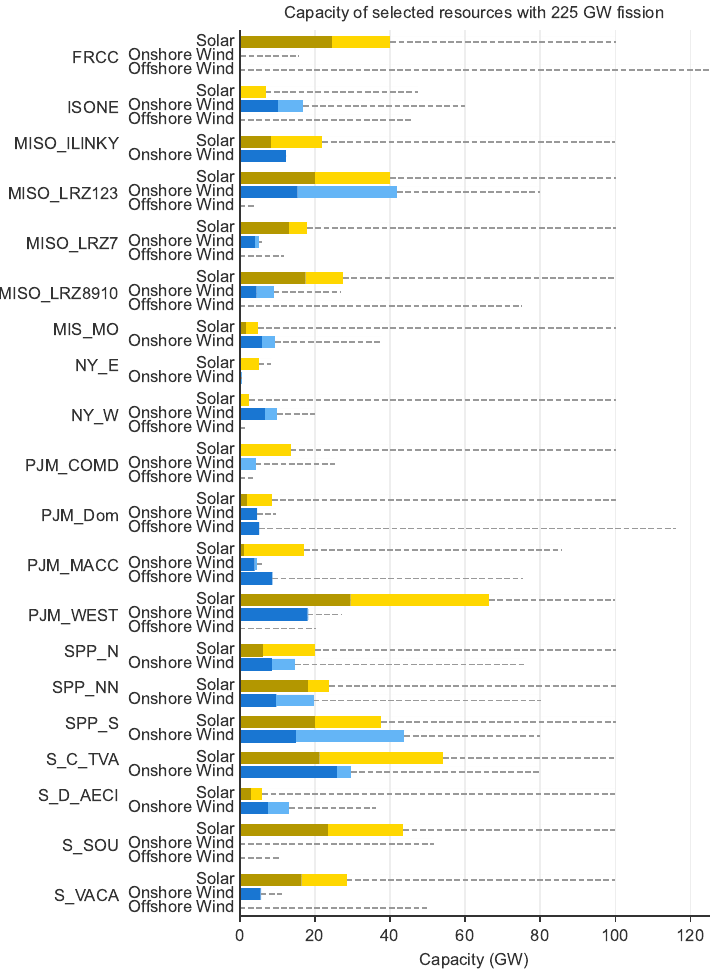}
\caption{Solar, onshore wind, and offshore wind capacity by zone in the reference scenario. Each bar shows the existing capacity in 2035 (darker, hashed), the capacity built 2035--2050 (solid), and the remaining capacity up to the maximum which could be built (dashed line).
 }\label{fig:buildout_clusters_thermal_r_225}
\end{center}
\end{figure}

\newpage
\begin{figure}[H]
\begin{center}
\includegraphics[width=1.0\textwidth]{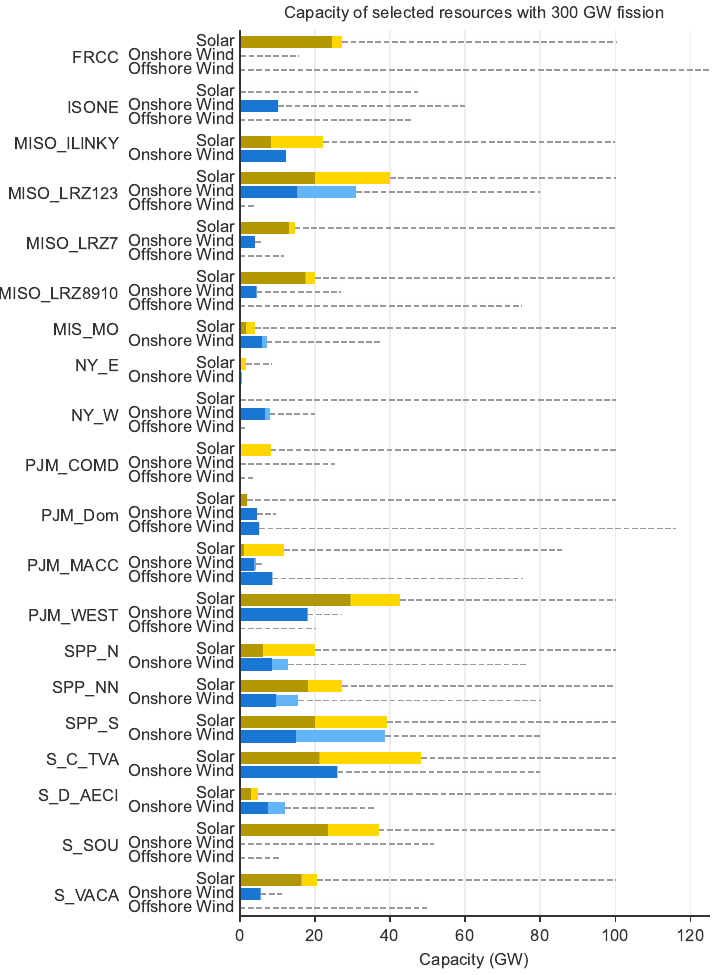}
\caption{Solar, onshore wind, and offshore wind capacity by zone in the reference scenario. Each bar shows the existing capacity in 2035 (darker, hashed), the capacity built 2035--2050 (solid), and the remaining capacity up to the maximum which could be built (dashed line).
 }\label{fig:buildout_clusters_thermal_r_300}
\end{center}
\end{figure}

\end{document}